\documentclass[twocolumn,astrosymb]{aastex62}

\PassOptionsToPackage{pdftex}{graphicx}
\usepackage{morefloats}
\usepackage{epstopdf}
\usepackage{amsmath}
\usepackage{amssymb}
\usepackage{latexsym}
\usepackage{multirow}
\usepackage{tensor}
\usepackage{enumitem}
\PassOptionsToPackage{pdfpagelabels}{hyperref}
\hypersetup{linkcolor=red,citecolor=blue,filecolor=cyan,urlcolor=teal}
\received{\today}
\revised{N/A}
\accepted{N/A}
\submitjournal{ApJS}
\shorttitle{ELECTRON PARTITION STATISTICS}
\shortauthors{Wilson III et al.}
\makeatletter
\renewcommand\@makefnmark{\hbox{\@textsuperscript{\normalfont\color{violet}\@thefnmark}}}
\makeatother
\newcommand{\totalnfitsall}{15,210}      
\newcommand{\totalnfitsups}{6546}        
\newcommand{\totalnfitsdns}{8664}        
\newcommand{\totalnfitslMf}{12,988}      
\newcommand{\totalnfitshMf}{2222}        
\newcommand{\totalnfitsQpe}{10,940}      
\newcommand{\totalnfitsQpa}{4270}        
\newcommand{\totalchbfitsA}{10,983}      
\newcommand{\totalncoreall}{14,418}      
\newcommand{\totalncoreups}{6112}        
\newcommand{\totalncoredns}{8306}        
\newcommand{\totalncorelMf}{12,405}      
\newcommand{\totalncorehMf}{2013}        
\newcommand{\totalncoreQpe}{10,387}      
\newcommand{\totalncoreQpa}{4031}        
\newcommand{\totalncoredff}{46}          
\newcommand{\totalnhaloall}{13,660}      
\newcommand{\totalnhaloups}{5734}        
\newcommand{\totalnhalodns}{7926}        
\newcommand{\totalnhalolMf}{11,738}      
\newcommand{\totalnhalohMf}{1922}        
\newcommand{\totalnhaloQpe}{9888}        
\newcommand{\totalnhaloQpa}{3772}        
\newcommand{\totalnbeamall}{11,578}      
\newcommand{\totalnbeamups}{4977}        
\newcommand{\totalnbeamdns}{6601}        
\newcommand{\totalnbeamlMf}{10,006}      
\newcommand{\totalnbeamhMf}{1572}        
\newcommand{\totalnbeamQpe}{8353}        
\newcommand{\totalnbeamQpa}{3225}        
\newcommand{\totalnbeamdff}{2145}        
\begin{document}

\title{Electron energy partition across interplanetary shocks: II.  Statistics}
\correspondingauthor{L.B. Wilson III}
\email{lynn.b.wilsoniii@gmail.com}

\author[0000-0002-4313-1970]{Lynn B. Wilson III}
\affiliation{NASA Goddard Space Flight Center, Heliophysics Science Division, Greenbelt, MD, USA.}

\author[0000-0002-4768-189X]{Li-Jen Chen}
\affiliation{NASA Goddard Space Flight Center, Heliophysics Science Division, Greenbelt, MD, USA.}

\author[0000-0002-6783-7759]{Shan Wang}
\affiliation{Astronomy Department, University of Maryland, College Park, Maryland, USA.}
\affiliation{NASA Goddard Space Flight Center, Heliophysics Science Division, Greenbelt, MD, USA.}

\author[0000-0003-0682-2753]{Steven J. Schwartz}
\affiliation{Laboratory for Atmospheric and Space Physics, University of Colorado, Boulder, Boulder, CO, USA.}

\author[0000-0002-2425-7818]{Drew L. Turner}
\affiliation{Space Sciences Department, The Aerospace Corporation, El Segundo, CA, USA.}

\author[0000-0002-7728-0085]{Michael L. Stevens}
\affiliation{Harvard-Smithsonian Center for Astrophysics, Harvard University, Cambridge, MA, USA.}

\author[0000-0002-7077-930X]{Justin C. Kasper}
\affiliation{University of Michigan, Ann Arbor, School of Climate and Space Sciences and Engineering, Ann Arbor, MI, USA.}

\author[0000-0003-2555-5953]{Adnane Osmane}
\affiliation{Department of Physics, University of Helsinki, Helsinki, Finland.}

\author[0000-0003-0939-8775]{Damiano Caprioli}
\affiliation{Department of Astronomy and Astrophysics, University of Chicago, Chicago, IL, USA.}

\author[0000-0002-1989-3596]{Stuart D. Bale}
\affiliation{Physics Department, University of California, Berkeley, CA 94720-7300, USA.}
\affiliation{Space Sciences Laboratory, University of California, Berkeley, CA 94720-7450, USA.}
\affiliation{The Blackett Laboratory, Imperial College London, London, SW7 2AZ, UK.}
\affiliation{School of Physics and Astronomy, Queen Mary University of London, London E1 4NS, UK.}

\author[0000-0002-1573-7457]{Marc P. Pulupa}
\affiliation{University of California Berkeley, Space Sciences Laboratory, Berkeley, CA, USA.}

\author[0000-0002-6536-1531]{Chadi S. Salem}
\affiliation{University of California Berkeley, Space Sciences Laboratory, Berkeley, CA, USA.}

\author[0000-0002-4288-5084]{Katherine A. Goodrich}
\affiliation{University of California Berkeley, Space Sciences Laboratory, Berkeley, CA, USA.}

\begin{abstract}
  A statistical analysis of \totalnfitsall~electron velocity distribution function (VDF) fits, observed within $\pm$2 hours of 52 interplanetary (IP) shocks by the \emph{Wind} spacecraft near 1 AU, is presented.  This is the second in a three-part series on electron VDFs near IP shocks.  The electron velocity moment statistics for the dense, low energy core, tenuous, hot halo, and field-aligned beam/strahl are a statistically significant list of values illustrated with both histograms and tabular lists for reference and baselines in future work.  The beam/strahl fit results in the upstream are currently the closest thing to a proper parameterization of the beam/strahl electron velocity moments in the ambient solar wind.  This work will also serve as a 1 AU baseline and reference for missions like \emph{Parker Solar Probe} and \emph{Solar Orbiter}.  The median density, temperature, beta, and temperature anisotropy values for the core(halo)[beam/strahl] components, with subscripts $ec$($eh$)[$eb$], of all fit results respectively are $n{\scriptstyle_{ec(h)[b]}}$ $\sim$ 11.3(0.36)[0.17] $cm^{-3}$, $T{\scriptstyle_{ec(h)[b], tot}}$  $\sim$ 14.6(48.4)[40.2] $eV$, $\beta{\scriptstyle_{ec(h)[b], tot}}$ $\sim$ 0.93(0.11)[0.05], and $\mathcal{A}{\scriptstyle_{ec(h)[b]}}$ $\sim$ 0.98(1.03)[0.93].  The nuanced details of the fitting method and data product description were published in Paper I and the detailed analysis of the results will be shown in Paper III.
\end{abstract}

\keywords{plasmas --- 
shock waves --- (Sun:) solar wind --- Sun: coronal mass ejections (CMEs)}

\phantomsection   
\section{Background and Motivation}  \label{sec:introduction}

\indent  The solar wind is a non-equilibrium, collisionless (or weakly collisional), ionized, kinetic gas that propagates away from the sun at supersonic speeds \citep[e.g.,][and references therein]{kasper06a, wilsoniii18b}.  The collisionless nature of the solar wind allows for anisotropic, non-Maxwellian, multi-component velocity distribution functions (VDFs) to exist for periods long enough to be observed by in situ spacecraft \citep[e.g.,][]{feldman75a, feldman78b, feldman79b, horaites18a, lin98a, phillips89a, phillips89b, scudder13a, stverak08a, stverak09a, wang12a, wicks16a}.  The consistent, though not ubiquitous, electron heat flux is evidence that the solar wind is not in thermodynamic equilibrium and the temperatures of species $s'$ and $s$ are not equal, i.e., $\left(T{\scriptstyle_{s'}}/T{\scriptstyle_{s}}\right){\scriptstyle_{tot}}$ $\neq$ 1, for $s'$ $\neq$ $s$ (see Appendix \ref{app:Definitions} for parameter definitions).  The temperature difference among particle species is consistently satisfied, which shows the solar wind is rarely in thermal equilibrium as well \citep[e.g.,][and references therein]{bame79a, feldman73b, feldman75a, feldman78b, feldman79b, kasper12a, kasper13a, maruca11a, maruca13a, pilipp90a, skoug00a, wilsoniii18b}.  Further, the recent observational evidence of inelastic collisions \citep[][]{wilsoniii19a}, which had been tangentially discussed under different circumstances in previous theoretical work \citep[e.g.,][]{scudder79a}, adds further evidence that the solar wind plasma is not in equilibrium.

\indent  The weakly collisional nature of the solar wind originally posed an issue as to whether shock waves could exist in such a medium \citep[e.g.,][]{coroniti70b, kellogg62a, petschek58, sagdeev66}.  The subsequent observations of a shock-like boundary upstream of the Earth's magnetosphere showed that the ramp thickness -- the spatial gradient scale length of the magnetic transition region -- is often a few $\lambda{\scriptstyle_{e}}$ up to $\lambda{\scriptstyle_{p}}$ \citep[e.g.,][and references therein]{hobara10a, mazelle10a}.  The collisional mean free path of a typical proton near Earth is roughly 1 astronomical unit (AU) whereas the typical corresponding thermal gyroradii ($\rho{\scriptstyle_{cp}}$) and/or inertial length ($\lambda{\scriptstyle_{p}}$) tend to satisfy $\sim$50--150 km \citep[e.g.,][and references therein]{wilsoniii18b}.  Thus, the shock ramp thickness is orders of magnitude smaller than the collisional mean free path which is why most astrophysical shocks are called collisionless.

\indent  The total distribution response -- characterized by velocity moments -- to a collisionless shock is often misleading \citep[e.g.,][]{wilsoniii13a} and not well correlated with any of the observable macroscopic shock parameters \citep[e.g.,][]{wilsoniii07a} except the change in bulk flow kinetic energy and some Mach number dependence \citep[e.g.,][]{masters11a, wilsoniii09a, wilsoniii10a}.  Further, recent high resolution observations show that the evolution of the electron VDF through a collisionless shock is not a trivial, uniform inflation of the distribution, but a multi-step process that deforms and redistributes/exchanges energy between the different electron components \citep[e.g.,][]{chen18a, goodrich18c, goodrich19a}.  However, there is no known way to quantify or parameterize these nuanced changes in a systematic way to examine a statistically significant set of shock crossings.  Further, although the details of the electron VDF evolution are not entirely captured by the velocity moments of the electron components, nearly all theories describing the evolution of electron VDFs rely upon either the velocity moments or a model velocity distribution function \citep[e.g.,][]{livadiotis15a, livadiotis17b, nicolaou18a, schunk75a, schunk77a, schwartz83b, schwartz88a, shizgal18a}.

\indent  Finally, there is a dearth of statistical results for suprathermal electron velocity moments in the solar wind, especially studies that separate the electron distribution into at least the three dominant components \citep[e.g.,][]{stverak09a}:  the cold, dense core with energies $E{\scriptstyle_{ec}}$ $\lesssim$ 15 eV, the hot, tenuous halo with $E{\scriptstyle_{eh}}$ $\gtrsim$ 20 eV, and the anti-sunward, field-aligned beam called the strahl with $E{\scriptstyle_{eb}}$ $\sim$few 10s of eV.  In the presence of strong collisionless shock waves, the strahl component can be contaminated with shock-reflected electrons.  Thus, this component will be referred to as the beam/strahl component because the shock-reflected and ambient strahl electrons cannot be separated.

\indent  In this second part (Paper II) of this three-part study, the statistical analysis of the fit results to the multi-component electron VDF analysis will be discussed.  The results are summarized for the 52 IP shocks observed by the \emph{Wind} spacecraft.  The notation, symbols, and data sets used herein are the same as those in \citet[][]{wilsoniii19a} (hereafter referred to as Paper I) and \citet[][]{wilsoniii19c} (hereafter referred to as Paper III).  One of the primary purposes of Paper II is provide statistical references for the three primary electron component velocity moments.  This is especially important for the beam/strahl component, as there have been very few studies providing details about the velocity moments near 1 AU.  This work will also serve as a 1 AU baseline and reference for missions like \emph{Parker Solar Probe} and \emph{Solar Orbiter}.

\indent  This paper is outlined as follows:  Section \ref{sec:DefinitionsDataSets} introduces the datasets, statistical analysis techniques and procedures, selection criteria, and velocity moment numerical integration; Section \ref{sec:StatsResultsElectronMoments} describes the statistical results through tables and histograms of the primary velocity moments examined herein; Section \ref{sec:CoulombCollisionRates} introduces and discusses Coulomb collision estimates; Section \ref{sec:ElectronHeatFlux} introduces and discusses the electron heat flux estimates; Section \ref{sec:SummaryofStatistics} summarizes the upstream only velocity moment statistics; and Section \ref{sec:Discussion} presents the discussion and conclusions.  We also include appendices that provide additional details for the reader on the parameter definitions (Appendix \ref{app:Definitions}), numerical velocity moment integration methodology (Appendix \ref{app:IntegratedVelocityMoments}), extra statistical tables and histograms (Appendix \ref{app:ExtraStatistics}), and a literature review of previous electron VDF studies in the near-Earth solar wind (Appendix \ref{app:PreviousElectronStudies}).

\section{Data Sets and Methodology}  \label{sec:DefinitionsDataSets}

\indent  As in Paper I, all data are observed by instruments on the \emph{Wind} spacecraft \citep{harten95a} near 1 AU.  The data utilized include quasi-static magnetic field vectors ($\mathbf{B}{\scriptstyle_{o}}$) from \emph{Wind}/MFI \citep[][]{lepping95}, electron and ion velocity distribution functions (VDFs) from \emph{Wind}/3DP \citep[][]{lin95a}, and proton and alpha-particle velocity moments from the \emph{Wind}/SWE Faraday Cups \citep[][]{kasper06a, ogilvie95}.  The instrument details are described in Paper I.  Parameters described with respect to $\mathbf{B}{\scriptstyle_{o}}$ are in a field-aligned coordinate basis using a subscript $j$ to denote the parallel ($j$ $=$ $\parallel$), the perpendicular ($j$ $=$ $\perp$), and total ($j$ $=$ $tot$) directions.  All electron parameters are shown with a subscript $s$ denoting the component (or sub-population) of the entire distribution where $s$ $=$ $ec$ for the core, $s$ $=$ $eh$ for the halo, $s$ $=$ $eb$ for the beam/strahl, and $s$ $=$ $e$ for the entire distribution.  The combined or mixed parameters (e.g., $\beta{\scriptstyle_{eff, j}}$) use the subscripts $s$ $=$ $eff$ for \emph{effective} and $s$ $=$ $int$ for \emph{integrated} parameters (see Appendix \ref{app:Definitions} for definitions).

\indent  The VDF fit results are taken from additional supplemental material in the form of two ASCII files\footnote{\textit{File 1}: a fit results file containing all results used in Paper I with post-fit constraint failures set to fill values, and \textit{File 2}: a fit constraint file containing all results regardless of post-fit constraints or other disqualifying criteria} found at \url{https://doi.org/10.5281/zenodo.2875806} \citep[][]{wilsoniii19k}.  In the following, data from tables show one-variable statistics of parameters from the electron VDF fit results, found within $\pm$2 hours of 52 IP shocks found in the \emph{Wind} shock database from the Harvard Smithsonian Center for Astrophysics\footnote{\url{https://www.cfa.harvard.edu/shocks/wi\_data/}} between 1995-02-26 and 2000-02-20 (for full list of event dates and times, see pdf file included with additional supplemental material \citep[][]{wilsoniii19k}).  The IP shocks examined were limited to fast-forward shocks that had burst mode electron VDFs within the chosen time range about each shock.

\indent  The statistics shown in the tables are relative to the \totalnfitsall~VDFs examined herein, of which \totalncoreall~had stable model fits ($f^{\left( core \right)}$) for the core, \totalnhaloall~had stable model fits ($f^{\left( halo \right)}$) for the halo, and \totalnbeamall~had stable model fits ($f^{\left( beam \right)}$) for the beam/strahl.  Note that all statistics presented herein are for stable fits with a fit flag for the respective component of two or higher.  The fit flags are defined in the appendices of Paper I and are provided in \textit{File 1} of the additional supplemental material \citep[][]{wilsoniii19k}.  Note that the software allows for solutions to be found for core only, the core and halo only, or the core and beam/strahl only.  However, there are post-fit constraints and post-fit checks (e.g., examine ratio of model to data for ``spiky'' fits that are unphysical) imposed on the results that can eliminate a fit component while leaving the other two alone, thus some VDFs in \textit{File 1} can have solutions to the core and beam/strahl or halo and beam/strahl.  The post-fit constraints are 1.5 $<$ $\kappa{\scriptstyle_{eh}}$ $\leq$ 20, 1.5 $<$ $\kappa{\scriptstyle_{eb}}$ $\leq$ 20, 0 $\leq$ $n{\scriptstyle_{eh}}/n{\scriptstyle_{ec}}$ $\leq$ 0.75, 0 $\leq$ $n{\scriptstyle_{eb}}/n{\scriptstyle_{ec}}$ $\leq$ 0.50, 0.0 $\leq$ $n{\scriptstyle_{eb}}/n{\scriptstyle_{eh}}$ $\leq$ 3.0, 11.4 eV $\leq$ $T{\scriptstyle_{eh, j}}$ $\leq$ 285 eV, and 11.4 eV $\leq$ $T{\scriptstyle_{eb, j}}$ $\leq$ 285 eV.  The justification and physical reasoning behind these constraints are discussed in detail in Paper I.

\indent  During the course of analysis it was found that some of the post-fit constraints were eliminating otherwise valid fit beam/strahl results.  Therefore, the combination of \textit{File 1} and \textit{File 2} from the supplemental material \citep[][]{wilsoniii19k} were used to reintroduce valid fit component results.  These inappropriately removed fit results were found by searching for the following:
\begin{itemize}[itemsep=0pt,parsep=0pt,topsep=0pt]
  \item  Fit Flag $\leq$ 0; AND
  \item  Fit Status $\leq$ 0 in \textit{File 1} AND Fit Status $>$ 0 in \textit{File 2}; AND
  \item  $\tilde{\chi}{\scriptstyle_{eb}}^{2}$ $\leq$ 10 [from \textit{File 2}]; AND
  \item  (2 $<$ $\kappa{\scriptstyle_{eb}}$ $<$ 20) AND (18 eV $<$ $T{\scriptstyle_{eb, j}}$ $<$ 300 eV) [from \textit{File 2}]; AND
  \item  0.1\% $\leq$ $\delta \mathcal{R}$ $<$ 80\%; AND
  \item  0 $<$ $\tilde{\chi}{\scriptstyle_{tot}}^{2}$ $<$ 100.
\end{itemize}
\noindent  This resulted in an additional \totalnbeamdff~beam/strahl fits.  There were an additional \totalncoredff~core fits that had fill values for $n{\scriptstyle_{ec}}$ in \textit{File 1} despite having otherwise valid fit parameters.  Thus, the totals will differ slightly from those reported in Paper I.

\indent  The following selection criteria were also defined, while still requiring the fit flag lower bound of two, to further differentiate the fit results as:
\begin{itemize}[itemsep=0pt,parsep=0pt,topsep=0pt]
  \item[] \textit{Criteria AT:} All VDFs satisfying: Fit Flag $\{c,h,b\}$ $\geq$ 2 and no violation of post-fit contraints;
  \item[] \textit{Criteria UP:} All VDFs satisfying \textit{Criteria AT} that were observed upstream of the IP shock ramp;
  \item[] \textit{Criteria DN:} All VDFs satisfying \textit{Criteria AT} that were observed downstream of the IP shock ramp;
  \item[] \textit{Criteria LM:} All VDFs satisfying \textit{Criteria AT} that were observed near IP shocks satisfying $\langle M{\scriptstyle_{f}} \rangle{\scriptstyle_{up}}$ $<$ 3;
  \item[] \textit{Criteria HM:} All VDFs satisfying \textit{Criteria AT} that were observed near IP shocks satisfying $\langle M{\scriptstyle_{f}} \rangle{\scriptstyle_{up}}$ $\geq$ 3;
  \item[] \textit{Criteria PE:} All VDFs satisfying \textit{Criteria AT} that were observed near IP shocks satisfying $\theta{\scriptstyle_{Bn}}$ $>$ 45$^{\circ}$; and
  \item[] \textit{Criteria PA:} All VDFs satisfying \textit{Criteria AT} that were observed near IP shocks satisfying $\theta{\scriptstyle_{Bn}}$ $\leq$ 45$^{\circ}$.
\end{itemize}
\noindent  The total number of VDFs for each criteria for each component type (e.g., core) are shown in Table \ref{tab:StatsStableFits} for reference.  Note that unlike the Earth's bow shock, most quasi-parallel IP shocks exhibit a much more well defined separation between upstream and downstream.  Thus, \textit{Criteria UP} and \textit{Criteria DN} are still distinguishable and valid for the IP shocks examined herein.  Despite the shock parameter-dependent selection criteria, the purpose of this work is not to analyze the effects of the shocks on the components.  These types of changes and dependencies will be presented in Paper III and are beyond the scope of this work.

\indent  The total/entire electron model VDF, $f{\scriptstyle_{s}}^{\left( mod \right)}$ $=$ $f^{\left( core \right)}$ $+$ $f^{\left( halo \right)}$ $+$ $f^{\left( beam \right)}$, is used to define integrated velocity moments such as the parallel electron heat flux, $q{\scriptstyle_{e, \parallel}}$, where the integration is performed using the Simpson's $\tfrac{1}{3}$ Rule algorithm.  These integrations are only performed on VDFs where a stable solution for all three components were found and satisfying selection criteria \textit{Criteria AT}.  There are \totalchbfitsA~VDFs that satisfy these criteria (see Appendix \ref{app:IntegratedVelocityMoments} for more details).

\begin{deluxetable}{| l | c | c | c | c | c | c | c |}
  \tabletypesize{\footnotesize}    
  \tablecaption{Statistic of Stable Fits by Criteria \label{tab:StatsStableFits}}
  \tablehead{\colhead{Type} & \colhead{AT} & \colhead{UP} & \colhead{DN} & \colhead{LM} & \colhead{HM} & \colhead{PE} & \colhead{PA}}
  \startdata
  \hline
  All  & \totalnfitsall & \totalnfitsups & \totalnfitsdns & \totalnfitslMf & \totalnfitshMf & \totalnfitsQpe & \totalnfitsQpa  \\
  Core & \totalncoreall & \totalncoreups & \totalncoredns & \totalncorelMf & \totalncorehMf & \totalncoreQpe & \totalncoreQpa  \\
  Halo & \totalnhaloall & \totalnhaloups & \totalnhalodns & \totalnhalolMf & \totalnhalohMf & \totalnhaloQpe & \totalnhaloQpa  \\
  Beam & \totalnbeamall & \totalnbeamups & \totalnbeamdns & \totalnbeamlMf & \totalnbeamhMf & \totalnbeamQpe & \totalnbeamQpa  \\
  \enddata
  \tablecomments{For symbol definitions, see Appendix \ref{app:Definitions}.}
\end{deluxetable}

\indent  In the following one-variable statistics and histogram distributions of $T{\scriptstyle_{s, j}}$, $n{\scriptstyle_{es}}$, $n{\scriptstyle_{es}} / n{\scriptstyle_{es'}}$, $\beta{\scriptstyle_{s, j}}$, $\left(T{\scriptstyle_{s'}}/T{\scriptstyle_{s}}\right){\scriptstyle_{j}}$, and $\left(T{\scriptstyle_{\perp}}/T{\scriptstyle_{\parallel}}\right){\scriptstyle_{es}}$ are presented (see Appendix \ref{app:Definitions} for parameter symbol definitions).  The minimum ($X{\scriptstyle_{min}}$), maximum ($X{\scriptstyle_{max}}$), mean ($\bar{X}$), median ($\tilde{X}$), lower quartile ($X{\scriptstyle_{25\%}}$), and upper quartile ($X{\scriptstyle_{75\%}}$) are presented in Tables \ref{tab:Temperatures}, \ref{tab:Density}, \ref{tab:Betas}, \ref{tab:TempRatios}, and \ref{tab:TempAnisotropies}.

\indent  The histograms shown Figures \ref{fig:Temperatures}, \ref{fig:Density}, \ref{fig:Betas}, \ref{fig:TempRatios}, and \ref{fig:TempAnisotropies} present the number of events normalized to the number of finite values for that parameter for the specified selection criteria (e.g., \textit{Criteria AT}).  In some histograms, one or more of the parameters are shown with multiplicative offsets to reduce the range of the horizontal axis.  All histograms were computed in linear space with uniform bin sizes for each parameter within any given panel.  In some of the histograms, isolated peaks appear that should be regarded with caution rather than as having a physically significant interpretation.  Some of these peaks arise because fit solutions contain results that lie on the boundary of an imposed constraint.  For a full list of limits and constraints, see ASCII files provided in the additional supplemental material \citep[][]{wilsoniii19k}.  Again, the justification and physical reasoning for imposing such constraints are explained in detail in Paper I.

\phantomsection   
\section{Statistics of Electron Moments}  \label{sec:StatsResultsElectronMoments}

\indent  In this section, the statistics of the electron velocity moments are presented in both tables of one-variable statistics and as histogram distributions.

\phantomsection   
\subsection{Electron Temperatures}  \label{subsec:ElectronTempsNew}

\indent  In this section one-variable statistics and distributions of $T{\scriptstyle_{s, j}}$ are introduced and discussed, for the core ($s$ $=$ $ec$), halo ($s$ $=$ $eh$), beam/strahl ($s$ $=$ $eb$), entire effective ($s$ $=$ $eff$), and entire integrated ($s$ $=$ $int$) distribution.  The solar wind is a non-equilibrium, weakly collisional, kinetic gas, thus the average kinetic energy in the species bulk flow rest frame more accurately describes the species temperature than a thermodynamic variable.  Therefore, the temperatures are shown in units of $eV$ rather than Kelvin.

\begin{figure}
  \centering
    {\includegraphics[trim = 0mm 0mm 0mm 0mm, clip, height=90mm]{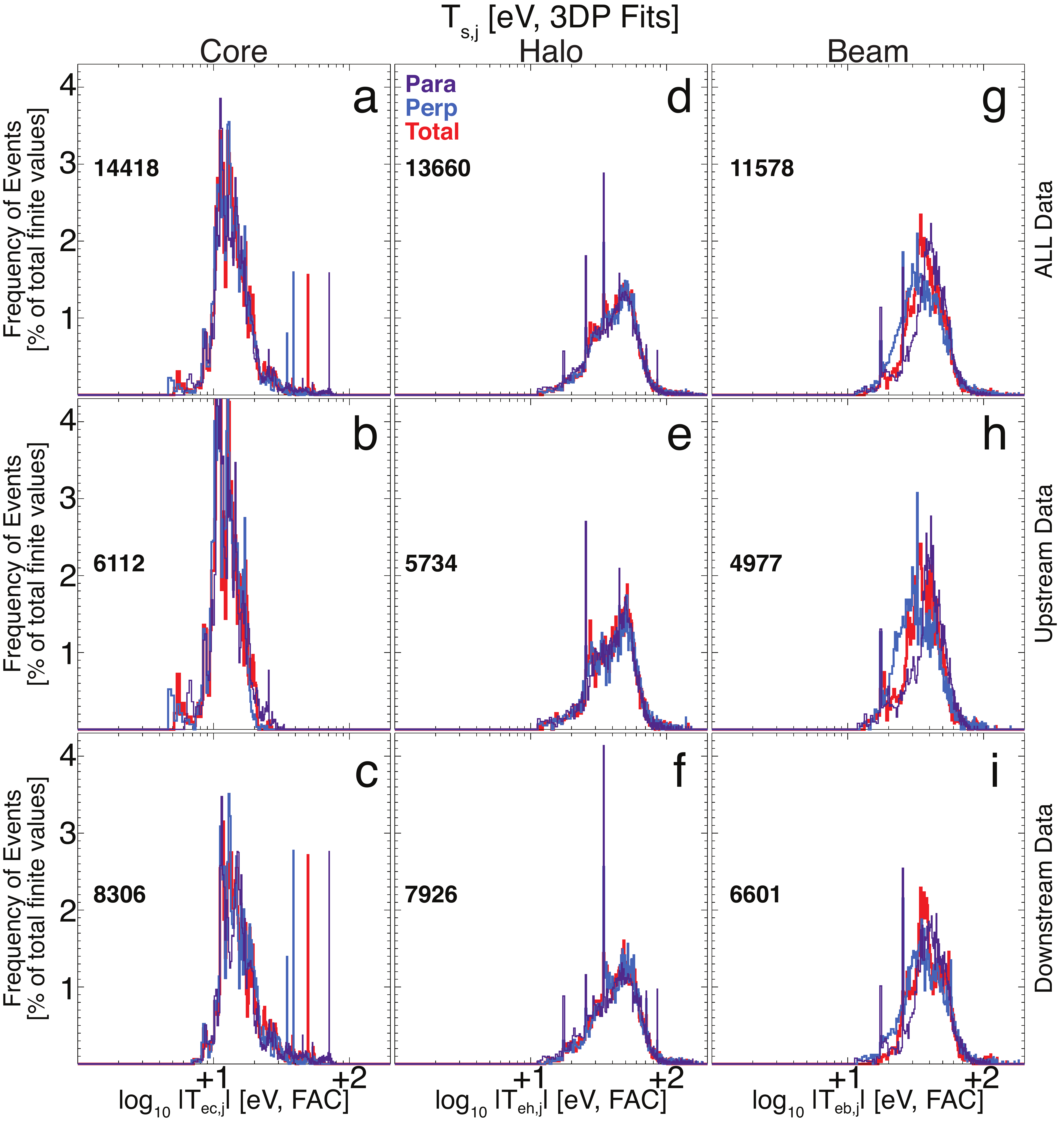}}
    \caption{Temperatures [eV] for different electron components in each column and for the different regions (i.e., rows) listed in Table \ref{tab:Temperatures}.  In each panel, there are three color-coded histograms for the different field-aligned components defined as follows: total (red); parallel (violet); and perpendicular (blue).  All histograms are normalized to the total number of finite points (i.e., black number in each panel) per parameter per component shown as a percentage.}
    \label{fig:Temperatures}
\end{figure}

\indent  Table \ref{tab:Temperatures} shows the one-variable statistics for $T{\scriptstyle_{s, j}}$ for \textit{Criteria AT} only.  Figure \ref{fig:Temperatures} shows the histograms of $T{\scriptstyle_{s, j}}$ for all time periods, upstream only, and downstream only.  For other selection criteria, Appendix \ref{app:ExtraStatistics} provides Table \ref{tab:ExtraTemperatures} and Figure \ref{fig:ExtraTemperatures}.

\startlongtable  
\begin{deluxetable}{| l | c | c | c | c | c | c |}
  \tabletypesize{\footnotesize}    
  \tablecaption{Temperature Parameters \label{tab:Temperatures}}
  \tablehead{\colhead{Temp. [eV]} & \colhead{$X{\scriptstyle_{min}}$}\tablenotemark{a} & \colhead{$X{\scriptstyle_{max}}$}\tablenotemark{b} & \colhead{$\bar{X}$}\tablenotemark{c} & \colhead{$\tilde{X}$}\tablenotemark{d} & \colhead{$X{\scriptstyle_{25\%}}$}\tablenotemark{e} & \colhead{$X{\scriptstyle_{75\%}}$}\tablenotemark{f}}
  \startdata
  \multicolumn{7}{ |c| }{\textit{Criteria AT: \totalnfitsall~VDFs}} \\
  \hline
  $T{\scriptstyle_{ec, \parallel}}$  & 5.67 & 89.1 & 19.1 & 15.0 & 12.1 & 19.1  \\
  $T{\scriptstyle_{ec, \perp}}$      & 4.75 & 62.8 & 16.4 & 14.5 & 12.0 & 17.9  \\
  $T{\scriptstyle_{ec, tot}}$        & 5.06 & 67.2 & 17.3 & 14.6 & 12.0 & 18.6  \\
  \hline
  $T{\scriptstyle_{eh, \parallel}}$  & 11.6 &  249 & 49.0 & 47.3 & 35.7 & 57.7  \\
  $T{\scriptstyle_{eh, \perp}}$      & 11.4 &  255 & 50.7 & 48.7 & 37.2 & 58.6  \\
  $T{\scriptstyle_{eh, tot}}$        & 11.6 &  222 & 50.2 & 48.4 & 37.4 & 58.1  \\
  \hline
  $T{\scriptstyle_{eb, \parallel}}$  & 11.5 & 280 & 44.2 & 42.8 & 36.2 & 51.4  \\
  $T{\scriptstyle_{eb, \perp}}$      & 11.7 & 277 & 42.6 & 39.2 & 30.6 & 50.0  \\
  $T{\scriptstyle_{eb, tot}}$        & 12.3 & 269 & 43.1 & 40.2 & 33.7 & 50.0  \\
  \hline
  $T{\scriptstyle_{eff, \parallel}}$ & 6.97 & 167 & 20.6 & 16.5 & 13.6 & 21.1  \\
  $T{\scriptstyle_{eff, \perp}}$     & 4.93 & 170 & 18.1 & 15.8 & 13.4 & 19.8  \\
  $T{\scriptstyle_{eff, tot}}$       & 5.61 & 169 & 18.9 & 16.0 & 13.6 & 20.5  \\
  \hline
  $T{\scriptstyle_{int, \parallel}}$ & 8.78 & 79.9 & 20.6 & 17.2 & 14.0 & 22.0  \\
  $T{\scriptstyle_{int, \perp}}$     & 8.09 & 69.6 & 17.2 & 15.6 & 13.4 & 19.6  \\
  $T{\scriptstyle_{int, tot}}$       & 8.41 & 69.1 & 18.3 & 16.1 & 13.7 & 20.4  \\
  \hline
  \enddata
  \tablenotetext{a}{minimum}
  \tablenotetext{b}{maximum}
  \tablenotetext{c}{mean}
  \tablenotetext{d}{median}
  \tablenotetext{e}{lower quartile}
  \tablenotetext{f}{upper quartile}
  \tablecomments{For symbol definitions, see Appendix \ref{app:Definitions}.}
\vspace{-20pt}
\end{deluxetable}

\indent  First note that the $T{\scriptstyle_{eff, j}}$ values in Tables \ref{tab:Temperatures} and \ref{tab:ExtraTemperatures} were computed using Equation \ref{eq:params_0} in Appendix \ref{app:Definitions}.  The same one-variable statistics for the integrated electron temperatures (see Appendix \ref{app:IntegratedVelocityMoments} for details), $T{\scriptstyle_{int, j}}$, are shown below $T{\scriptstyle_{eff, j}}$ in Table \ref{tab:Temperatures}.  The integrated temperature one-variable statistics are all within a few percent of the effective values, except $X{\scriptstyle_{min}}$ and $X{\scriptstyle_{max}}$, as further evidenced by the statistical differences illustrated in Appendix \ref{app:IntegratedVelocityMoments}.  Thus, while the effective temperatures calculated from the fit results statistically represent the true temperature of the total VDF, the component values are of more interest as particle dynamics are intrinsically energy and pitch-angle dependent.

\indent  The $T{\scriptstyle_{ec, j}}$ values change across the shock, which is expected since shocks heat and compress the media through which they propagate.  The magnitude of the changes are most dramatic on the higher temperature end of the histograms shown in Figures \ref{fig:Temperatures} and \ref{fig:ExtraTemperatures}, which have significant high end tails for every temperature component except for \textit{Criteria UP}.  In fact, the profile of the \textit{Criteria UP} histogram in Figure \ref{fig:Temperatures} is similar to that of the total electron temperature in the solar wind reported in \citet[][]{wilsoniii18b}.  Thus, the \textit{Criteria UP} core parameters appear to be consistent with the ambient solar wind on a statistical basis.

\indent  The three large spikes in the $T{\scriptstyle_{ec, j}}$ histograms in Figure \ref{fig:Temperatures}c are entirely due to the following selection criteria \textit{Criteria DN}, \textit{Criteria LM}, and \textit{Criteria PE}.  That is, they appear downstream of low Mach number, quasi-perpendicular shocks.  However, the tail itself on top of which these spikes are superposed is present in the downstream of all shock types, but dominated by low and high Mach number, quasi-perpendicular shocks.  That is, quasi-parallel shocks seem to be limited in generating large downstream core temperatures.  The small peaks to the left of the main peak in Figure \ref{fig:Temperatures}a are isolated to \textit{Criteria UP}, \textit{Criteria LM}, and \textit{Criteria PE} shocks, as shown in Figures \ref{fig:Temperatures}b and \ref{fig:ExtraTemperatures}b and \ref{fig:ExtraTemperatures}d.

\indent  The $T{\scriptstyle_{eh, j}}$ histograms are less symmetric and show a skewness toward lower values.  The \textit{Criteria UP} histograms show a bimodal distribution that is dominated by \textit{Criteria LM} and \textit{Criteria PE} shocks.  The \textit{Criteria HM} shocks show a different bimodal distribution, i.e., peaks at different values, and generally higher values of $T{\scriptstyle_{eh, j}}$.  Interestingly, the histograms for \textit{Criteria UP} and \textit{Criteria DN} share the same higher temperature peak but the latter lacks the lower temperature peak.  This leads to the one-variable statistics values being slightly larger for \textit{Criteria DN}, but only slightly.  The biggest difference in one-variable statistics is shown in Table \ref{tab:ExtraTemperatures} between \textit{Criteria LM} and \textit{Criteria HM} shocks.  This is somewhat expected as stronger shocks are predicted to be more efficient particle accelerators and the efficiency increases with increasing particle energy \citep[e.g.,][]{caprioli14a, park15a}.  Otherwise, the halo temperature histograms and one-variable statistics are remarkably stable between the different selection criteria.

\indent  The $T{\scriptstyle_{eb, j}}$ histograms are even more stable among the selection criteria in one-variable statistics with the only clear differences occurring between \textit{Criteria PA} and \textit{Criteria PE} shocks, but it's a rather weak difference compared to other electron VDF parameters discussed in this work.  This seems to contradict a clear difference in the $T{\scriptstyle_{eb, \perp}}$ histogram profiles among the various selection criteria, which is clearly different in Figures \ref{fig:Temperatures}h and \ref{fig:Temperatures}i and Figures \ref{fig:ExtraTemperatures}l--\ref{fig:ExtraTemperatures}o.  What is likely contributing to the lower $T{\scriptstyle_{eb, \perp}}$ values in the \textit{Criteria UP} histograms is shock-reflected electrons, which are more field-aligned than the nominal solar wind strahl.  This would skew the normal anisotropy in the beam component toward lower $T{\scriptstyle_{eb, \perp}}$ and higher $T{\scriptstyle_{eb, \parallel}}$ values.  The most dramatic difference between $T{\scriptstyle_{eb, \perp}}$ and $T{\scriptstyle_{eb, \parallel}}$ histograms is for \textit{Criteria PA} shocks seen in Figure \ref{fig:ExtraTemperatures}o.  This is apparent in the one-variable statistics values in Table \ref{tab:ExtraTemperatures}.

\indent  In summary, it is difficult to diagnose the source of the differences and similarities for each electron component temperature between opposing selection criteria because the populations can change components and sometimes overlap.  For instance, upstream core electrons can become energized by a shock and enter what is modeled as the halo in the downstream.  It is not possible to distinguish between the two or track particles, obviously, but it is possible to gain a statistical basis for the partition of random kinetic energy between the three electron components examined herein.  In short, the core electrons receive the largest amount of energy across the IP shocks examined, the halo respond well to high Mach number shocks, and the beam/strahl only show clear differences between quasi-parallel and quasi-perpendicular in $T{\scriptstyle_{eb, \perp}}$ and $T{\scriptstyle_{eb, \parallel}}$.  A detailed examination of the dependencies of $T{\scriptstyle_{s, j}}$ on various macroscopic shock parameters will be presented in Paper III.

\phantomsection   
\subsection{Number Densities}  \label{subsec:ElectronDensNew}

\indent  In this section one-variable statistics and distributions of $n{\scriptstyle_{s}}$ and $n{\scriptstyle_{s}} / n{\scriptstyle_{s'}}$ are introduced and discussed, where $s$ $=$ $ec$, $eh$, $eb$, $eff$, and $int$ for the electrons and $s$ $=$ $p$ (protons), $\alpha$ (alpha-particles), and $i$ (all ions) for the ions.

\startlongtable  
\begin{deluxetable}{| l | c | c | c | c | c | c |}
  \tabletypesize{\footnotesize}    
  \tablecaption{Density Parameters \label{tab:Density}}
  \tablehead{\colhead{$n{\scriptstyle_{s}}$ [$cm^{-3}$]} & \colhead{$X{\scriptstyle_{min}}$}\tablenotemark{a} & \colhead{$X{\scriptstyle_{max}}$} & \colhead{$\bar{X}$} & \colhead{$\tilde{X}$} & \colhead{$X{\scriptstyle_{25\%}}$} & \colhead{$X{\scriptstyle_{75\%}}$}}
  \startdata
  \multicolumn{7}{ |c| }{\textit{Criteria AT: \totalnfitsall~VDFs}} \\
  \hline
  $n{\scriptstyle_{p}}$                         &    0.10 & 76.2 & 14.8 & 11.7 &  6.43 & 21.5  \\
  $n{\scriptstyle_{\alpha}}$                    &    0.02 & 4.75 & 0.45 & 0.28 &  0.13 & 0.66  \\
  $n{\scriptstyle_{i}}$                         &    0.18 & 98.8 & 15.5 & 11.5 &  7.19 & 19.9  \\
  $n{\scriptstyle_{ec}}$                        &    0.30 & 55.3 & 13.8 & 11.3 &  6.55 & 19.4  \\
  $n{\scriptstyle_{eh}}$                        &   0.002 & 6.87 & 0.51 & 0.36 &  0.20 & 0.63  \\
  $n{\scriptstyle_{eb}}$                        &  0.0009 & 3.50 & 0.24 & 0.17 &  0.10 & 0.29  \\
  $n{\scriptstyle_{eff}}$                       &   0.004 & 56.9 & 14.4 & 11.9 &  6.92 & 20.3  \\
  $n{\scriptstyle_{int}}$                       &    0.39 & 56.9 & 14.6 & 12.1 &  7.30 & 20.4  \\
  $n{\scriptstyle_{eh}} / n{\scriptstyle_{ec}}$ &  0.0002 & 0.30 & 0.05 & 0.03 &  0.02 & 0.06  \\
  $n{\scriptstyle_{eb}} / n{\scriptstyle_{ec}}$ & 0.00003 & 0.30 & 0.03 & 0.02 & 0.008 & 0.04  \\
  $n{\scriptstyle_{eb}} / n{\scriptstyle_{eh}}$ &   0.002 & 9.86 & 0.82 & 0.50 &  0.24 & 0.96  \\
  \hline
  \enddata
  \tablenotetext{a}{Header symbols match that of Table \ref{tab:Temperatures}}
  \tablecomments{For symbol definitions, see Appendix \ref{app:Definitions}.}
\vspace{-20pt}
\end{deluxetable}

\indent  Table \ref{tab:Density} shows the one-variable statistics for $n{\scriptstyle_{s}}$ and $n{\scriptstyle_{s}} / n{\scriptstyle_{s'}}$ (for electrons and ions) for all time periods only (see Table \ref{tab:ExtraDensity} and Figure \ref{fig:ExtraBetas} in Appendix \ref{app:ExtraStatistics} for other selection criteria).  Figure \ref{fig:ExtraDensity} shows the histograms of $n{\scriptstyle_{s}}$ (ions and electrons) and $n{\scriptstyle_{s}} / n{\scriptstyle_{s'}}$ (electrons only).

\begin{figure}
  \centering
    {\includegraphics[trim = 0mm 0mm 0mm 0mm, clip, height=90mm]{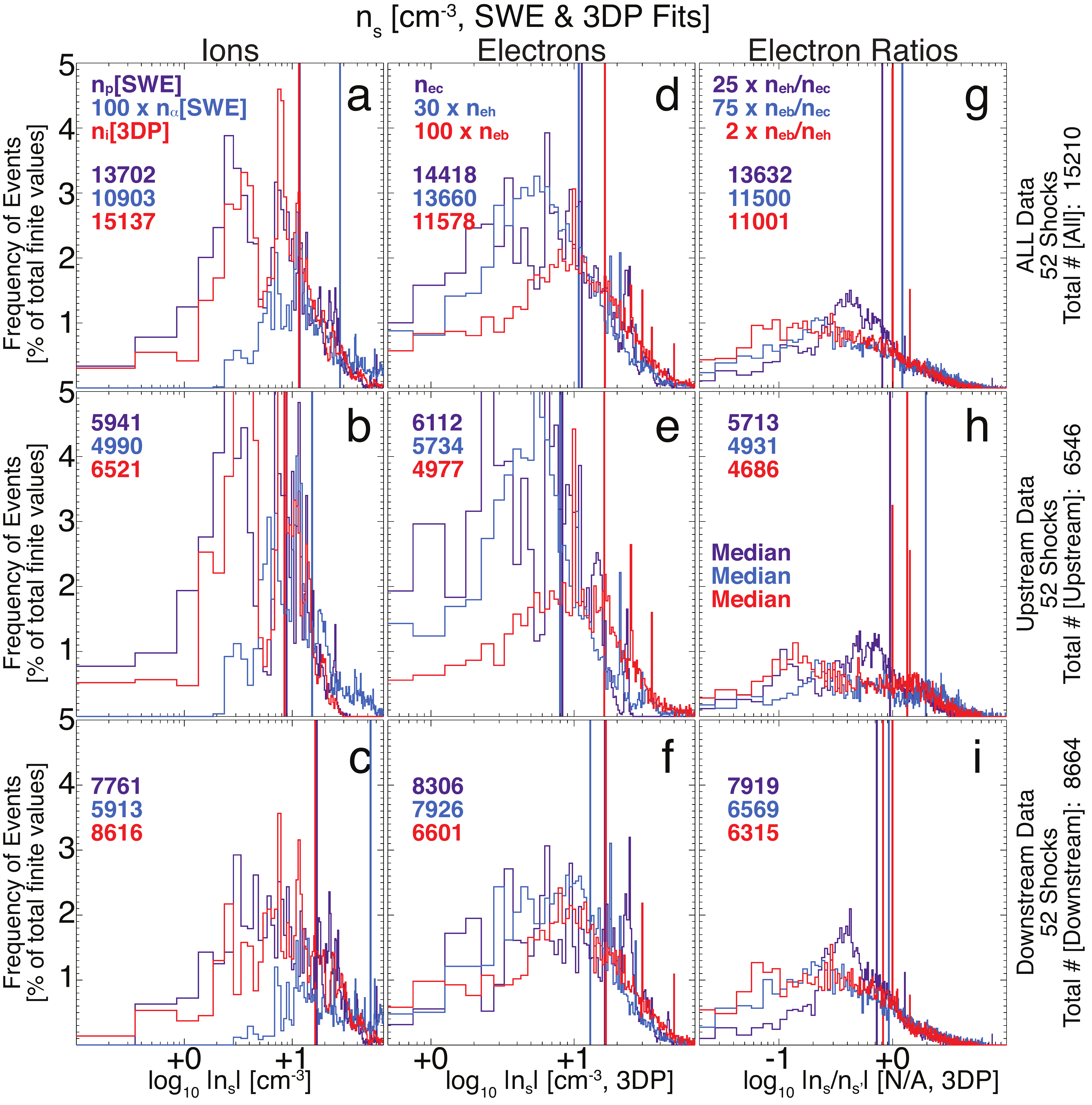}}
    \caption{Densities [$cm^{-3}$] and density ratios for different ion and electron components as a percentage of the total number of finite points (i.e., color-coded numbers in each panel).  The format is similar to Figure \ref{fig:Temperatures} with the row organization but the columns differ.  The first column here shows proton (violet) and alpha-particle (blue) density from \emph{Wind}/SWE and total ion density from \emph{Wind}/3DP (red).  The second column shows $n{\scriptstyle_{es}}$ for the core (violet), halo (blue), and beam/strahl (red) components.  The third column shows $n{\scriptstyle_{s}} / n{\scriptstyle_{s'}}$ for the halo-to-core (violet), beam-to-core (blue), and beam-to-halo (red) density ratios.  The corresponding one-variable statistics for the electron parameters are shown in Table \ref{tab:Density}.  Note that the $n{\scriptstyle_{\alpha}}$, $n{\scriptstyle_{eh}}$, $n{\scriptstyle_{eb}}$, and all three $n{\scriptstyle_{s}} / n{\scriptstyle_{s'}}$ values were offset by constant factors (shown in panels a, d, and g) to reduce the horizontal axis dynamic range.}
    \label{fig:Density}
\end{figure}

\indent  Note that the $n{\scriptstyle_{eff}}$ values in Table \ref{tab:Density} were computed by summing the fit results, i.e., $n{\scriptstyle_{eff}}$ $=$ $n{\scriptstyle_{ec}}$ $+$ $n{\scriptstyle_{eh}}$ $+$ $n{\scriptstyle_{eb}}$.  The same one-variable statistics for the integrated electron densities (see Appendix \ref{app:IntegratedVelocityMoments} for details), $n{\scriptstyle_{e, int}}$, are shown just below $n{\scriptstyle_{eff}}$ in Table \ref{tab:Density}.  As one can see, these results are consistent with the summed moment values shown in Table \ref{tab:Density}.

\indent  The ion densities in Figures \ref{fig:Density} and \ref{fig:ExtraDensity} are included as a reference, though not the focus of this work.  The histograms of $n{\scriptstyle_{p}}$ and $n{\scriptstyle_{i}}$ are both bimodal and peak at roughly the same values, showing consistency between the two independent measurements from \emph{Wind} SWE and 3DP.  The slight offset toward higher values for $n{\scriptstyle_{i}}$ results from it including the alpha-particle densities, i.e., it is the total ion number density.  Note that the peaks of $n{\scriptstyle_{p}}$ and $n{\scriptstyle_{i}}$ are both near the same values as the bimodal peaks in $n{\scriptstyle_{ec}}$, adding evidence to the accuracy of the fit results already presented in Paper I.  The ion densities are not the focus and futher discussion is beyond the scope of this work.

\indent  The $n{\scriptstyle_{ec}}$ values change across the shock ramp, as expected since a shock compresses the fluid density and the core is representative of the bulk of the electron VDF.  The magnitude of the change between \textit{Criteria UP} and \textit{Criteria DN} (Figures \ref{fig:Density}e and \ref{fig:Density}f) is consistent with those for $n{\scriptstyle_{p}}$ and $n{\scriptstyle_{i}}$ (Figures \ref{fig:Density}b and \ref{fig:Density}c) and those expected from the Rankine-Hugoniot conservation relations, within uncertainties, for each event \citep[see supplemental PDF][for list of compression ratios]{wilsoniii19k}.  The only selection criteria difference that may be somewhat surprising is that between \textit{Criteria PE} and \textit{Criteria PA} shocks.  All one-variable statistic values of $n{\scriptstyle_{ec}}$, except $X{\scriptstyle_{min}}$, are larger for \textit{Criteria PE} than \textit{Criteria PA} shocks.  This effect is clearly dominated by the \textit{Criteria DN} values as evidenced by the similar profiles in Figures \ref{fig:Density}e and \ref{fig:ExtraDensity}j and between Figures \ref{fig:Density}f and \ref{fig:ExtraDensity}i.  Again, this is not tremendously surprising as the density compression ratio for quasi-parallel shocks is lower than that for quasi-perpendicular.  In summary, the core electron densities behave as one would expect across IP shocks.

\indent  The $n{\scriptstyle_{eh}}$ values also show compression across the shock, but to a lesser extent than $n{\scriptstyle_{ec}}$.  Although the one-variable statistics for $n{\scriptstyle_{ec}}$ did not show a tremendous difference between \textit{Criteria LM} and \textit{Criteria HM} shocks, $n{\scriptstyle_{eh}}$ is clearly higher for \textit{Criteria HM} shocks.  This may result from the higher temperatures observed at \textit{Criteria HM} shocks, causing some core electrons to be included in the halo fits or it may indicate that the halo responds more to stronger shocks.  The latter is likely as stronger shocks are more efficient at accelerating particles and the efficiency increases with increasing particle energy \citep[e.g.,][]{caprioli14a, park15a}.  That is, stronger shocks produce more suprathermal electrons which result in larger $n{\scriptstyle_{eh}}$ fit values.

\indent  The $n{\scriptstyle_{eb}}$ values are effectively the same between \textit{Criteria UP} and \textit{Criteria DN} and only slightly different between \textit{Criteria PE} and \textit{Criteria PA} shocks.  The $n{\scriptstyle_{eb}}$ values do show larger values at \textit{Criteria HM} than \textit{Criteria LM} shocks, but again the differences are small compared to those for $n{\scriptstyle_{ec}}$ and $n{\scriptstyle_{eh}}$.  Thus, the beam/strahl electron densities do not seem to be strongly dependent upon any macroscopic shock parameter or upon the shock region.  This might result from their nearly field-aligned pitch-angle distribution, which reduces the effects of magnetic field gradients on their dynamics.

\indent  In summary, similar to the $T{\scriptstyle_{s, j}}$ the core shows the strongest dependence on \textit{Criteria UP} versus \textit{Criteria DN} and all other selection criteria.  The beam/strahl densities are also somewhat indifferent to the selection criteria, much like the associated temperatures with the halo showing mostly weak dependencies.

\phantomsection   
\subsection{Electron Betas}  \label{subsec:ElectronBetasNew}

\indent  In this section one-variable statistics and distributions of plasma betas, $\beta{\scriptstyle_{s, j}}$, are introduced and discussed, where $s$ $=$ $ec$, $eh$, $eb$, and $eff$ and $j$ $=$ $\parallel$ (parallel), $\perp$ (perpendicular), and $tot$ (total).

\startlongtable  
\begin{deluxetable}{| l | c | c | c | c | c | c |}
  \tabletypesize{\footnotesize}    
  \tablecaption{Electron Beta Parameters \label{tab:Betas}}
  \tablehead{\colhead{$\beta{\scriptstyle_{s,j}}$ [N/A]} & \colhead{$X{\scriptstyle_{min}}$}\tablenotemark{a} & \colhead{$X{\scriptstyle_{max}}$} & \colhead{$\bar{X}$} & \colhead{$\tilde{X}$} & \colhead{$X{\scriptstyle_{25\%}}$} & \colhead{$X{\scriptstyle_{75\%}}$}}
  \startdata
  \multicolumn{7}{ |c| }{\textit{Criteria AT: \totalnfitsall~VDFs}} \\
  \hline
  $\beta{\scriptstyle_{ec, \parallel}}$  &    0.05 & 3313 & 3.62 & 0.97 & 0.58 & 2.03  \\
  $\beta{\scriptstyle_{ec, \perp}}$      &    0.04 & 3268 & 3.51 & 0.91 & 0.49 & 2.01  \\
  $\beta{\scriptstyle_{ec, tot}}$        &    0.05 & 3283 & 3.54 & 0.93 & 0.52 & 2.01  \\
  $\beta{\scriptstyle_{eh, \parallel}}$  &  0.0001 &  375 & 0.48 & 0.10 & 0.05 & 0.22  \\
  $\beta{\scriptstyle_{eh, \perp}}$      &  0.0008 &  378 & 0.49 & 0.11 & 0.05 & 0.22  \\
  $\beta{\scriptstyle_{eh, tot}}$        &  0.0009 &  377 & 0.49 & 0.11 & 0.05 & 0.22  \\
  $\beta{\scriptstyle_{eb, \parallel}}$  & 0.00002 & 33.7 & 0.13 & 0.05 & 0.02 & 0.11  \\
  $\beta{\scriptstyle_{eb, \perp}}$      & 0.00003 & 46.4 & 0.12 & 0.05 & 0.02 & 0.10  \\
  $\beta{\scriptstyle_{eb, tot}}$        & 0.00003 & 42.0 & 0.12 & 0.05 & 0.02 & 0.10  \\
  $\beta{\scriptstyle_{eff, \parallel}}$ &  0.0009 & 3721 & 4.15 & 1.12 & 0.69 & 2.28  \\
  $\beta{\scriptstyle_{eff, \perp}}$     &  0.0010 & 3693 & 4.04 & 1.04 & 0.60 & 2.25  \\
  $\beta{\scriptstyle_{eff, tot}}$       &  0.0009 & 3702 & 4.08 & 1.06 & 0.63 & 2.26  \\
  \hline
  \enddata
  \tablenotetext{a}{Header symbols match that of Table \ref{tab:Temperatures}}
  \tablecomments{For symbol definitions, see Appendix \ref{app:Definitions}.}
\vspace{-20pt}
\end{deluxetable}

\begin{figure}
  \centering
    {\includegraphics[trim = 0mm 0mm 0mm 0mm, clip, height=90mm]{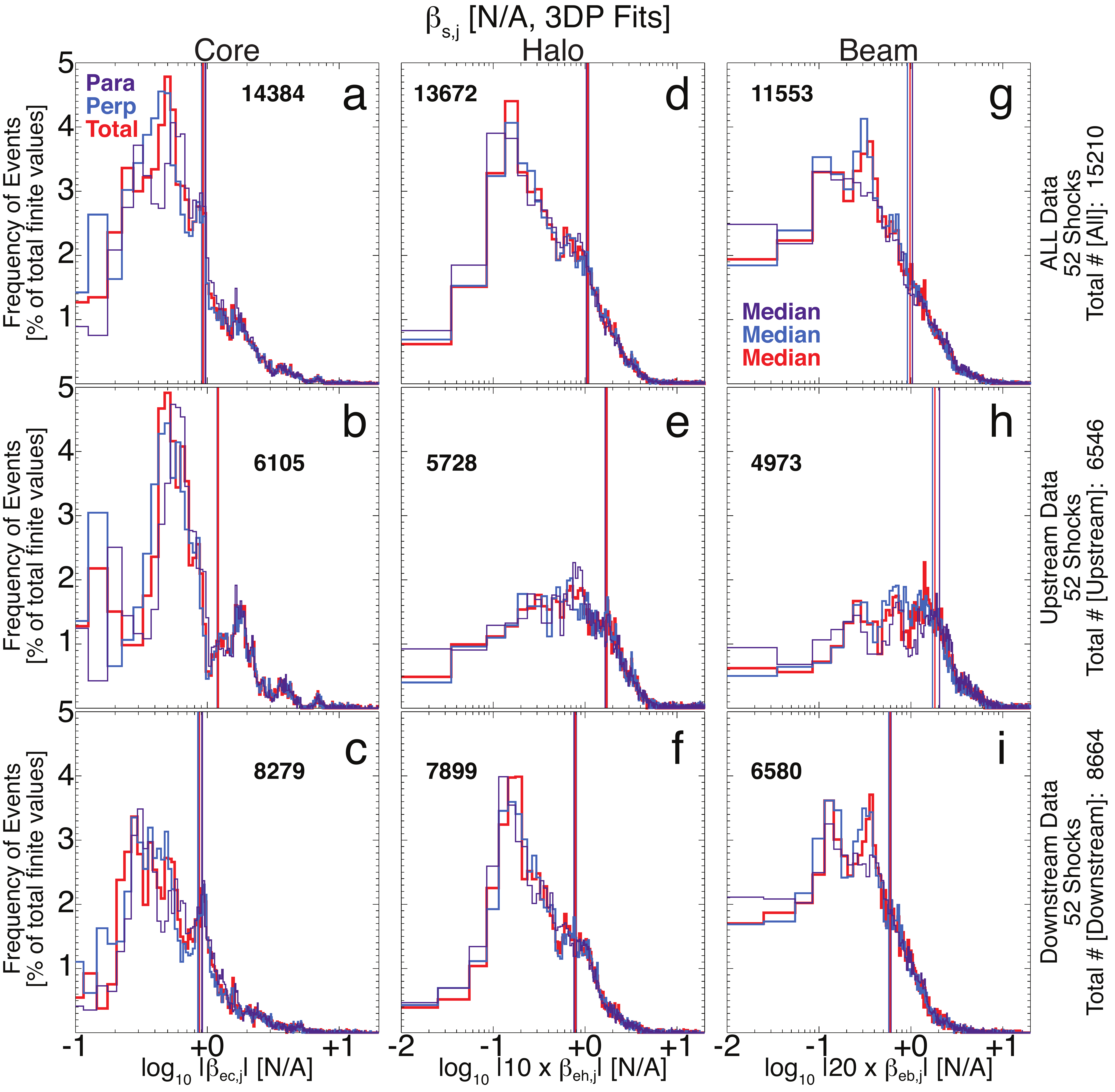}}
    \caption{The same format as Figures \ref{fig:Temperatures} and \ref{fig:Density} except for electron betas [N/A].  Note that all $\beta{\scriptstyle_{eh, j}}$ and $\beta{\scriptstyle_{eb, j}}$ values were offset by constant factors of 10 and 20, respectively, to reduce the horizontal axis dynamic range.}
    \label{fig:Betas}
\end{figure}

\indent  Table \ref{tab:Betas} shows the one-variable statistics for $\beta{\scriptstyle_{s, j}}$ for all time periods only.  Figure \ref{fig:Betas} shows the histograms of $\beta{\scriptstyle_{s, j}}$ (see Table \ref{tab:ExtraBetas} and Figure \ref{fig:ExtraBetas} in Appendix \ref{app:ExtraStatistics} for other selection criteria).

\indent  The $\beta{\scriptstyle_{ec, j}}$ values are much more stable than the $\beta{\scriptstyle_{eh, j}}$ or $\beta{\scriptstyle_{eb, j}}$ between the different selection criteria, but even so the one-variable statistic values can differ by over 100\%.  The $\beta{\scriptstyle_{ec, j}}$ histograms in Figure \ref{fig:Betas} show a bimodal distribution for selection criteria \textit{Criteria UP} and \textit{Criteria DN}, but the peaks occur at lower values for \textit{Criteria DN}.  The bimodal profile of the $\beta{\scriptstyle_{ec, j}}$ histograms for \textit{Criteria UP} occurs upstream of low Mach number, quasi-parallel shocks whereas the bimodal profile for \textit{Criteria DN} occurs downstream of high Mach number, quasi-perpendicular shocks.

\indent  Unlike $\beta{\scriptstyle_{ec, j}}$, the histograms for \textit{Criteria UP} and \textit{Criteria DN} are completely different in profile for both $\beta{\scriptstyle_{eh, j}}$ and $\beta{\scriptstyle_{eb, j}}$.  That is, the \textit{Criteria UP} histograms for both $\beta{\scriptstyle_{eh, j}}$ and $\beta{\scriptstyle_{eb, j}}$ are broad with weak peaks while the \textit{Criteria DN} histograms show similar profiles to those for selection criteria \textit{Criteria AT}.  For both suprathermal components, the \textit{Criteria UP} histograms are skewed toward higher values than the \textit{Criteria DN} histograms.  When looking at the other selection criteria histograms shown in Figure \ref{fig:ExtraBetas} (in Appendix \ref{app:ExtraStatistics}), the profile of the \textit{Criteria AT} $\beta{\scriptstyle_{s, j}}$ histograms are clearly dominated by the low Mach number and quasi-perpendicular shock results, which is likely due to the significantly larger fraction of VDFs satisfying selection criteria \textit{Criteria LM} and \textit{Criteria PE}.  However, there is no clear selection criteria differences in Figure \ref{fig:ExtraBetas} to explain the upstream/downstream histogram differences in Figure \ref{fig:Betas}.  Thus, the difference appears to solely rely upon the region of observation near the shock, not the shock strength or geometry.  Yet despite the apparent lack of dependence on the shock parameters, the one-variable statistic values can differ by over 300\% between any two opposing selection criteria for both $\beta{\scriptstyle_{eh, j}}$ and $\beta{\scriptstyle_{eb, j}}$.

\indent  Therefore, the $\beta{\scriptstyle_{ec, j}}$ values are more stable between any two opposing selection criteria than either $\beta{\scriptstyle_{eh, j}}$ or $\beta{\scriptstyle_{eb, j}}$ and both $\beta{\scriptstyle_{eh, j}}$ and $\beta{\scriptstyle_{eb, j}}$ depend upon all selection criteria.  That is, the histogram profiles and one-variable statistics can be wildly different between \textit{Criteria UP} and \textit{Criteria DN}, \textit{Criteria LM} and \textit{Criteria HM}, and \textit{Criteria PE} and \textit{Criteria PA}.  A detailed examination of the changes and dependencies in $\beta{\scriptstyle_{s, j}}$ will be explored in greater detail in Paper III and is beyond the scope of this work.

\phantomsection   
\subsection{Electron Temperature Ratios}  \label{subsec:ElectronTempRatiosNew}

\indent  In this section one-variable statistics and distributions of the electron temperature ratios (see Appendix \ref{app:Definitions} for parameter definitions) for the core ($s$ $=$ $ec$), halo ($s$ $=$ $eh$), beam/strahl ($s$ $=$ $eb$), and entire effective ($s$ $=$ $eff$) are presented.

\startlongtable  
\begin{deluxetable}{| l | c | c | c | c | c | c |}
  \tabletypesize{\footnotesize}    
  \tablecaption{Electron Temperature Ratio Parameters \label{tab:TempRatios}}
  \tablehead{\colhead{Ratio} & \colhead{$X{\scriptstyle_{min}}$}\tablenotemark{a} & \colhead{$X{\scriptstyle_{max}}$} & \colhead{$\bar{X}$} & \colhead{$\tilde{X}$} & \colhead{$X{\scriptstyle_{25\%}}$} & \colhead{$X{\scriptstyle_{75\%}}$}}
  \startdata
  \multicolumn{7}{ |c| }{\textit{Criteria AT: \totalnfitsall~VDFs}} \\
  \hline
  $\tensor*{ \mathcal{T}  }{^{eh}_{ec}}{\scriptstyle_{\parallel}}$ & 0.17 & 17.9 & 3.11 & 3.02 & 2.21 & 3.97  \\
  $\tensor*{ \mathcal{T}  }{^{eh}_{ec}}{\scriptstyle_{    \perp}}$ & 0.41 & 17.9 & 3.40 & 3.27 & 2.45 & 4.10  \\
  $\tensor*{ \mathcal{T}  }{^{eh}_{ec}}{\scriptstyle_{      tot}}$ & 0.34 & 16.3 & 3.29 & 3.20 & 2.39 & 4.04  \\
  $\tensor*{ \mathcal{T}  }{^{eb}_{ec}}{\scriptstyle_{\parallel}}$ & 0.23 & 25.7 & 2.86 & 2.85 & 2.09 & 3.54  \\
  $\tensor*{ \mathcal{T}  }{^{eb}_{ec}}{\scriptstyle_{    \perp}}$ & 0.46 & 24.6 & 2.88 & 2.69 & 2.12 & 3.38  \\
  $\tensor*{ \mathcal{T}  }{^{eb}_{ec}}{\scriptstyle_{      tot}}$ & 0.42 & 25.0 & 2.86 & 2.73 & 2.22 & 3.35  \\
  $\tensor*{ \mathcal{T}  }{^{eb}_{eh}}{\scriptstyle_{\parallel}}$ & 0.15 & 6.12 & 1.06 & 0.93 & 0.69 & 1.30  \\
  $\tensor*{ \mathcal{T}  }{^{eb}_{eh}}{\scriptstyle_{    \perp}}$ & 0.13 & 7.11 & 0.95 & 0.81 & 0.61 & 1.11  \\
  $\tensor*{ \mathcal{T}  }{^{eb}_{eh}}{\scriptstyle_{      tot}}$ & 0.17 & 6.08 & 0.97 & 0.85 & 0.66 & 1.14  \\
  $\tensor*{ \mathcal{T} }{^{eh}_{eff}}{\scriptstyle_{\parallel}}$ & 0.17 & 17.4 & 2.80 & 2.73 & 2.03 & 3.57  \\
  $\tensor*{ \mathcal{T} }{^{eh}_{eff}}{\scriptstyle_{    \perp}}$ & 0.43 & 16.8 & 3.03 & 2.95 & 2.24 & 3.69  \\
  $\tensor*{ \mathcal{T} }{^{eh}_{eff}}{\scriptstyle_{      tot}}$ & 0.37 & 15.2 & 2.94 & 2.88 & 2.17 & 3.64  \\
  $\tensor*{ \mathcal{T} }{^{eb}_{eff}}{\scriptstyle_{\parallel}}$ & 0.24 & 19.6 & 2.56 & 2.57 & 1.89 & 3.17  \\
  $\tensor*{ \mathcal{T} }{^{eb}_{eff}}{\scriptstyle_{    \perp}}$ & 0.46 & 18.7 & 2.57 & 2.37 & 1.91 & 3.02  \\
  $\tensor*{ \mathcal{T} }{^{eb}_{eff}}{\scriptstyle_{      tot}}$ & 0.43 & 19.0 & 2.57 & 2.44 & 1.99 & 3.03  \\
  \hline
  \enddata
  \tablenotetext{a}{Header symbols match that of Table \ref{tab:Temperatures}}
  \tablecomments{For symbol definitions, see Appendix \ref{app:Definitions}.}
\end{deluxetable}

\begin{figure}
  \centering
    {\includegraphics[trim = 0mm 0mm 0mm 0mm, clip, height=90mm]{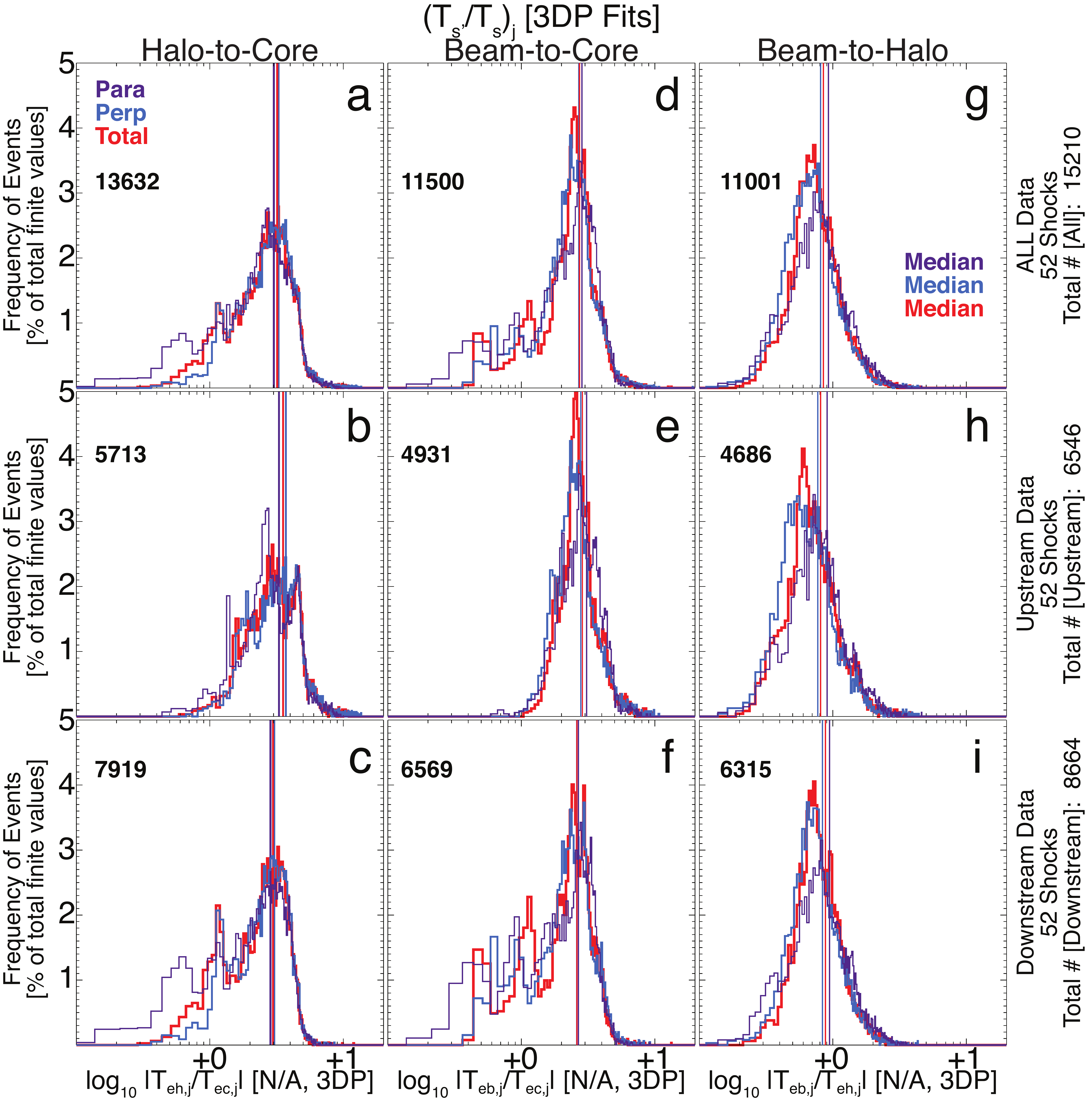}}
    \caption{The same format as Figures \ref{fig:Temperatures} and \ref{fig:Density} except for electron temperature ratios [N/A].}
    \label{fig:TempRatios}
\end{figure}

\indent  Table \ref{tab:TempRatios} shows the one-variable statistics for $\tensor*{ \mathcal{T} }{^{s'}_{s}}{\scriptstyle_{j}}$ $=$ $\left(T{\scriptstyle_{s'}}/T{\scriptstyle_{s}}\right){\scriptstyle_{j}}$ for all time periods only.  Figure \ref{fig:TempRatios} shows the histograms of $\tensor*{ \mathcal{T} }{^{s'}_{s}}{\scriptstyle_{j}}$ (see Table \ref{tab:ExtraTempRatios} and Figure \ref{fig:ExtraTempRatios} in Appendix \ref{app:ExtraStatistics} for other selection criteria).

\indent  In Figures \ref{fig:TempRatios} and \ref{fig:ExtraTempRatios} one can see that the temperature ratios dependent upon the core (i.e., first two columns) show a tail toward lower values clearly occurring in the downstream (i.e., \textit{Criteria DN}).  This is largely because the halo and beam/strahl are less dependent upon the region than the strength and geometry.  The large tails appear to be predominantly at shocks satisfying \textit{Criteria DN} and \textit{Criteria PA} (i.e., $\theta{\scriptstyle_{Bn}}$ $\leq$ 45$^{\circ}$ shocks) for $\tensor*{ \mathcal{T}  }{^{eh}_{ec}}{\scriptstyle_{j}}$.  There are tails for both \textit{Criteria LM} and \textit{Criteria HM} shocks, but are more important in \textit{Criteria HM} shocks.  Notice that $\tensor*{ \mathcal{T}  }{^{eh}_{ec}}{\scriptstyle_{\perp}}$ is bimodal but $\tensor*{ \mathcal{T}  }{^{eh}_{ec}}{\scriptstyle_{\parallel}}$ is trimodal for \textit{Criteria DN}.  This two- versus three-peak histogram form appears as well for \textit{Criteria PA}, suggesting the profile results from quasi-parallel shocks and occurs in the downstream.

\indent  Interestingly, the tails at small values for $\tensor*{ \mathcal{T}  }{^{eb}_{ec}}{\scriptstyle_{j}}$ are more nuanced.  Again, they occur in the downstream but for both \textit{Criteria LM} and \textit{Criteria HM} in addition to both \textit{Criteria PE} and \textit{Criteria PA} shocks.  The nuance is that there are clear peaks at low values for \textit{Criteria HM} and \textit{Criteria PA} shocks near $\sim$0.4--0.5 and $\sim$0.9--1.0, respectively.  For reference, the dominant peak of the histograms are up in the $\sim$1.8--3.0 range for all selection criteria for $\tensor*{ \mathcal{T}  }{^{eb}_{ec}}{\scriptstyle_{j}}$.  The $\tensor*{ \mathcal{T}  }{^{eb}_{ec}}{\scriptstyle_{\parallel}}$ histograms are bimodal for both \textit{Criteria HM} and \textit{Criteria PA} shocks.  The $\tensor*{ \mathcal{T}  }{^{eb}_{ec}}{\scriptstyle_{\perp}}$ histograms are both bimodal for \textit{Criteria PA} shocks but trimodal for \textit{Criteria HM} shocks.

\indent  The $\tensor*{ \mathcal{T}  }{^{eb}_{eh}}{\scriptstyle_{j}}$ histograms are more stable between the various selection criteria.  One can see that $\tensor*{ \mathcal{T}  }{^{eb}_{eh}}{\scriptstyle_{\perp}}$ consistently has a peak at smaller values than $\tensor*{ \mathcal{T}  }{^{eb}_{eh}}{\scriptstyle_{\parallel}}$ for all selection criteria except \textit{Criteria HM}.

\phantomsection   
\subsection{Electron Temperature Anisotropies}  \label{subsec:ElectronAnisotropiesNew}

\indent  In this section one-variable statistics and distributions of the electron temperature anisotropy (see Appendix \ref{app:Definitions} for parameter definitions) for the core ($s$ $=$ $ec$), halo ($s$ $=$ $eh$), beam/strahl ($s$ $=$ $eb$), and entire effective ($s$ $=$ $eff$) are presented.

\startlongtable  
\begin{deluxetable}{| l | c | c | c | c | c | c |}
  \tabletypesize{\footnotesize}    
  \tablecaption{Electron Temperature Anisotropy Parameters \label{tab:TempAnisotropies}}
  \tablehead{\colhead{Anisotropy} & \colhead{$X{\scriptstyle_{min}}$}\tablenotemark{a} & \colhead{$X{\scriptstyle_{max}}$} & \colhead{$\bar{X}$} & \colhead{$\tilde{X}$} & \colhead{$X{\scriptstyle_{25\%}}$} & \colhead{$X{\scriptstyle_{75\%}}$}}
  \startdata
  \multicolumn{7}{ |c| }{\textit{Criteria AT: \totalnfitsall~VDFs}} \\
  \hline
  $\mathcal{A}{\scriptstyle_{ec}}$  & 0.38 & 1.56 & 0.93 & 0.98 & 0.90 & 1.01  \\
  $\mathcal{A}{\scriptstyle_{eh}}$  & 0.24 & 15.0 & 1.06 & 1.03 & 0.95 & 1.12  \\
  $\mathcal{A}{\scriptstyle_{eb}}$  & 0.13 & 15.2 & 1.00 & 0.93 & 0.78 & 1.11  \\
  $\mathcal{A}{\scriptstyle_{eff}}$ & 0.35 & 2.80 & 0.93 & 0.98 & 0.91 & 1.01  \\
  \hline
  \enddata
  \tablenotetext{a}{Header symbols match that of Table \ref{tab:Temperatures}}
  \tablecomments{For symbol definitions, see Appendix \ref{app:Definitions}.}
\end{deluxetable}

\begin{figure}
  \centering
    {\includegraphics[trim = 0mm 0mm 0mm 0mm, clip, height=90mm]{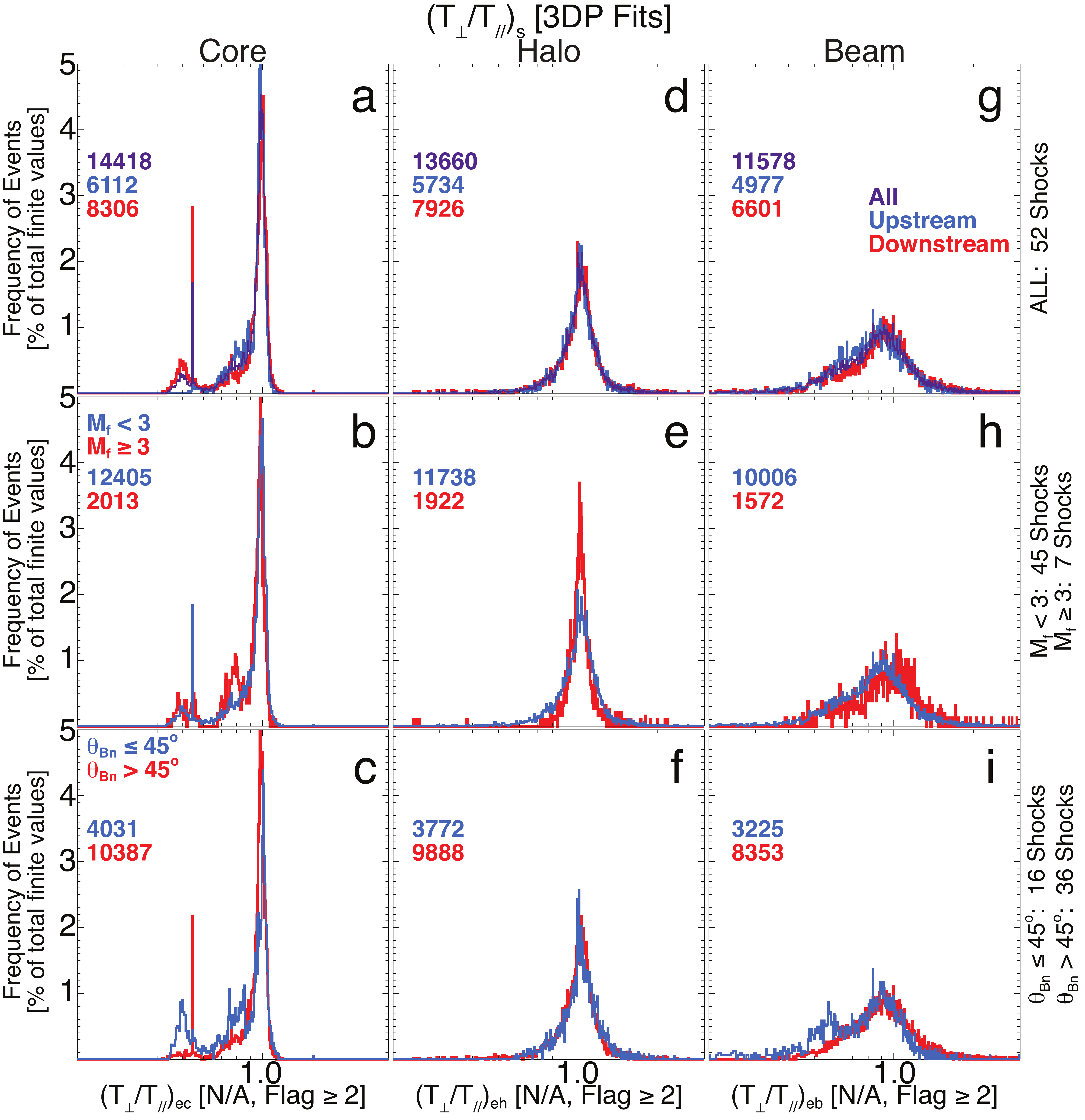}}
    \caption{Temperature anisotropies [N/A] for different electron components in each column for different regions and shock parameters (i.e., color-coded labels by row).  The top row shows all (violet), upstream (blue), and downstream (red) anisotropies.  The middle row shows low Mach number (blue) and high Mach number (red) anisotropies.  The bottom row shows quasi-paralell (blue) and quasi-perpendicular (red) anisotropies.}
    \label{fig:TempAnisotropies}
\end{figure}

\indent  Table \ref{tab:TempAnisotropies} shows the one-variable statistics for $\mathcal{A}{\scriptstyle_{s}}$ $=$ $\left(T{\scriptstyle_{\perp}}/T{\scriptstyle_{\parallel}}\right){\scriptstyle_{s}}$ for all time periods only (see Table \ref{tab:ExtraTempAnisotropies} in Appendix \ref{app:ExtraStatistics} for other selection criteria).  Figure \ref{fig:TempAnisotropies} shows the histograms of $\mathcal{A}{\scriptstyle_{s}}$.  Note that Figure \ref{fig:TempAnisotropies} differs from previous histograms herein because the smaller number of parameters allows for the presentation of all selection criteria to be plotted simultaneously for all three electron components.

\indent  A quick examination of Figure \ref{fig:TempAnisotropies} shows an obvious bimodal distribution in $\mathcal{A}{\scriptstyle_{ec}}$ for all selection criteria.  The smaller $\mathcal{A}{\scriptstyle_{ec}}$ peak corresponds to stronger parallel than perpendicular heating downstream of shocks (i.e., \textit{Criteria DN}), as evidenced by the red line in Figure \ref{fig:TempAnisotropies}a.  The bimodal dependence appears to be more strongly dependent upon $\theta{\scriptstyle_{Bn}}$ than $\langle M{\scriptstyle_{f}} \rangle{\scriptstyle_{up}}$, where the peak near $\sim$0.5 is clearly dominant for \textit{Criteria PA} shocks in Figure \ref{fig:TempAnisotropies}c.  The $\langle M{\scriptstyle_{f}} \rangle{\scriptstyle_{up}}$ appears to be a little more complicated as the distribution is trimodal in Figure \ref{fig:TempAnisotropies}b for \textit{Criteria HM} shocks.  The 5$^{th}$ and 95$^{th}$ percentile values for $\mathcal{A}{\scriptstyle_{ec}}$ are $\sim$0.42 and $\sim$1.21, respectively.  From the quartiles, one can see that only $\sim$25\% fell below $\sim$0.85 or above $\sim$0.99.  Note that $\mathcal{A}{\scriptstyle_{eff}}$ is dominated by the core and so has similar dependencies to that of $\mathcal{A}{\scriptstyle_{ec}}$.  In summary, the core electrons tend toward isotropy and only appear to strongly deviate from that downstream of high Mach number and/or quasi-parallel shocks.

\indent  The halo shows a larger total range of $\mathcal{A}{\scriptstyle_{eh}}$ and more values satisfying $\mathcal{A}{\scriptstyle_{eh}}$ $>$ 1.0, but the distributions are strongly peaked near unity as shown in Figures \ref{fig:TempAnisotropies}d--f.  The distributions show little or no dependence on $\theta{\scriptstyle_{Bn}}$ in Figure \ref{fig:TempAnisotropies}f and Table \ref{tab:ExtraTempAnisotropies} but there does appear to be stronger tails for \textit{Criteria LM} shocks in Figure \ref{fig:TempAnisotropies}e and Table \ref{tab:ExtraTempAnisotropies}.  That is, there is a statistically larger range of $\mathcal{A}{\scriptstyle_{eh}}$ for low Mach number shocks.  However, the general shape of the histogram distributions in Figures \ref{fig:TempAnisotropies}d--\ref{fig:TempAnisotropies}f are the same for each selection criteria suggesting the shock itself has little to do with affecting or regulating the halo temperature anisotropy.  This could imply some other mechanism is responsible, as suggested in previous work, like whistler and/or firehose modes \citep[][]{robergclark18b, tong19a, vasko19a, wilsoniii13a}.  Thus, this may suggest only instabilities and/or turbulence significantly affect the halo temperature anisotropy and electron heat flux in the solar wind, not the IP shocks.

\indent  Finally, the distributions of $\mathcal{A}{\scriptstyle_{eb}}$ seem to show more variation and dependence on the macroscopic shock parameters.  One can see that $\mathcal{A}{\scriptstyle_{eb}}$ also exhibits a bimodal distribution for \textit{Criteria PA} shocks in Figure \ref{fig:TempAnisotropies}i, similar to $\mathcal{A}{\scriptstyle_{ec}}$ though the peaks are at different locations.  The distribution also appears to skew toward smaller $\mathcal{A}{\scriptstyle_{eb}}$ for \textit{Criteria LM} shocks than the converse in Figure \ref{fig:TempAnisotropies}h.  That is, higher Mach number shocks have statistically larger $\mathcal{A}{\scriptstyle_{eb}}$ than the converse.  The same is true for quasi-perpendicular shocks than the converse.  That is, high Mach number, quasi-perpendicular shocks show larger $\mathcal{A}{\scriptstyle_{eb}}$ than the converse suggesting perpendicular scattering is more efficient in these shocks for the beam/strahl component.  It is not clear whether the shock is directly responsible for these differences or if the responsible mechanism finds the environment surrounding these types of shocks more conducive for existence and/or affecting the beam/strahl electrons.  A possible explanation for the larger anisotropy near high Mach number and/or quasi-perpendicular shocks is that the beam/strahl component is more likely contaminated with foreshock electrons, which would have larger pitch-angles near the shock due to processes like fast Fermi acceleration \citep[e.g.,][]{kraussvarban89b, leroy84a, wu84b} and/or shock drift acceleration \citep[e.g.,][]{ball01a, lever01, vandas01a}.  However, these same mechanisms could only generate field-aligned beams far upstream of the shock, along the quasi-static magnetic field similar to the terrestrial electron foreshock edge \citep[e.g.,][]{anderson79a, anderson81a}.

\indent  A slightly different view of the temperature anisotropy statistics can be seen in Figure \ref{fig:ExtraTempAnisotropies} in Appendix \ref{app:ExtraStatistics}.  The anisotropies of each electron component are plotted versus the parallel electron beta of each electron component.  Note that the results in the diagonal panels are consistent with previous observations \citep[e.g.,][]{adrian16a, stverak08a}.  However, a detailed examination of the changes in $\mathcal{A}{\scriptstyle_{s}}$ is beyond the scope of this work and will be examined in Paper III.

\phantomsection   
\section{Coulomb Collision Rates}  \label{sec:CoulombCollisionRates}

\indent  In this section one-variable statistics of the Coulomb collision rates (see Appendix \ref{app:Definitions} for parameter definitions) between the electron components -- core ($s$ $=$ $ec$), halo ($s$ $=$ $eh$), beam/strahl ($s$ $=$ $eb$) -- and protons ($s$ $=$ $p$) and alpha-particles ($s$ $=$ $\alpha$) are presented.

\indent  Calculating the Coulomb collision rates between different electron components and different species is important for verifying that indeed a variation or range of parameters are not solely due to differences in solar wind.  Using Equations \ref{eq:coulomb_coll_0} -- \ref{eq:coulomb_coll_10}, the collision rates, $\nu{\scriptstyle_{ss'}}$, between species $s$ and $s'$ can be approximated for the different selection criteria discussed herein.

\startlongtable  
\begin{deluxetable}{| l | c | c | c | c | c | c |}
  \tabletypesize{\footnotesize}    
  \tablecaption{Coulomb Collision Rates [\# per week] \label{tab:CoulombCollisionRates}}
  \tablehead{\colhead{$\nu{\scriptstyle_{ss'}}$} & \colhead{$X{\scriptstyle_{min}}$}\tablenotemark{a} & \colhead{$X{\scriptstyle_{max}}$} & \colhead{$\bar{X}$} & \colhead{$\tilde{X}$} & \colhead{$X{\scriptstyle_{25\%}}$} & \colhead{$X{\scriptstyle_{75\%}}$}}
  \startdata
  \multicolumn{7}{ |c| }{\textit{Criteria AT: \totalnfitsall~VDFs}} \\
  \hline
  $\nu{\scriptstyle_{p \alpha}}$      &  0.0003 & 0.52 & 0.02 & 0.01 & 0.008 & 0.02  \\
  $\nu{\scriptstyle_{ebb}}$           & 0.00007 & 1.42 & 0.03 & 0.02 & 0.008 & 0.03  \\
  $\nu{\scriptstyle_{ehh}}$           & 0.00006 & 3.14 & 0.05 & 0.03 &  0.01 & 0.06  \\
  $\nu{\scriptstyle_{ehb}}$           & 0.00008 & 8.76 & 0.07 & 0.03 &  0.02 & 0.07  \\
  $\nu{\scriptstyle_{eh \alpha}}$     &  0.0008 & 1.53 & 0.08 & 0.04 &  0.02 & 0.10  \\
  $\nu{\scriptstyle_{eb \alpha}}$     &   0.003 & 1.55 & 0.09 & 0.05 &  0.02 & 0.12  \\
  $\nu{\scriptstyle_{\alpha \alpha}}$ &   0.002 & 1.60 & 0.12 & 0.06 &  0.02 & 0.14  \\
  $\nu{\scriptstyle_{ec \alpha}}$     &   0.009 & 3.11 & 0.35 & 0.23 &  0.12 & 0.48  \\
  $\nu{\scriptstyle_{p p}}$           &  0.0001 & 3.97 & 0.53 & 0.30 &  0.10 & 0.74  \\
  $\nu{\scriptstyle_{ehp}}$           &   0.004 & 16.1 & 0.63 & 0.49 &  0.23 & 0.80  \\
  $\nu{\scriptstyle_{ebp}}$           &   0.004 & 8.78 & 0.70 & 0.52 &  0.29 & 0.90  \\
  $\nu{\scriptstyle_{ehc}}$           &    0.02 & 49.3 & 2.37 & 1.75 &  0.79 & 2.79  \\
  $\nu{\scriptstyle_{ebc}}$           &    0.02 & 57.8 & 4.68 & 2.18 &  1.08 & 3.90  \\
  $\nu{\scriptstyle_{ecp}}$           &   0.009 & 14.7 & 2.99 & 2.64 &  1.17 & 4.27  \\
  $\nu{\scriptstyle_{ecc}}$           &    0.05 & 22.3 & 5.45 & 4.80 &  2.23 & 7.81  \\
  \hline
  \enddata
  \tablenotetext{a}{Header symbol definitions match that of Table \ref{tab:Temperatures}}
  \tablecomments{For symbol definitions, see Appendix \ref{app:Definitions}.}
\vspace{-20pt}
\end{deluxetable}

\indent  Table \ref{tab:CoulombCollisionRates} shows the one-variable statistics for $\nu{\scriptstyle_{ss'}}$ [\# $week^{-1}$]\footnote{Divide by 604,800 to convert to \# $s^{-1}$.} for all time periods only sorted, from smallest to largest, by the $\tilde{X}$ values.  The values for the other selection criteria can be found in Appendix \ref{app:ExtraStatistics} in Table \ref{tab:ExtraCoulombCollisionRates}.  The median values all fall below 8$\times$10$^{-6}$ \# $s^{-1}$ while the upper quartile values below 1$\times$10$^{-4}$ \# $s^{-1}$, consistent with previous statistical work \citep[e.g.,][]{wilsoniii18b}.

\indent  The $\tilde{X}$ values for the rms mean free path (Equation \ref{eq:coulomb_coll_11}) range from $\sim$0.57 AU (astronomical unit) for $\lambda{\scriptstyle_{pp}}^{mpf}$ to $\sim$869 AU for $\lambda{\scriptstyle_{ebb}}^{mpf}$ (the smallest $\bar{X}$ for all rates is $\sim$5 AU).  Note that proton-proton interactions are the only ones that have $\lambda{\scriptstyle_{ss'}}^{mpf}$ $<$ 1.0 AU.  Further, the medians that satisfy $\leq$ 5.0 AU are, from smallest to largest, $\lambda{\scriptstyle_{pp}}^{mpf}$ $\sim$ 0.57 AU, $\lambda{\scriptstyle_{\alpha \alpha}}^{mpf}$ $\sim$ 1.36 AU, $\lambda{\scriptstyle_{ecc}}^{mpf}$ $\sim$ 1.92 AU, and $\lambda{\scriptstyle_{ecp}}^{mpf}$ $\sim$ 2.23 AU.

\indent  Note that although the values of $\nu{\scriptstyle_{ss'}}$ with either $s$ $=$ $c$ or $s'$ $=$ $c$ tend to be larger than the rates not involving the core electrons, they are still very slow.  For instance, the largest $\nu{\scriptstyle_{ss'}}$ value is between beam/strahl and core electrons at $\sim$58/week but that is still only $\sim$10$^{-4}$ \# $s^{-1}$, i.e., only $\sim$8 collisions per day.  Further, $\sim$75\% of all $\nu{\scriptstyle_{ebc}}$ values are at or below $\sim$0.56 \# $day^{-1}$.  If Coulomb collision rates between core electrons and any other species were higher, the core would relax to a bi-Maxwellian.  However, it is interesting that $\sim$80.5\% satisfied 2.00 $\leq$ $s{\scriptstyle_{ec}}$ $\leq$ 2.05 despite the low collision rates with core electrons.  This may imply some remnant property of the solar atmosphere where collision rates are much higher or where preferential heating takes place \citep[e.g.,][]{kasper17a, kasper19a, marsch06a}.

\phantomsection   
\section{Electron Heat Flux}  \label{sec:ElectronHeatFlux}

\indent  In this section one-variable statistics of the parallel electron heat flux, $q{\scriptstyle_{e, \parallel}}$ (see Appendix \ref{app:Definitions} for parameter definitions), for the entire model electron VDF fits and the normalized heat flux, $q{\scriptstyle_{e, \parallel}} / q{\scriptstyle_{e o}}$, are presented.  The integration performed to compute $q{\scriptstyle_{e, \parallel}}$ also required the existence of stable solutions for all three electron components (see Appendix \ref{app:IntegratedVelocityMoments} for details).  There are \totalchbfitsA~VDFs that satisfy these criteria.

\begin{figure}
  \centering
    {\includegraphics[trim = 0mm 0mm 0mm 0mm, clip, height=70mm]{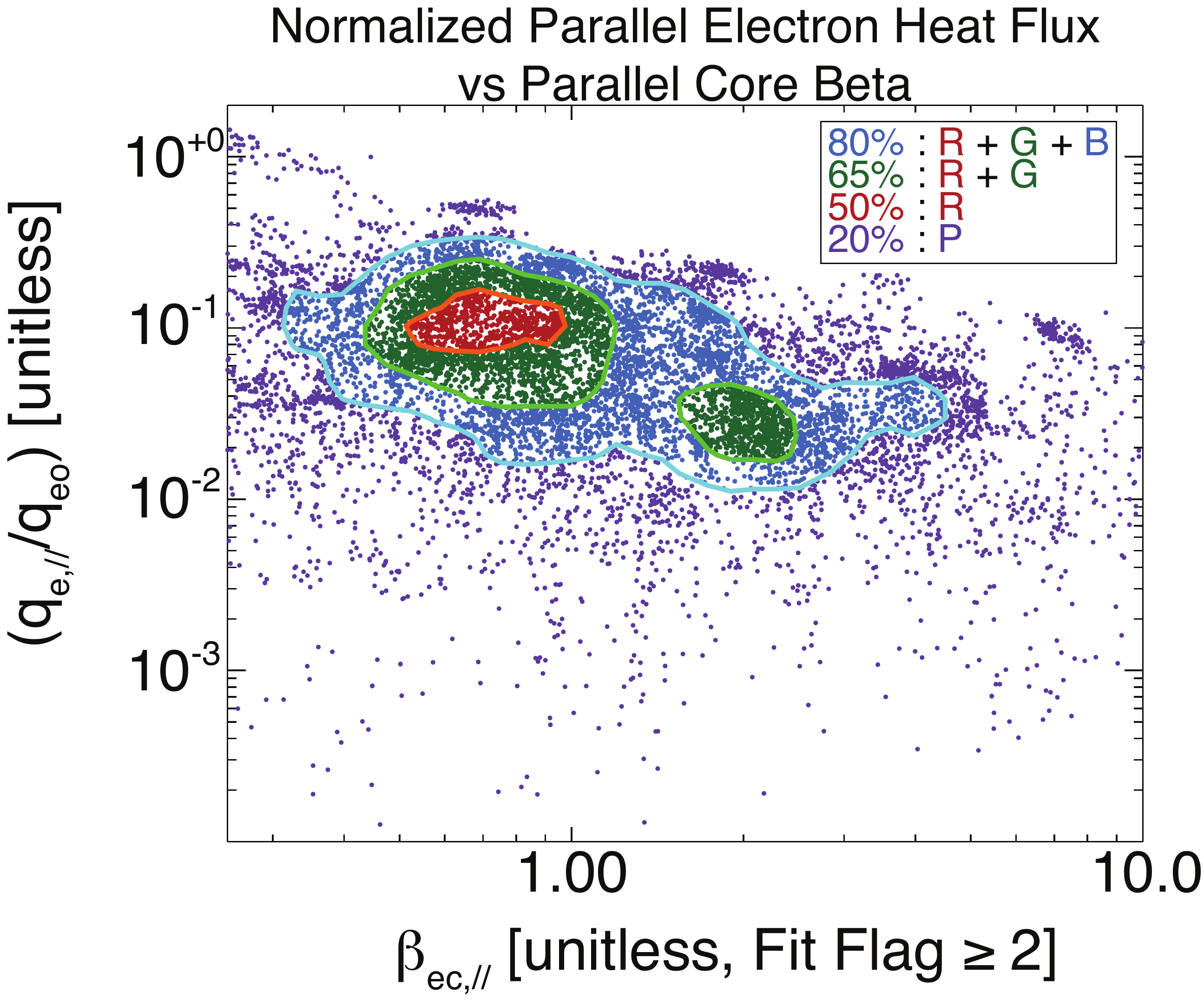}}
    \caption{Normalized parallel electron heat flux, $q{\scriptstyle_{e, \parallel}} / q{\scriptstyle_{e o}}$ [N/A], versus parallel core electron beta, $\beta{\scriptstyle_{ec, \parallel}}$ [N/A].  The color-coded contours (legend in upper right-hand corner) are generated from a two-dimensional histogram of the scatter plot data, where contour levels are defined by fractions of the maximum histogram value.  For instance, the green contour represents the convex hull of the points within the histogram bins that had histogram bin values greater than at least 35\% of the maximum histogram value.}
    \label{fig:HeatFluxvsCoreParaBeta}
\end{figure}

\indent  Figure \ref{fig:HeatFluxvsCoreParaBeta} shows a scatter plot of $q{\scriptstyle_{e, \parallel}} / q{\scriptstyle_{e o}}$ versus $\beta{\scriptstyle_{ec, \parallel}}$ for selection criteria \textit{Criteria AT}.  The color-coded contours indicate the regions of the highest density of points in the scatter plot.  The legend in upper right-hand corner indicates roughly the approximate fraction of points within in each contour, e.g., the fraction within the cyan contour is $\sim$80\% of the total \totalchbfitsA~points shown.  

\indent  The one-variable statistics for $q{\scriptstyle_{e, \parallel}}$ and $q{\scriptstyle_{e, \parallel}} / q{\scriptstyle_{e o}}$ are shown as $X{\scriptstyle_{25\%}}$--$X{\scriptstyle_{75\%}}$($\bar{X}$)[$\tilde{X}$] and given by:
\begin{itemize}[itemsep=0pt,parsep=0pt,topsep=0pt]
  \item  $q{\scriptstyle_{e, \parallel}}$ $\sim$ 2.39--7.51(6.00)[4.11] $\mu W \ m^{-2}$
  \item  $q{\scriptstyle_{e, \parallel}} / q{\scriptstyle_{e o}}$ $\sim$ 2.56--12.3(9.33)[5.84] \%
\end{itemize}

\noindent  The normalized magnitudes and $\beta{\scriptstyle_{ec, \parallel}}^{-1}$ trend are consistent with previous results \citep[e.g.,][]{bale13a, lacombe14a, tong18b, tong19a, tong19b, wilsoniii13a}.  However, it is worth noting that the $\beta{\scriptstyle_{ec, \parallel}}^{-1}$ trend in the $q{\scriptstyle_{e, \parallel}} / q{\scriptstyle_{e o}}$ may result from the fact that $q{\scriptstyle_{e o}}$ $\propto$ $B{\scriptstyle_{o}}^{2} \ V{\scriptstyle_{Tec, \parallel}} \ \beta{\scriptstyle_{ec, \parallel}}$, that is $q{\scriptstyle_{e o}}$ can be written in terms of $\beta{\scriptstyle_{ec, \parallel}}$.  Although it's beyond the scope of this work, the electron heat flux is a known source of free energy for several wave modes and of the \totalchbfitsA~VDFs with heat flux values, nearly 90\% were found to be unstable to the whistler heat flux instability \citep[e.g.,][]{gary94a, gary99a}.  This will be examined in more detail in Paper III.

\phantomsection   
\section{Summary of Upstream Statistics}  \label{sec:SummaryofStatistics}

\indent  Recall that the primary purpose of this second of three parts is to provide a statistical baseline for reference of the velocity moment values under different conditions.  One of the benefits of this large data set is that the \textit{Criteria UP} results offer a useful baseline for comparison with quiescent solar wind studies.  Further, in the process of the literature review, a dearth of velocity moment results for the beam/strahl component were found (e.g., see Appendix \ref{app:PreviousElectronStudies}).  Therefore, the \textit{Criteria UP} results can be referenced as approximate values for the solar wind\footnote{This should emphasize that in the absence of a better dataset, the upstream only results presented herein are the only statistically significant set of beam/strahl velocity moments (of which the authors are aware).}.

\indent  From the parameter lists and tables in Appendix \ref{app:ExtraStatistics} (and results in Paper I), one can see that the \textit{Criteria UP} values, reported as $\tilde{X} \ \substack{X{\scriptstyle_{75\%}} \\ X{\scriptstyle_{25\%}}}$, for the electron component velocity moment parameters\footnote{The values of $\lvert V{\scriptstyle_{os, \parallel}} \rvert$ exclude magntiudes below 1 km/s prior to calculating the one variable statistics.} are:
\begin{itemize}[itemsep=0pt,parsep=0pt,topsep=0pt]
  \item[]  \textit{Core}
  \begin{itemize}[itemsep=0pt,parsep=0pt,topsep=0pt]
    \item  $n{\scriptstyle_{ec}}$ $\sim$ $8.29 \ \substack{12.6 \\ 4.35}$ $cm^{-3}$;
    \item  $T{\scriptstyle_{ec, tot}}$  $\sim$ $13.0 \ \substack{15.5 \\ 10.9}$ $eV$;
    \item  $\beta{\scriptstyle_{ec, tot}}$ $\sim$ $1.21 \ \substack{2.34 \\ 0.59}$;
    \item  $\mathcal{A}{\scriptstyle_{ec}}$  $\sim$ $0.98 \ \substack{1.00 \\ 0.91}$;
    \item  $\lvert V{\scriptstyle_{oec, \parallel}} \rvert$ $\sim$ $25.0 \ \substack{40.0 \\ 11.4}$ km/s;
    \item  $\kappa{\scriptstyle_{ec}}$ $\sim$ $7.92 \ \substack{10.1 \\ 5.44}$;
    \item  $s{\scriptstyle_{ec}}$ $\sim$ $2.00 \ \substack{2.03 \\ 2.00}$;
  \end{itemize}
  \item[]  \textit{Halo}
  \begin{itemize}[itemsep=0pt,parsep=0pt,topsep=0pt]
    \item  $n{\scriptstyle_{eh}}$ $\sim$ $0.27 \ \substack{0.49 \\ 0.17}$ $cm^{-3}$;
    \item  $T{\scriptstyle_{eh, tot}}$  $\sim$ $47.2 \ \substack{55.7 \\ 36.3}$ $eV$;
    \item  $\beta{\scriptstyle_{eh, tot}}$ $\sim$ $0.17 \ \substack{0.32 \\ 0.08}$;
    \item  $\mathcal{A}{\scriptstyle_{eh}}$  $\sim$ $1.03 \ \substack{1.12 \\ 0.95}$;
    \item  $\lvert V{\scriptstyle_{oeh, \parallel}} \rvert$ $\sim$ $940 \ \substack{1647 \\ 401}$ km/s;
    \item  $\kappa{\scriptstyle_{eh}}$ $\sim$ $4.10 \ \substack{4.83 \\ 3.25}$;
  \end{itemize}
  \item[]  \textit{Beam/Strahl}
  \begin{itemize}[itemsep=0pt,parsep=0pt,topsep=0pt]
    \item  $n{\scriptstyle_{eb}}$ $\sim$ $0.16 \ \substack{0.28 \\ 0.10}$ $cm^{-3}$;
    \item  $T{\scriptstyle_{eb, tot}}$  $\sim$ $38.8 \ \substack{46.6 \\ 32.4}$ $eV$;
    \item  $\beta{\scriptstyle_{eb, tot}}$ $\sim$ $0.09 \ \substack{0.16 \\ 0.05}$;
    \item  $\mathcal{A}{\scriptstyle_{eb}}$  $\sim$ $0.90 \ \substack{1.07 \\ 0.75}$;
    \item  $\lvert V{\scriptstyle_{oeb, \parallel}} \rvert$ $\sim$ $2110 \ \substack{3000 \\ 1400}$ km/s;
    \item  $\kappa{\scriptstyle_{eb}}$ $\sim$ $3.84 \ \substack{4.67 \\ 3.26}$;
  \end{itemize}
  \item[]  \textit{Other}
  \begin{itemize}[itemsep=0pt,parsep=0pt,topsep=0pt]
    \item  $n{\scriptstyle_{eff}}$ $\sim$ $8.63 \ \substack{13.7 \\ 4.76}$ $cm^{-3}$;
    \item  $T{\scriptstyle_{eff, tot}}$  $\sim$ $14.6 \ \substack{17.9 \\ 12.5}$ $eV$;
    \item  $\beta{\scriptstyle_{eff, tot}}$ $\sim$ $1.42 \ \substack{2.63 \\ 0.76}$;
    \item  $\mathcal{A}{\scriptstyle_{eff}}$ $\sim$ $0.97 \ \substack{1.00 \\ 0.92}$;
    \item  $n{\scriptstyle_{eh}} / n{\scriptstyle_{ec}}$ $\sim$ $3.8 \ \substack{7.4 \\ 2.2}$ \%;
    \item  $n{\scriptstyle_{eb}} / n{\scriptstyle_{ec}}$ $\sim$ $2.8 \ \substack{4.3 \\ 1.2}$ \%;
    \item  $n{\scriptstyle_{eb}} / n{\scriptstyle_{eh}}$ $\sim$ $69.3 \ \substack{113 \\ 31.2}$ \%;
    \item  $\tensor*{ \mathcal{T}  }{^{eh}_{ec}}{\scriptstyle_{      tot}}$ $\sim$ $3.54 \ \substack{4.59 \\ 2.69}$;
    \item  $\tensor*{ \mathcal{T}  }{^{eb}_{ec}}{\scriptstyle_{      tot}}$ $\sim$ $2.87 \ \substack{3.59 \\ 2.45}$;
    \item  $\tensor*{ \mathcal{T}  }{^{eb}_{eh}}{\scriptstyle_{      tot}}$ $\sim$ $0.81 \ \substack{1.12 \\ 0.62}$;
    \item  $\tensor*{ \mathcal{T}  }{^{eh}_{eff}}{\scriptstyle_{      tot}}$ $\sim$ $3.07 \ \substack{3.98 \\ 2.33}$;
    \item  $\tensor*{ \mathcal{T}  }{^{eb}_{eff}}{\scriptstyle_{      tot}}$ $\sim$ $2.46 \ \substack{3.05 \\ 2.14}$;
    \item  $\nu{\scriptstyle_{ehc}}$ $\sim$ $\left( 2.31 \ \substack{3.85 \\ 1.09} \right) \times 10^{-6}$ $\# \ s^{-1}$;
    \item  $\nu{\scriptstyle_{ebc}}$ $\sim$ $\left( 2.74 \ \substack{5.79 \\ 1.43} \right) \times 10^{-6}$ $\# \ s^{-1}$;
    \item  $\nu{\scriptstyle_{ecp}}$ $\sim$ $\left( 3.60 \ \substack{5.84 \\ 1.91} \right) \times 10^{-6}$ $\# \ s^{-1}$; and
    \item  $\nu{\scriptstyle_{ecc}}$ $\sim$ $\left( 6.45 \ \substack{11.0 \\ 3.31} \right) \times 10^{-6}$ $\# \ s^{-1}$.
  \end{itemize}
\end{itemize}

\indent  The majority of the literature on the strahl electrons focus entirely on the pitch-angle width versus energy and/or radial distance from the sun \citep[e.g.,][]{anderson12a, kajdic16a, gurgiolo12a, gurgiolo16a, horaites18a, pagel07, walsh13a} or they compute the total heat flux of the distribution \citep[e.g.,][]{crooker03a, crooker08a, pagel05a, pagel05b}.  A few studies examined density ratios among the various components \citep[e.g.,][]{maksimovic05a, stverak09a} and some have extrapolated an effective temperature \citep[e.g.,][]{tao16a, tao16b} from a limited energy range measurement.  In only one study, of which the authors are aware, have the beam/strahl velocity moments been presented for multiple distributions \citep[i.e.,][]{vinas10a}.  However, this study only presented results from a single, short duration interval.  Therefore, the upstream only velocity moment results for the beam/strahl component presented herein is the closest to a statistically significant presentation of those parameters in the solar wind near 1 AU to date.

\indent  Note that although the \textit{Criteria UP} values for $T{\scriptstyle_{ec, j}}$ and $T{\scriptstyle_{eff, j}}$ are slightly higher than those reported for the total electron temperature in a recent large, long-term statistical study of the solar wind under various conditions \citep[e.g.,][]{wilsoniii18b}, they are still well within the total range reported therein.  The upstream values are also consistent with numerous other previous solar wind observations near 1 AU (e.g., see Appendix \ref{app:PreviousElectronStudies}).  The $T{\scriptstyle_{eb, j}}$ values and histograms in Tables \ref{tab:Temperatures} and \ref{tab:ExtraTemperatures} and Figure \ref{fig:Temperatures} are perhaps the most novel as there have been so few studies examining the beam/strahl velocity moment parameters.  In fact, the few studies that have examined the velocity moments of the beam/strahl either limited the energy range (e.g., $\geq$100 eV) and thus only had effective moments \citep[e.g.,][]{tao16a, tao16b} or they performed a limited case study \citep[e.g.,][]{vinas10a}.  The one statistical study of solar wind parameters that examined a three component electron VDF used a truncated model function to exclude contributions from data below a cutoff energy \citep[][]{stverak09a}, which limited their analysis to the kappa values and number densities.  Other studies focusing on the beam/strahl component discuss only the pitch-angle angular width versus energy and/or radial distance from the sun \citep[e.g.,][]{anderson12a, graham17a, graham18a, horaites18a}.  While these studies are clearly important and relevant to understanding the origin of the strahl and its relation to solar wind acceleration, the angular width is difficult to translate into more commonly used parameters for modeling like number density, drift velocity, or temperature.  Although the \textit{Criteria UP} values of $T{\scriptstyle_{eb, j}}$ have a large range spanning from $\sim$12 eV to $\sim$280 eV, the majority fall in the more modest range of $\sim$29--50 eV, consistent with the few previous studies that examined the beam/strahl temperature.

\indent  Again, the beam/strahl-dependent density \textit{Criteria UP} values are novel in that there is little previous work on this topic \citep[e.g.,][]{maksimovic05a, stverak09a}.  The values of $n{\scriptstyle_{eb}}$ are relatively unaffected by the shock in that there is little-to-no change in the one-variable statistics between \textit{Criteria UP} and \textit{Criteria DN} values.  It is likely that the beam/strahl electrons are less susceptible to the effects of the shock or they stream so quickly that there is little connection between those observed upstream and those downstream, other than the influence of shock-reflected electrons in the upstream.  The beam/strahl drifts from Paper I easily exceed almost all IP shock speeds, so it is unlikely that beam/strahl electrons starting downstream could not overtake an IP shock.  It is just as unlikely that beam/strahl electrons starting upstream could not outrun an IP shock.  Therefore, the most likely conclusion is that only a narrow region near the shock ramp would exhibit shock-parameter-dependent effects on the beam/strahl fit results.  This will be investigated in detail in Paper III.

\indent  The electron plasma beta \textit{Criteria UP} values for $\beta{\scriptstyle_{ec, j}}$ and $\beta{\scriptstyle_{eff, j}}$ are consistent with a recent large, long-term statistical study of the solar wind under various conditions \citep[e.g.,][]{wilsoniii18b} and numerous other previous solar wind observations near 1 AU \citep[e.g.,][]{adrian16a, bale13a}.  Although the $\beta{\scriptstyle_{ec, j}}$ and $\beta{\scriptstyle_{eff, j}}$ \textit{Criteria UP} values have maxima in excess of 800 and 950, respectively, at least $\sim$75\% fall below $\sim$2.4 and $\sim$2.7, respectively.  That is, the upstream only core and effective electron betas are typically consistent with low beta plasmas relative to, e.g., the intracluster medium where $\beta{\scriptstyle_{e}}$ $\sim$ 100 \citep[e.g.,][]{robergclark16a, robergclark18b}.  Similarly, $\beta{\scriptstyle_{eh, j}}$ and $\beta{\scriptstyle_{eb, j}}$ both tend to fall below $\sim$3.2 and $\sim$1.8, respectively.  Further, the variation between any two components of $\beta{\scriptstyle_{eh, j}}$ for any one-variable statistics value is remarkably small, with all except $X{\scriptstyle_{min}}$ falling within a few percent of each other.  The differences for $\beta{\scriptstyle_{eb, j}}$ show slightly more variation but are still small.  Note that there is only one study \citep[i.e.,][]{vinas10a}, of which the authors are aware, that quantified $\beta{\scriptstyle_{eb, j}}$ and our \textit{Criteria UP} values are consistent with those previous results.

\indent  Similar to other velocity moments discussed above, there are no direct comparisons, of which the authors are aware, of the beam/strahl temperatures with the core or halo individually.  However, the \textit{Criteria UP} $\tensor*{ \mathcal{T}  }{^{eh}_{ec}}{\scriptstyle_{j}}$ values are consistent with previous results in the solar wind \citep[][]{feldman75a, skoug00a} and near IP shocks \citep[][]{wilsoniii09a, wilsoniii12c}.

\indent  The range between the 5$^{th}$ and 95$^{th}$ percentiles for the $\mathcal{A}{\scriptstyle_{s}}$ \textit{Criteria UP} values is smallest for the core and effective anisotropies and this holds even when examining the range between the quartiles.  All $\bar{X}$ values satisfying \textit{Criteria UP} fall within $\sim$6\% of unity and all $\tilde{X}$ within $\sim$4\% of unity.  This is rather obvious from Figure \ref{fig:TempAnisotropies} in that the core and halo components are sharply peaked near unity while the beam/strahl exhibits a broader distribution but still peaked near unity.

\phantomsection   
\section{Discussion}  \label{sec:Discussion}

\indent  The statistical analysis of \totalnfitsall~electron VDFs observed by the \emph{Wind} spacecraft within $\pm$2 hours of 52 IP shocks is presented.  Tables of one-variable statistics combined with histograms separated by the seven selection criteria used herein provide a comprehensive summary of the properties of the electron VDFs in and around IP shocks near 1 AU.  The fit results satisfying the \textit{Criteria UP} criteria are the only currently available dataset of beam/strahl velocity moment values near 1 AU.  The net electron heat flux and the two-particle collision rates between all electron components and protons and alpha-particles are also provided.

\indent  From Tables \ref{tab:Temperatures}--\ref{tab:CoulombCollisionRates} (and results in Paper I), one can see that the \textit{Criteria AT} values, reported as $\tilde{X} \ \substack{X{\scriptstyle_{75\%}} \\ X{\scriptstyle_{25\%}}}$, for the electron component velocity moment parameters\footnote{The values of $\lvert V{\scriptstyle_{os, \parallel}} \rvert$ exclude magntiudes below 1 km/s prior to calculating the one variable statistics.} are:
\begin{itemize}[itemsep=0pt,parsep=0pt,topsep=0pt]
  \item[]  \textit{Core}
  \begin{itemize}[itemsep=0pt,parsep=0pt,topsep=0pt]
    \item  $n{\scriptstyle_{ec}}$ $\sim$ $11.3 \ \substack{19.4 \\ 6.55}$ $cm^{-3}$;
    \item  $T{\scriptstyle_{ec, tot}}$  $\sim$ $14.6 \ \substack{18.6 \\ 12.0}$ $eV$;
    \item  $\beta{\scriptstyle_{ec, tot}}$ $\sim$ $0.93 \ \substack{2.01 \\ 0.52}$;
    \item  $\mathcal{A}{\scriptstyle_{ec}}$  $\sim$ $0.98 \ \substack{1.01 \\ 0.90}$;
    \item  $\lvert V{\scriptstyle_{oec, \parallel}} \rvert$ $\sim$ $29.1 \ \substack{49.7 \\ 14.2}$ km/s;
    \item  $\kappa{\scriptstyle_{ec}}$ $\sim$ $7.92 \ \substack{10.1 \\ 5.44}$;
    \item  $s{\scriptstyle_{ec}}$ $\sim$ $2.00 \ \substack{2.04 \\ 2.00}$;
    \item  $p{\scriptstyle_{ec}}$ $\sim$ $3.00 \ \substack{4.00 \\ 2.20}$;
    \item  $q{\scriptstyle_{ec}}$ $\sim$ $2.00 \ \substack{2.38 \\ 2.00}$;
  \end{itemize}
  \item[]  \textit{Halo}
  \begin{itemize}[itemsep=0pt,parsep=0pt,topsep=0pt]
    \item  $n{\scriptstyle_{eh}}$ $\sim$ $0.36 \ \substack{0.63 \\ 0.20}$ $cm^{-3}$;
    \item  $T{\scriptstyle_{eh, tot}}$  $\sim$ $48.4 \ \substack{58.1 \\ 37.4}$ $eV$;
    \item  $\beta{\scriptstyle_{eh, tot}}$ $\sim$ $0.11 \ \substack{0.22 \\ 0.05}$;
    \item  $\mathcal{A}{\scriptstyle_{eh}}$  $\sim$ $1.03 \ \substack{1.12 \\ 0.95}$;
    \item  $\lvert V{\scriptstyle_{oeh, \parallel}} \rvert$ $\sim$ $901 \ \substack{1692 \\ 362}$ km/s;
    \item  $\kappa{\scriptstyle_{eh}}$ $\sim$ $4.37 \ \substack{5.31 \\ 3.57}$;
  \end{itemize}
  \item[]  \textit{Beam/Strahl}
  \begin{itemize}[itemsep=0pt,parsep=0pt,topsep=0pt]
    \item  $n{\scriptstyle_{eb}}$ $\sim$ $0.17 \ \substack{0.29 \\ 0.10}$ $cm^{-3}$;
    \item  $T{\scriptstyle_{eb, tot}}$  $\sim$ $40.2 \ \substack{50.0 \\ 33.7}$ $eV$;
    \item  $\beta{\scriptstyle_{eb, tot}}$ $\sim$ $0.05 \ \substack{0.10 \\ 0.02}$;
    \item  $\mathcal{A}{\scriptstyle_{eb}}$  $\sim$ $0.93 \ \substack{1.11 \\ 0.78}$;
    \item  $\lvert V{\scriptstyle_{oeb, \parallel}} \rvert$ $\sim$ $2282 \ \substack{3000 \\ 1400}$ km/s;
    \item  $\kappa{\scriptstyle_{eb}}$ $\sim$ $4.17 \ \substack{5.11 \\ 3.41}$;
  \end{itemize}
  \item[]  \textit{Other}
  \begin{itemize}[itemsep=0pt,parsep=0pt,topsep=0pt]
    \item  $n{\scriptstyle_{eff}}$ $\sim$ $11.9 \ \substack{20.3 \\ 6.92}$ $cm^{-3}$;
    \item  $T{\scriptstyle_{eff, tot}}$  $\sim$ $16.0 \ \substack{20.5 \\ 13.6}$ $eV$;
    \item  $\beta{\scriptstyle_{eff, tot}}$ $\sim$ $1.06 \ \substack{2.26 \\ 0.63}$;
    \item  $\mathcal{A}{\scriptstyle_{eff}}$ $\sim$ $0.97 \ \substack{1.01 \\ 0.91}$;
    \item  $n{\scriptstyle_{eh}} / n{\scriptstyle_{ec}}$ $\sim$ $3.2 \ \substack{6.1 \\ 1.8}$ \%;
    \item  $n{\scriptstyle_{eb}} / n{\scriptstyle_{ec}}$ $\sim$ $1.8 \ \substack{3.6 \\ 0.7}$ \%;
    \item  $n{\scriptstyle_{eb}} / n{\scriptstyle_{eh}}$ $\sim$ $50.0 \ \substack{96.2 \\ 24.4}$ \%;
    \item  $\tensor*{ \mathcal{T}  }{^{eh}_{ec}}{\scriptstyle_{      tot}}$ $\sim$ $3.19 \ \substack{4.04 \\ 2.39}$;
    \item  $\tensor*{ \mathcal{T}  }{^{eb}_{ec}}{\scriptstyle_{      tot}}$ $\sim$ $2.73 \ \substack{3.35 \\ 2.22}$;
    \item  $\tensor*{ \mathcal{T}  }{^{eb}_{eh}}{\scriptstyle_{      tot}}$ $\sim$ $0.85 \ \substack{1.14 \\ 0.66}$;
    \item  $\tensor*{ \mathcal{T}  }{^{eh}_{eff}}{\scriptstyle_{      tot}}$ $\sim$ $2.88 \ \substack{3.64 \\ 2.17}$;
    \item  $\tensor*{ \mathcal{T}  }{^{eb}_{eff}}{\scriptstyle_{      tot}}$ $\sim$ $2.44 \ \substack{3.03 \\ 1.99}$;
    \item  $\nu{\scriptstyle_{ehc}}$ $\sim$ $\left( 2.90 \ \substack{4.61 \\ 1.30} \right) \times 10^{-6}$ $\# \ s^{-1}$;
    \item  $\nu{\scriptstyle_{ebc}}$ $\sim$ $\left( 3.60 \ \substack{6.45 \\ 1.78} \right) \times 10^{-6}$ $\# \ s^{-1}$;
    \item  $\nu{\scriptstyle_{ecp}}$ $\sim$ $\left( 4.36 \ \substack{7.07 \\ 1.93} \right) \times 10^{-6}$ $\# \ s^{-1}$; and
    \item  $\nu{\scriptstyle_{ecc}}$ $\sim$ $\left( 7.93 \ \substack{12.9 \\ 3.69} \right) \times 10^{-6}$ $\# \ s^{-1}$.
  \end{itemize}
\end{itemize}

\indent  Although the detailed analysis of the electron VDF fit parameters on the macroscopic shock properties are beyond the scope of this work and included in Paper III, some statistical dependencies are discussed herein.  The dependencies of $T{\scriptstyle_{s, j}}$ and $n{\scriptstyle_{es}}$ on the selection criteria are weak for the halo and beam/strahl but clear for the core and consistent with expectations.  That is, the core is heated and compressed in the downstream compared to the upstream.

\indent  Although the individual $T{\scriptstyle_{s, j}}$ did not show significant variations between the selection criteria, the $\tensor*{ \mathcal{T} }{^{s'}_{s}}{\scriptstyle_{j}}$ did show some strong dependencies on the selection criteria.  The core-dependent ratios (i.e., $s$ $=$ $c$) show some rather dramatic differences in the histogram profiles even though the primary peaks are relatively constant (except for \textit{Criteria HM} shocks).  In contract, the $\tensor*{ \mathcal{T} }{^{eb}_{eh}}{\scriptstyle_{j}}$ histograms are much more stable in profile and one-variable statistics values between any two opposing selection criteria.  There are still differences in the histograms of \textit{Criteria PA} and \textit{Criteria HM} shocks, but they are more subtle than those for either of the core-dependent ratios.

\indent  The $\mathcal{A}{\scriptstyle_{ec}}$ histograms are primarily different in the lower value tails between any two opposing selection criteria.  For instance, the histograms are tripolar for \textit{Criteria PA} and \textit{Criteria HM} shocks, but only bipolar for \textit{Criteria LM} shocks and effectively monopolar for \textit{Criteria PE} shocks.  Note that the $\mathcal{A}{\scriptstyle_{ec}}$ values do not deviate to values much larger than unity, i.e., the core is more often oblate in the parallel than perpendicular directions.  In contrast, the $\mathcal{A}{\scriptstyle_{eh}}$ histograms are monopolar and peaked near unity with large tails on both sides of unity, i.e., the halo can be oblate in both the parallel and perpendicular directions but tends towards near isotropy.  Finally, the $\mathcal{A}{\scriptstyle_{eb}}$ show an even broader range of values and only the \textit{Criteria PA} values show a bipolar distribution.  The primary peak for all selection criteria is near $\sim$0.9 except for \textit{Criteria HM}, which is closer to unity.  Although the $\mathcal{A}{\scriptstyle_{eb}}$ histograms have long tails on both sides of the primary peaks, the distributions are skewed toward smaller values.

\indent  Lastly, the $\beta{\scriptstyle_{s, j}}$ values showed dramatic differences between opposing selection criteria with one-variable statistic values differing by upwards of 100\%.  Not only do the peaks change between opposing selection criteria, the histogram profiles show remarkable differences as well.  The core beta values are more stable than either the halo or beam/strahl, but even the core shows significant differences.  Thus, the electron component betas seem to exhibit the most striking dependencies on macroscopic shock parameters.

\indent  The fit results were also used to calculate the two-particle Coulomb collision rates, all of which had median values below 8$\times$10$^{-6}$ \# $s^{-1}$, or less than $\sim$5 collisions per week (for \textit{Criteria AT}).  When calculating the collisional mean free paths, the only two-particle collision rates with median values less than one astronomical unit are for proton-proton collisions.  Next the alpha-alpha and core-core values are $<$ 2 AU followed by beam-core, proton-alpha, and halo-core satisfying $<$ 8 AU.  The rest of the median values all satisfy $>$ 15 AU.  That is, the median distance before a collision occurs for most species is nearly the orbital radius of Uranus.  The bottom five (or largest five) median mean free paths all satisfy $>$ 185 AU.  That is, the median location before experiencing a collision is outside the heliosphere for the most tenuous of the species examined herein, i.e., alpha-particles, halo, and beam/strahl electrons.  Yet despite the low particle-particle collision rates, most of the core exponents (i.e., $\sim$80.5\%) satisfy 2.00 $\leq$ $s{\scriptstyle_{ec}}$ $\leq$ 2.05, which are self-similar VDFs that are visually indistinguishable from Maxwellians (see Paper I for details).  This seems to suggest some remnant property of the solar atmosphere where collision rates are much higher \citep[e.g.,][]{kasper17a, kasper19a, marsch06a}.

\indent  The parallel electron heat flux was also calculated for VDFs with stable solutions for all three electron components.  The magnitudes and normalized values had the following ranges $q{\scriptstyle_{e, \parallel}}$ $\sim$ 10$^{-6}$--76 $\mu W \ m^{-2}$ and $q{\scriptstyle_{e, \parallel}} / q{\scriptstyle_{e o}}$ $\sim$ 10$^{-5}$--190 \%.  However, 95\% of the magnitudes and normalized values satisfied $\leq$16.2 $\mu W \ m^{-2}$ and $\leq$24.4 \%, respectively, consistent with previous work \citep[e.g.,][]{bale13a, lacombe14a, tong18b, tong19a, tong19b, wilsoniii13a}.  A preliminary examination found that as many as $\sim$90\% of the observed electron VDFs are unstable to the whistler heat flux instability \citep[e.g.,][]{gary94a, gary99a}, however  instability analysis is beyond the scope of this work and will be discussed in Paper III.

\indent  The results presented herein provide a statistically significant list of values and histogram distributions for reference and baselines in future work.  The beam/strahl fit results for selection criteria \textit{Criteria UP} are currently the closest thing to a proper parameterization of the beam/strahl electron velocity moments in the ambient solar wind.  These results are useful for multiple modeling and simulation studies in addition to comparison with inaccessible regions like the intracluster medium.  Further, these results will provide a statistical baseline for the \emph{Parker Solar Probe} mission and the future \emph{Solar Orbiter} and IMAP missions.

\acknowledgments
\noindent  The authors thank A.F.- Vi{\~n}as and D.A. Roberts for useful discussions of basic plasma physics and C. Markwardt for helpful feedback on the usage nuances of his MPFIT software.  The work was supported by the International Space Science Institute's (ISSI) International Teams programme.  L.B.W. was partially supported by \emph{Wind} MO\&DA grants and a Heliophysics Innovation Fund (HIF) grant.  L.-J.C. and S.W. were partially supported by the MMS mission in addition to NASA grants 80NSSC18K1369 and 80NSSC17K0012, NSF grants AGS-1619584 and AGS-1552142, and DOE grant DESC0016278.  D.L.T. was partially supported by NASA grant NNX16AQ50G.  M.L.S. was partially supported by grants NNX14AT26G and NNX13AI75G.  J.C.K. was partially supported by NASA grants NNX14AR78G and 80NSSC18K0986.  D.C. was partially supported by grants NNX17AG30G, GO8-19110A, 80NSSC18K1726, 80NSSC18K1218, and NSF grant 1714658.  S.J.S. was partially supported by the MMS/FIELDS investigation.  C.S.S. was partially supported by NASA grant NNX16AI59G and NSF SHINE grant 1622498.  S.D.B. and C.S.S. were partially supported by NASA grant NNX16AP95G.  S.D.B. acknowledges the support of the Leverhulme Trust Visiting Professor program.  M.P.P. and K.A.G. were supported by Parker Solar Probe instrument funds.

\appendix
\phantomsection   
\section{Definitions and Notation}  \label{app:Definitions}

\indent  As in Paper I, this appendix the symbols and notation used throughout will be defined.  All direction-dependent parameters we use the subscript $j$ to represent the direction where $j$ $=$ $tot$ for the entire distribution, $j$ $=$ $\parallel$ for the the parallel direction, and $j$ $=$ $\perp$ for the perpendicular direction, where parallel/perpendicular is with respect to the quasi-static magnetic field vector, $\mathbf{B}{\scriptstyle_{o}}$ [nT].  The use of the generic subscript $s$ to denote the particle species (e.g., electrons, protons, etc.) or the component of a single particle species (e.g., electron core).  For the electron components, the subscript will be $s$ $=$ $ec$ for the core, $s$ $=$ $eh$ for the halo, $s$ $=$ $eb$ for the beam/strahl, and $s$ $=$ $eff$ for the effective, and $s$ $=$ $e$ for the total/entire population.  Below are the symbol/parameters definitions:
\begin{itemize}[itemsep=0pt,parsep=0pt,topsep=0pt]
  \item[]  \textit{one-variable statistics}
  \begin{itemize}[itemsep=0pt,parsep=0pt,topsep=0pt]
    \item  $X{\scriptstyle_{min}}$ $\equiv$ minimum
    \item  $X{\scriptstyle_{max}}$ $\equiv$ maximum
    \item  $\bar{X}$ $\equiv$ mean
    \item  $\tilde{X}$ $\equiv$ median
    \item  $X{\scriptstyle_{25\%}}$ $\equiv$ lower quartile
    \item  $X{\scriptstyle_{75\%}}$ $\equiv$ upper quartile
    \item  $\sigma$ $\equiv$ standard deviation
    \item  $\sigma^{2}$ $\equiv$ variance
  \end{itemize}
  \item[]  \textit{fundamental parameters}
  \begin{itemize}[itemsep=0pt,parsep=0pt,topsep=0pt]
    \item  $\varepsilon{\scriptstyle_{o}}$ $\equiv$ permittivity of free space
    \item  $\mu{\scriptstyle_{o}}$ $\equiv$ permeability of free space
    \item  $c$ $\equiv$ speed of light in vacuum [$km \ s^{-1}$] $=$ $\left( \varepsilon{\scriptstyle_{o}} \ \mu{\scriptstyle_{o}} \right)^{-1/2}$
    \item  $k{\scriptstyle_{B}}$ $\equiv$ the Boltzmann constant [$J \ K^{-1}$]
    \item  $e$ $\equiv$ the fundamental charge [$C$]
  \end{itemize}
  \item[]  \textit{plasma parameters}
  \begin{itemize}[itemsep=0pt,parsep=0pt,topsep=0pt]
    \item  $n{\scriptstyle_{s}}$ $\equiv$ the number density [$cm^{-3}$] of species $s$
    \item  $m{\scriptstyle_{s}}$ $\equiv$ the mass [$kg$] of species $s$
    \item  $Z{\scriptstyle_{s}}$ $\equiv$ the charge state of species $s$
    \item  $q{\scriptstyle_{s}}$ $\equiv$ the charge [$C$] of species $s$ $=$ $Z{\scriptstyle_{s}} \ e$
    \item  $T{\scriptstyle_{s, j}}$ $\equiv$ the scalar temperature [$eV$] of the j$^{th}$ component of species $s$
    \item  $\tensor*{ \mathcal{T}{\scriptstyle_{j}} }{^{s'}_{s}}$ $=$ $\left(T{\scriptstyle_{s'}}/T{\scriptstyle_{s}}\right){\scriptstyle_{j}}$ $\equiv$ the temperature ratio [N/A] of species $s$ and $s'$ of the j$^{th}$ component
    \item  $\mathcal{A}{\scriptstyle_{s}}$ $=$ $\left(T{\scriptstyle_{\perp}}/T{\scriptstyle_{\parallel}}\right){\scriptstyle_{s}}$ $\equiv$ the temperature anisotropy [N/A] of species $s$
    \item  $V{\scriptstyle_{Ts, j}}$ $\equiv$ the most probable thermal speed [$km \ s^{-1}$] of a one-dimensional velocity distribution (see Equation \ref{eq:params_2})
    \item  $\mathbf{V}{\scriptstyle_{os}}$ $\equiv$ the drift velocity [$km \ s^{-1}$] of species $s$ in the plasma bulk flow rest frame
    \item  $C{\scriptstyle_{s}}$ $\equiv$ the sound or ion-acoustic sound speed [$km \ s^{-1}$] \citep[see supplemental pdf file][for definitions]{wilsoniii19k}
    \item  $V{\scriptstyle_{A}}$ $\equiv$ the Alfv\'{e}n speed [$km \ s^{-1}$] \citep[see supplemental pdf file][for definitions]{wilsoniii19k}
    \item  $V{\scriptstyle_{f}}$ $\equiv$ the fast mode speed [$km \ s^{-1}$] \citep[see supplemental pdf file][for definitions]{wilsoniii19k}
    \item  $\Omega{\scriptstyle_{cs}}$ $\equiv$ the angular cyclotron frequency [$rad \ s^{-1}$] (see Equation \ref{eq:params_3})
    \item  $\omega{\scriptstyle_{ps}}$ $\equiv$ the angular plasma frequency [$rad \ s^{-1}$] (see Equation \ref{eq:params_4})
    \item  $\lambda{\scriptstyle_{De}}$ $\equiv$ the electron Debye length [$m$] (see Equation \ref{eq:params_5})
    \item  $\rho{\scriptstyle_{cs}}$ $\equiv$ the thermal gyroradius [$km$] (see Equation \ref{eq:params_6})
    \item  $\lambda{\scriptstyle_{s}}$ $\equiv$ the inertial length [$km$] (see Equation \ref{eq:params_7})
    \item  $\beta{\scriptstyle_{s, j}}$ $\equiv$ the plasma beta [N/A] of the j$^{th}$ component of species $s$ (see Equations \ref{eq:params_8} and \ref{eq:params_9})
    \item  $\kappa{\scriptstyle_{s}}$ $\equiv$ the kappa exponent of species $s$ \citep[e.g., see][for definition in model fit equation]{wilsoniii19a}
    \item  $s{\scriptstyle_{s}}$ $\equiv$ the symmetric self-similar exponent of species $s$ \citep[e.g., see][for definition in model fit equation]{wilsoniii19a}
    \item  $p{\scriptstyle_{s}}$($q{\scriptstyle_{s}}$) $\equiv$ the parallel(perpendicular) asymmetric self-similar exponent of species $s$ \citep[e.g., see][for definition in model fit equation]{wilsoniii19a}
    \item  $\phi{\scriptstyle_{sc}}$ $\equiv$ the scalar, quasi-static spacecraft potential [eV] \citep[e.g.,][]{pulupa14a, scime94a} (see Appendices of Paper I for more details)
    \item  $E{\scriptstyle_{min}}$ $\equiv$ the minimum energy bin midpoint value [eV] of an electrostatic analyzer \citep[e.g., see Appendices in][]{wilsoniii17c, wilsoniii18b}
    \item  $q{\scriptstyle_{e, \parallel}}$ $=$ $\tfrac{m{\scriptstyle_{e}}}{2} \ \int \ d^{3}v \ f{\scriptstyle_{e}}^{\left( mod \right)} v{\scriptstyle_{\parallel}} \ v^{2}$ $\equiv$ the parallel electron heat flux [$\mu W \ m^{-2}$] of the entire electron VDF model, $f{\scriptstyle_{e}}^{\left( mod \right)}$ $=$ $f^{\left( core \right)}$ $+$ $f^{\left( halo \right)}$ $+$ $f^{\left( beam \right)}$
    \item  $q{\scriptstyle_{e o}}$ $=$ $\tfrac{3}{2} \ m{\scriptstyle_{e}} \ n{\scriptstyle_{e}} \ V{\scriptstyle_{Tec, \parallel}}^{3}$ $\equiv$ the free-streaming limit electron heat flux [$\mu W \ m^{-2}$] \citep[e.g.,][]{gary99a}
  \end{itemize}
\end{itemize}

\noindent  Similar to Paper I, the variables that rely upon multiple parameters are given in the following equations:

\begin{subequations}
  \begin{align}
    T{\scriptstyle_{eff, j}} & = \frac{ \sum_{s} n{\scriptstyle_{s}} \ T{\scriptstyle_{s, j}} }{ \sum_{s} n{\scriptstyle_{s}} } \label{eq:params_0} \\
    T{\scriptstyle_{s, tot}} & = \frac{1}{3} \left( T{\scriptstyle_{s, \parallel}} + 2 \ T{\scriptstyle_{s, \perp}} \right) \label{eq:params_1} \\
    V{\scriptstyle_{Ts, j}} & = \sqrt{ \frac{ 2 \ k{\scriptstyle_{B}} \ T{\scriptstyle_{s, j}} }{ m{\scriptstyle_{s}} } } \label{eq:params_2} \\
    \Omega{\scriptstyle_{cs}} & = \frac{ q{\scriptstyle_{s}} \ B{\scriptstyle_{o}} }{ m{\scriptstyle_{s}} } \label{eq:params_3} \\
    \omega{\scriptstyle_{ps}} & = \sqrt{ \frac{ n{\scriptstyle_{s}} \ q{\scriptstyle_{s}}^{2} }{ \varepsilon{\scriptstyle_{o}} \ m{\scriptstyle_{s}} } } \label{eq:params_4} \\
    \lambda{\scriptstyle_{De}} & = \frac{ V{\scriptstyle_{Te, tot}} }{ \sqrt{ 2 } \ \omega{\scriptstyle_{pe}} } = \sqrt{ \frac{ \varepsilon{\scriptstyle_{o}} \ k{\scriptstyle_{B}} \ T{\scriptstyle_{e, tot}} }{ n{\scriptstyle_{e}} \ e^{2} } } \label{eq:params_5} \\
    \rho{\scriptstyle_{cs}} & = \frac{ V{\scriptstyle_{Ts, tot}} }{ \Omega{\scriptstyle_{cs}} } \label{eq:params_6} \\
    \lambda{\scriptstyle_{s}} & = \frac{ c }{ \omega{\scriptstyle_{ps}} } \label{eq:params_7} \\
    \beta{\scriptstyle_{s, j}} & = \frac{ 2 \mu{\scriptstyle_{o}} n{\scriptstyle_{s}} k{\scriptstyle_{B}} T{\scriptstyle_{s, j}} }{ \lvert \mathbf{B}{\scriptstyle_{o}} \rvert^{2} } \label{eq:params_8} \\
    \beta{\scriptstyle_{eff, j}} & = \frac{ 2 \mu{\scriptstyle_{o}} n{\scriptstyle_{eff}} k{\scriptstyle_{B}} T{\scriptstyle_{eff, j}} }{ \lvert \mathbf{B}{\scriptstyle_{o}} \rvert^{2} } \label{eq:params_9} \\
    \intertext{where $n{\scriptstyle_{eff}}$ is defined as:}
    n{\scriptstyle_{eff}} & = \sum_{s} \ n{\scriptstyle_{es}} \label{eq:params_10}
  \end{align}
\end{subequations}

\indent  Following the format from \citet[][]{wilsoniii18b}, one can calculate estimates of Coulomb collision rates\footnote{Note that the rates are for isotropic Maxwellian velocity distributions and would change for kappa and self-similar depending on the exponent values \citep[e.g.,][]{marsch85a}.} \citep[e.g.,][]{hernandez85a, hinton84a, krall73a, schunk75a, schunk77a, spitzer53a}, $\nu{\scriptstyle_{ss'}}$, between species $s$ and $s'$ given by:

\begin{subequations}
  \begin{align}
    \nu{\scriptstyle_{ss'}} & = \frac{ C{\scriptstyle_{ss'}} \ n{\scriptstyle_{s'}} }{ V{\scriptstyle_{Tss'}}^{3} } \ \ln{\Lambda{\scriptstyle_{ss'}}} \label{eq:coulomb_coll_0} \\
    C{\scriptstyle_{ss'}} & = \frac{ A{\scriptstyle_{ss'}} \ e^{4} }{ 3 \left( 4 \pi \varepsilon{\scriptstyle_{o}} \right)^{2} \ \mu{\scriptstyle_{ss'}}^{2} } \label{eq:coulomb_coll_1} \\
    \Lambda{\scriptstyle_{ss'}} & \simeq \frac{ \left( 4 \pi \varepsilon{\scriptstyle_{o}} \right) \ \mu{\scriptstyle_{ss'}} \ V{\scriptstyle_{Tss'}}^{2} }{ \sqrt{2} \ Z{\scriptstyle_{s}} \ Z{\scriptstyle_{s'}} \ e^{2} } \left[ \left( \frac{ \omega{\scriptstyle_{ps}} }{ V{\scriptstyle_{Ts, tot}} } \right)^{2} + \left( \frac{ \omega{\scriptstyle_{ps'}} }{ V{\scriptstyle_{Ts', tot}} } \right)^{2} \right]^{-1/2} \label{eq:coulomb_coll_2} \\
    V{\scriptstyle_{Tss'}} & = \sqrt{ V{\scriptstyle_{Ts, tot}}^{2} + V{\scriptstyle_{Ts', tot}}^{2} } \label{eq:coulomb_coll_3} \\
    \mu{\scriptstyle_{ss'}} & = \frac{ m{\scriptstyle_{s}} \ m{\scriptstyle_{s'}} }{ \left( m{\scriptstyle_{s}} + m{\scriptstyle_{s'}} \right) } \label{eq:coulomb_coll_4}
  \end{align}
\end{subequations}

\noindent  where the species-dependent integration constants $A{\scriptstyle_{ss'}}$ are given by:

\begin{subequations}
  \begin{align}
    A{\scriptstyle_{ee}} & = 4 \sqrt{ 2 \pi } \label{eq:coulomb_coll_5} \\
    A{\scriptstyle_{pp}} & = 4 \sqrt{ 2 \pi } \label{eq:coulomb_coll_6} \\
    A{\scriptstyle_{\alpha \alpha}} & = 64 \sqrt{ 2 \pi } \label{eq:coulomb_coll_7} \\
    A{\scriptstyle_{ep}} & = 2 \sqrt{ 4 \pi } \label{eq:coulomb_coll_8} \\
    A{\scriptstyle_{e \alpha}} & = 8 \sqrt{ 4 \pi } \label{eq:coulomb_coll_9} \\
    A{\scriptstyle_{p \alpha}} & = 8 \sqrt{ 2 \pi } \label{eq:coulomb_coll_10}
  \end{align}
\end{subequations}

\noindent  Then the particle rms mean free path is given by:

\begin{equation}
  \label{eq:coulomb_coll_11}
  \lambda{\scriptstyle_{ss'}}^{mpf} = \frac{ V{\scriptstyle_{Tss'}} }{ \nu{\scriptstyle_{ss'}} }
\end{equation}

\indent  For the macroscopic shock parameters, the values are averaged over asymptotic regions away from the shock transition region.

\begin{itemize}[itemsep=0pt,parsep=0pt,topsep=0pt]
  \item[]  \textit{shock parameters}
  \begin{itemize}[itemsep=0pt,parsep=0pt,topsep=0pt]
    \item  subscripts $up$ and $dn$ $\equiv$ denote the upstream (i.e., before the shock arrives time-wise at the spacecraft for a forward shock) and downstream (i.e., the shocked region)
    \item  $\langle Q \rangle{\scriptstyle_{j}}$ $\equiv$ the average of parameter $Q$ over the $j^{th}$ shock region, where $j$ $=$ $up$ or $dn$
    \item  $\Delta Q$ $=$ $\langle Q \rangle{\scriptstyle_{dn}}$ - $\langle Q \rangle{\scriptstyle_{up}}$ $\equiv$ the change in the asymptotic average of parameter $Q$ over the $j^{th}$ shock region
    \item  $\mathcal{R}{\scriptstyle_{ns}}$ $=$ $\langle n{\scriptstyle_{s}} \rangle{\scriptstyle_{dn}}$/$\langle n{\scriptstyle_{s}} \rangle{\scriptstyle_{up}}$ $\equiv$ the shock compression ratio of species $s$
    \item  $\mathcal{R}{\scriptstyle_{Ts,j}}$ $=$ $\langle T{\scriptstyle_{s,j}} \rangle{\scriptstyle_{dn}}$/$\langle T{\scriptstyle_{s,j}} \rangle{\scriptstyle_{up}}$ $\equiv$ the downstream-to-upstream $j^{th}$ component temperature ratio of species $s$
    \item  $\mathbf{n}{\scriptstyle_{sh}}$ $\equiv$ the shock normal unit vector [N/A]
    \item  $\theta{\scriptstyle_{Bn}}$ $\equiv$ the shock normal angle\footnote{The acute reference angle between $\langle \mathbf{B}{\scriptstyle_{o}} \rangle{\scriptstyle_{up}}$ and $\mathbf{n}{\scriptstyle_{sh}}$.} [deg]
    \item  $\langle \lvert V{\scriptstyle_{shn}} \rvert \rangle{\scriptstyle_{j}}$ $\equiv$ the $j^{th}$ region average shock normal speed [$km \ s^{-1}$] in the spacecraft frame
    \item  $\langle \lvert U{\scriptstyle_{shn}} \rvert \rangle{\scriptstyle_{j}}$ $\equiv$ the $j^{th}$ region average shock normal speed [$km \ s^{-1}$] in the shock rest frame (i.e., the speed of the flow relative to the shock)
    \item  $\langle M{\scriptstyle_{A}} \rangle{\scriptstyle_{j}}$ $\equiv$ the $j^{th}$ region average Alfv\'{e}nic Mach number [N/A] $=$ $\langle \lvert U{\scriptstyle_{shn}} \rvert \rangle{\scriptstyle_{j}} / \langle V{\scriptstyle_{A}} \rangle{\scriptstyle_{j}}$
    \item  $\langle M{\scriptstyle_{f}} \rangle{\scriptstyle_{j}}$ $\equiv$ the $j^{th}$ region average fast mode Mach number [N/A] $=$ $\langle \lvert U{\scriptstyle_{shn}} \rvert \rangle{\scriptstyle_{j}} / \langle V{\scriptstyle_{f}} \rangle{\scriptstyle_{j}}$
    \item  $M{\scriptstyle_{cr}}$ $\equiv$ the first critical Mach number [N/A]
  \end{itemize}
\end{itemize}

\noindent  These definitions are used throughout.

\phantomsection   
\section{Integrated Velocity Moments}  \label{app:IntegratedVelocityMoments}

\indent  This appendix provides details regarding the numerical integration of the total model fit VDFs to determine velocity moments and comparison with summed velocity moments.  Note that velocity moments of the components can be summed to find totals for the entire VDF against which one can compare an integrated equivalent.  The comparison is performed as a sanity check.

\indent  The n$^{th}$ moment of a velocity distribution function, $f\left( \mathbf{x}, \mathbf{v}, t \right)$, is generically defined as the expectation value of the n$^{th}$ order of a dynamical function, $g\left( \mathbf{x}, \mathbf{v} \right)$, given by:

\begin{equation}
  \label{eq:velmoms_0}
  \langle g^{n}\left( \mathbf{x}, \mathbf{v} \right) \rangle \equiv \int \ d^{3}v \ g^{n}\left( \mathbf{x}, \mathbf{v} \right) \ f\left( \mathbf{x}, \mathbf{v}, t \right)
\end{equation}

\noindent  where the zeroth moment is the volume density (e.g., number density), the first relates to peak of the distribution (e.g., the bulk flow velocity), the second to the width of the peak (e.g., random kinetic energy density or pressure tensor), the third to the skewness (e.g., heat flux tensor), the fourth to the kurtosis, etc.

\indent  For velocity moment calculations of in situ spacecraft measurements, the dynamical function is the velocity coordinate $\mathbf{v}$ and the spatial and temporal dependence drop out resulting in $f\left( \mathbf{x}, \mathbf{v}, t \right)$ $\rightarrow$ $f\left( \mathbf{v} \right)$ for each VDF observed\footnote{Note that the pressure and heat flux tensors should be computed in the species rest frame, thus the dynamical function is the peculiar velocity or $\mathbf{v} - \mathbf{V}{\scriptstyle_{os}}$}.  The total/entire electron model VDF, $f{\scriptstyle_{s}}^{\left( mod \right)}$ $=$ $f^{\left( core \right)}$ $+$ $f^{\left( halo \right)}$ $+$ $f^{\left( beam \right)}$, is constructed from the valid fit parameters discussed in Section \ref{sec:DefinitionsDataSets} only for VDFs with stable solutions for all three components.  The integrals are calculated in the core electron rest frame, thus the only relevant heat flux component is the parallel, $q{\scriptstyle_{e, \parallel}}$, because the suprathermal electrons have no finite perpendicular drift velocities (e.g., see Paper I).

\indent  The integrals are numerically approximated using the Simpson's $\tfrac{1}{3}$ Rule algorithm.  Since the models are gyrotropic, the two perpendicular velocities are symmetric reducing the three dimensional integrals to two dimensional integrals\footnote{$d^{3}v$ $\rightarrow$ $\pi \ \lvert v{\scriptstyle_{\perp}} \lvert \ dv{\scriptstyle_{\perp}} \ dv{\scriptstyle_{\parallel}}$}.  Some simple bench-marking tests revealed that the range of the regular velocity grid coordinates was more important than the number of grid points for reducing the difference between the ``known'' and integrated value of any given velocity moment.  It was found that the minimum threshold for grid range and density while simultaneously reducing the computational time to keep the percent difference within less than a percent was $\pm$80,000 km/s and 301x301 points.  The velocity grid is constructed in linear space because tests of logarithmically spaced velocity coordinates resulted in larger percent differences.

\indent  For brevity the percent difference between the summed and integrated velocity moment parameters is defined as $\Delta Q{\scriptstyle_{i2f}}$ $=$ $\lvert  Q{\scriptstyle_{int}} - Q{\scriptstyle_{eff}} \rvert/Q{\scriptstyle_{eff}} \times 100\%$, where the subscript $eff$ is for \emph{effective} and $int$ is for \emph{integrated}.  For instance, the percent difference between the summed and integrated electron density is given as $\Delta n{\scriptstyle_{i2f}}$ $=$ $\lvert n{\scriptstyle_{int}} - n{\scriptstyle_{eff}} \rvert/n{\scriptstyle_{eff}} \times 100\%$ (the $e$ for electron is assumed, since only electron VDFs are integrated).  The one-variable statistics of these percent differences are shown as $X{\scriptstyle_{25\%}}$--$X{\scriptstyle_{75\%}}$($\bar{X}$)[$\tilde{X}$] and given by:
\begin{itemize}[itemsep=0pt,parsep=0pt,topsep=0pt]
  \item  $\Delta n{\scriptstyle_{i2f}}$ $\sim$ 0.001\%--0.66\%(0.74\%)[0.002\%]
  \item  $\Delta \lvert \mathbf{V}{\scriptstyle_{oe,\parallel}} \rvert{\scriptstyle_{i2f}}$ $\sim$ 0.001\%--848\%(3068\%)[3.14\%]
  \item  $\Delta T{\scriptstyle_{e \parallel, i2f}}$ $\sim$ 1.21\%--6.59\%(4.68\%)[3.04\%]
  \item  $\Delta T{\scriptstyle_{e \perp, i2f}}$ $\sim$ 0.003\%--1.94\%(2.28\%)[0.012\%]
  \item  $\Delta T{\scriptstyle_{e   tot, i2f}}$ $\sim$ 0.74\%--25.0\%(26.3\%)[2.47\%]
\end{itemize}
\noindent  The large values of the parallel drift percent difference are dominated by outliers, as evidenced by the small median value.  Thus, the numerical integration results are within expected uncertainties/errors.

\indent  These tests were performed to verify the accuracy of the integrated $q{\scriptstyle_{e, \parallel}}$ values since there is no properly summed value from the original fit parameter sets.  Further, the inaccuracy of this type of numerical integration increases with increasing velocity moment \citep[e.g.,][]{gershman15a, paschmann98a, song97a}, thus why the errors in the lowest moments were minimized prior to calculation of the heat flux.

\phantomsection   
\section{Extra Statistics}  \label{app:ExtraStatistics}

\indent  This appendix presents additions to the statistics and tables presented in the main paper.  The tables are referenced as supplements for the statistics in the paper.

\begin{figure*}
  \centering
    {\includegraphics[trim = 0mm 0mm 0mm 0mm, clip, height=180mm]{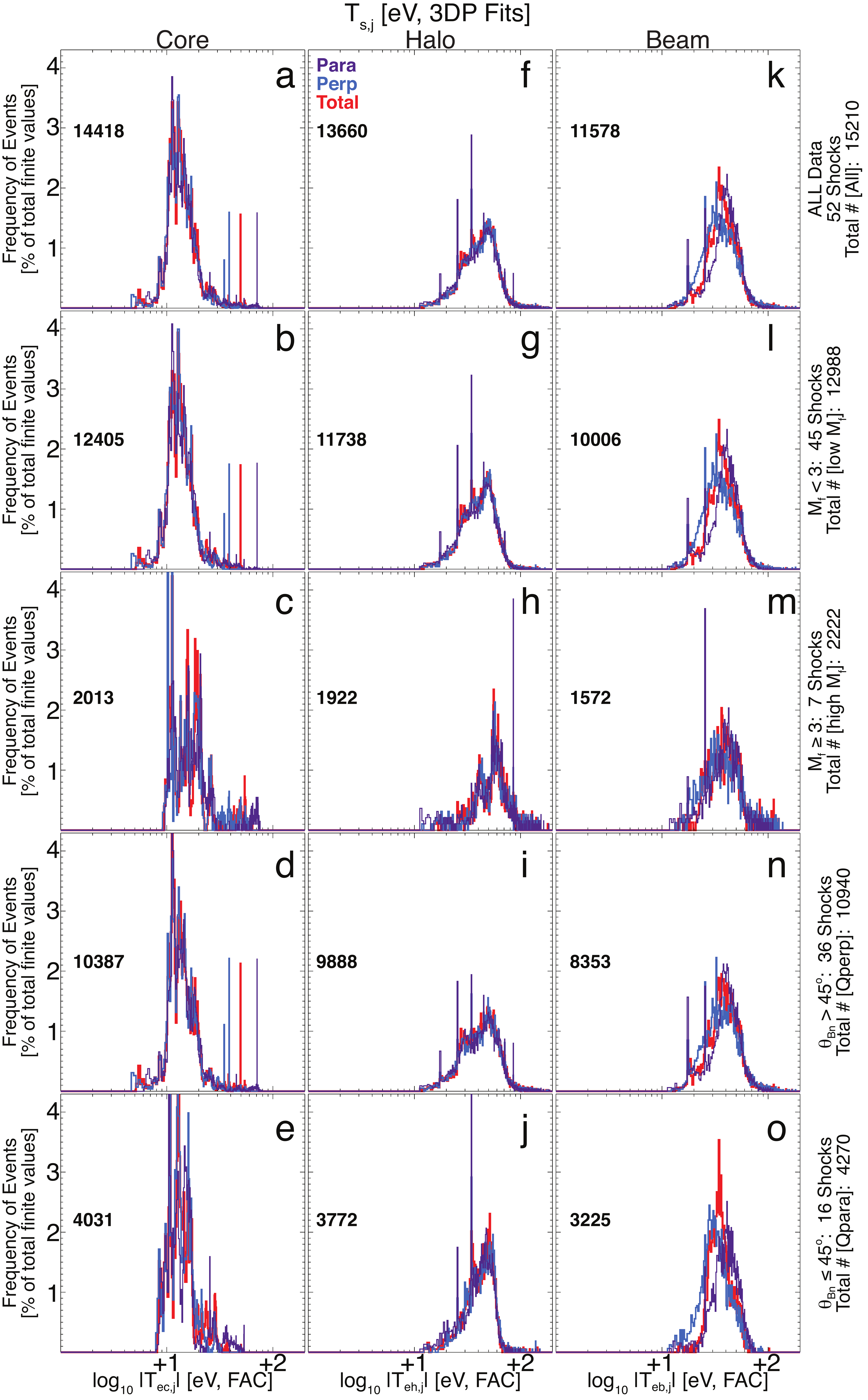}}
    \caption{Temperatures [eV] for different electron components in each column with a similar format to that of Figure \ref{fig:Temperatures} but for other selection criteria.  The first row is the same as that in Figure \ref{fig:Temperatures}, but the next four are, in the following order, low Mach number, high Mach number, quasi-perpendicular, and quasi-parallel shocks.  Similar to Figure \ref{fig:Temperatures}, the total number of finite points for each panel are the same for each color-coded line (label in panel f).  To the far right are the total number of VDFs analyzed for each criteria for reference.}
    \label{fig:ExtraTemperatures}
\end{figure*}

\startlongtable  
\begin{deluxetable*}{| l | c | c | c | c | c | c || c | c | c | c | c | c |}
  \tabletypesize{\footnotesize}    
  \tablecaption{Temperature Parameters \label{tab:ExtraTemperatures}}
  \tablehead{\colhead{Temp. [eV]} & \colhead{$X{\scriptstyle_{min}}$}\tablenotemark{a} & \colhead{$X{\scriptstyle_{max}}$} & \colhead{$\bar{X}$} & \colhead{$\tilde{X}$} & \colhead{$X{\scriptstyle_{25\%}}$} & \colhead{$X{\scriptstyle_{75\%}}$} & \colhead{$X{\scriptstyle_{min}}$} & \colhead{$X{\scriptstyle_{max}}$} & \colhead{$\bar{X}$} & \colhead{$\tilde{X}$} & \colhead{$X{\scriptstyle_{25\%}}$} & \colhead{$X{\scriptstyle_{75\%}}$}}
  \startdata
  & \multicolumn{6}{ c }{\textit{Criteria UP: \totalnfitsups~VDFs}} & \multicolumn{6}{ ||c| }{\textit{Criteria DN: \totalnfitsdns~VDFs}} \\
  \hline
  $T{\scriptstyle_{ec, \parallel}}$  & 5.67 & 36.7 & 14.1 & 13.2 & 11.2 & 16.2 & 7.34 & 89.1 & 22.7 & 16.5 & 13.6 & 24.7  \\
  $T{\scriptstyle_{ec, \perp}}$      & 4.75 & 26.5 & 13.0 & 12.8 & 10.9 & 15.1 & 7.16 & 62.8 & 19.0 & 16.3 & 13.2 & 20.7  \\
  $T{\scriptstyle_{ec, tot}}$        & 5.06 & 27.3 & 13.4 & 13.0 & 10.9 & 15.5 & 7.22 & 67.2 & 20.2 & 16.4 & 13.4 & 22.6  \\
  $T{\scriptstyle_{eh, \parallel}}$  & 11.6 &  188 & 46.6 & 46.3 & 35.7 & 54.9 & 11.7 &  249 & 50.8 & 48.2 & 35.7 & 60.5  \\
  $T{\scriptstyle_{eh, \perp}}$      & 11.5 &  204 & 48.5 & 47.5 & 35.7 & 56.9 & 11.4 &  255 & 52.4 & 49.6 & 38.1 & 60.1  \\
  $T{\scriptstyle_{eh, tot}}$        & 11.6 &  180 & 47.9 & 47.2 & 36.3 & 55.7 & 11.7 &  222 & 51.8 & 49.0 & 38.0 & 59.8  \\
  $T{\scriptstyle_{eb, \parallel}}$  & 12.1 &  280 & 43.0 & 42.2 & 36.4 & 49.9 & 11.5 &  201 & 45.1 & 43.6 & 36.0 & 52.4  \\
  $T{\scriptstyle_{eb, \perp}}$      & 11.9 &  264 & 39.8 & 36.6 & 28.7 & 46.2 & 11.7 &  277 & 44.7 & 41.3 & 32.3 & 53.1  \\
  $T{\scriptstyle_{eb, tot}}$        & 12.3 &  269 & 40.9 & 38.8 & 32.4 & 46.6 & 14.3 &  238 & 44.9 & 41.6 & 34.8 & 52.4  \\
  $T{\scriptstyle_{eff, \parallel}}$ & 6.97 & 95.6 & 16.3 & 14.9 & 12.8 & 18.6 & 8.41 &  167 & 23.8 & 17.6 & 14.6 & 26.3  \\
  $T{\scriptstyle_{eff, \perp}}$     & 4.93 &  111 & 15.1 & 14.4 & 12.3 & 17.3 & 8.17 &  170 & 20.3 & 17.3 & 14.2 & 22.4  \\
  $T{\scriptstyle_{eff, tot}}$       & 5.61 &  106 & 15.5 & 14.6 & 12.5 & 17.9 & 8.35 &  169 & 21.4 & 17.4 & 14.4 & 24.3  \\
  \hline
  & \multicolumn{6}{ c }{\textit{Criteria LM: \totalnfitslMf~VDFs}} & \multicolumn{6}{ ||c| }{\textit{Criteria HM: \totalnfitshMf~VDFs}} \\
  \hline
  $T{\scriptstyle_{ec, \parallel}}$  & 5.67 & 84.0 & 17.9 & 14.7 & 11.9 & 18.2 & 9.49 & 89.1 & 26.3 & 19.5 & 14.7 & 26.3  \\
  $T{\scriptstyle_{ec, \perp}}$      & 4.75 & 56.0 & 15.6 & 14.0 & 11.9 & 17.1 & 9.23 & 62.8 & 22.0 & 18.5 & 14.2 & 25.2  \\
  $T{\scriptstyle_{ec, tot}}$        & 5.06 & 61.2 & 16.3 & 14.2 & 11.9 & 17.5 & 9.45 & 67.2 & 23.4 & 19.0 & 14.4 & 25.8  \\
  $T{\scriptstyle_{eh, \parallel}}$  & 11.7 &  249 & 46.2 & 45.7 & 34.8 & 55.0 & 11.6 &  221 & 66.2 & 60.6 & 49.8 & 74.7  \\
  $T{\scriptstyle_{eh, \perp}}$      & 11.4 &  255 & 47.8 & 47.2 & 35.9 & 56.4 & 11.5 &  226 & 68.7 & 60.6 & 49.4 & 78.5  \\
  $T{\scriptstyle_{eh, tot}}$        & 11.7 &  203 & 47.3 & 46.9 & 36.2 & 55.6 & 11.6 &  222 & 67.9 & 60.6 & 49.1 & 76.5  \\
  $T{\scriptstyle_{eb, \parallel}}$  & 11.5 &  201 & 43.9 & 42.8 & 36.4 & 51.1 & 12.1 &  280 & 46.3 & 43.2 & 34.6 & 53.2  \\
  $T{\scriptstyle_{eb, \perp}}$      & 11.7 &  275 & 41.7 & 38.7 & 30.4 & 49.3 & 11.9 &  277 & 48.4 & 42.8 & 32.1 & 55.4  \\
  $T{\scriptstyle_{eb, tot}}$        & 13.8 &  238 & 42.4 & 39.9 & 33.6 & 49.4 & 12.3 &  269 & 47.7 & 42.9 & 34.1 & 53.0  \\
  $T{\scriptstyle_{eff, \parallel}}$ & 6.97 & 78.8 & 19.3 & 16.0 & 13.4 & 20.0 & 11.3 &  167 & 28.5 & 21.8 & 16.6 & 29.7  \\
  $T{\scriptstyle_{eff, \perp}}$     & 4.93 & 89.9 & 17.1 & 15.3 & 13.1 & 19.0 & 10.3 &  170 & 24.4 & 20.3 & 16.5 & 28.6  \\
  $T{\scriptstyle_{eff, tot}}$       & 5.61 & 80.6 & 17.8 & 15.5 & 13.3 & 19.3 & 11.1 &  169 & 25.8 & 20.7 & 16.5 & 29.1  \\
  \hline
  & \multicolumn{6}{ c }{\textit{Criteria PE: \totalnfitsQpe~VDFs}} & \multicolumn{6}{ ||c| }{\textit{Criteria PA: \totalnfitsQpa~VDFs}} \\
  \hline
  $T{\scriptstyle_{ec, \parallel}}$  & 5.67 & 89.1 & 19.3 & 14.9 & 12.1 & 19.1 & 7.93 & 55.8 & 18.5 & 15.2 & 12.2 & 19.1  \\
  $T{\scriptstyle_{ec, \perp}}$      & 4.75 & 62.8 & 16.9 & 14.6 & 12.0 & 18.5 & 7.78 & 50.3 & 15.2 & 13.7 & 12.0 & 17.0  \\
  $T{\scriptstyle_{ec, tot}}$        & 5.06 & 67.2 & 17.7 & 14.7 & 12.0 & 18.7 & 7.88 & 51.4 & 16.3 & 14.2 & 12.3 & 17.6  \\
  $T{\scriptstyle_{eh, \parallel}}$  & 11.6 &  249 & 50.8 & 48.9 & 36.0 & 60.8 & 11.7 &  171 & 44.5 & 44.8 & 35.1 & 52.5  \\
  $T{\scriptstyle_{eh, \perp}}$      & 11.4 &  226 & 52.1 & 49.6 & 37.4 & 61.0 & 11.9 &  255 & 47.2 & 46.8 & 36.9 & 54.9  \\
  $T{\scriptstyle_{eh, tot}}$        & 11.6 &  222 & 51.7 & 49.2 & 37.6 & 60.5 & 12.0 &  203 & 46.3 & 46.4 & 36.8 & 53.5  \\
  $T{\scriptstyle_{eb, \parallel}}$  & 11.5 &  280 & 44.5 & 42.8 & 36.4 & 51.6 & 12.1 &  201 & 43.6 & 42.8 & 35.6 & 51.0  \\
  $T{\scriptstyle_{eb, \perp}}$      & 12.6 &  277 & 44.8 & 41.5 & 31.9 & 52.4 & 11.7 &  207 & 37.0 & 34.3 & 28.4 & 43.7  \\
  $T{\scriptstyle_{eb, tot}}$        & 13.8 &  269 & 44.7 & 41.8 & 34.5 & 51.7 & 12.3 &  151 & 39.2 & 36.9 & 32.2 & 44.9  \\
  $T{\scriptstyle_{eff, \parallel}}$ & 6.97 &  167 & 20.9 & 16.3 & 13.6 & 21.1 & 7.93 & 65.0 & 19.8 & 16.7 & 13.6 & 20.9  \\
  $T{\scriptstyle_{eff, \perp}}$     & 4.93 &  170 & 18.7 & 15.9 & 13.5 & 20.3 & 8.22 & 66.7 & 16.7 & 15.2 & 13.3 & 18.2  \\
  $T{\scriptstyle_{eff, tot}}$       & 5.61 &  169 & 19.4 & 16.1 & 13.6 & 20.6 & 8.13 & 66.0 & 17.7 & 15.6 & 13.6 & 18.7  \\
  \hline
  \enddata
  \tablenotetext{a}{Header symbols match that of Table \ref{tab:Temperatures}}
  \tablecomments{For symbol definitions, see Appendix \ref{app:Definitions}.}
\end{deluxetable*}

\clearpage
\startlongtable  
\begin{deluxetable*}{| l | c | c | c | c | c | c || c | c | c | c | c | c |}
  \tabletypesize{\footnotesize}    
  \tablecaption{Density Parameters \label{tab:ExtraDensity}}
  \tablehead{\colhead{$n{\scriptstyle_{s}}$ [$cm^{-3}$]} & \colhead{$X{\scriptstyle_{min}}$}\tablenotemark{a} & \colhead{$X{\scriptstyle_{max}}$} & \colhead{$\bar{X}$} & \colhead{$\tilde{X}$} & \colhead{$X{\scriptstyle_{25\%}}$} & \colhead{$X{\scriptstyle_{75\%}}$} & \colhead{$X{\scriptstyle_{min}}$} & \colhead{$X{\scriptstyle_{max}}$} & \colhead{$\bar{X}$} & \colhead{$\tilde{X}$} & \colhead{$X{\scriptstyle_{25\%}}$} & \colhead{$X{\scriptstyle_{75\%}}$}}
  \startdata
  & \multicolumn{6}{ c }{\textit{Criteria UP: \totalnfitsups~VDFs}} & \multicolumn{6}{ ||c| }{\textit{Criteria DN: \totalnfitsdns~VDFs}} \\
  \hline
  $n{\scriptstyle_{p}}$                          &    0.10 & 42.1 & 9.88 & 8.87 &  3.98 & 13.6 &    0.62 & 76.2 & 18.7 & 16.9 &  7.94 & 25.8  \\
  $n{\scriptstyle_{\alpha}}$                     &    0.02 & 2.25 & 0.24 & 0.15 &  0.10 & 0.29 &    0.03 & 4.75 & 0.64 & 0.53 &  0.23 & 0.88  \\
  $n{\scriptstyle_{i}}$                          &    0.18 & 66.3 & 9.24 & 8.47 &  4.49 & 12.2 &    0.43 & 98.8 & 20.2 & 16.3 &  8.92 & 27.8  \\
  $n{\scriptstyle_{ec}}$                         &    0.30 & 26.9 & 9.07 & 8.29 &  4.35 & 12.6 &    0.63 & 55.3 & 17.3 & 16.4 &  8.48 & 24.3  \\
  $n{\scriptstyle_{eh}}$                         &   0.002 & 4.45 & 0.42 & 0.27 &  0.17 & 0.49 &   0.002 & 6.87 & 0.58 & 0.43 &  0.26 & 0.70  \\
  $n{\scriptstyle_{eb}}$                         &  0.0009 & 2.29 & 0.23 & 0.16 &  0.10 & 0.28 &   0.001 & 3.50 & 0.24 & 0.18 &  0.10 & 0.30  \\
  $n{\scriptstyle_{eff}}$                        &   0.004 & 27.4 & 9.56 & 8.63 &  4.76 & 13.7 &   0.004 & 56.9 & 17.9 & 17.0 &  9.05 & 24.9  \\
  $n{\scriptstyle_{eh}} / n{\scriptstyle_{ec}}$  &  0.0006 & 0.30 & 0.05 & 0.04 &  0.02 & 0.07 &  0.0002 & 0.30 & 0.04 & 0.03 &  0.02 & 0.05  \\
  $n{\scriptstyle_{eb}} / n{\scriptstyle_{ec}}$  & 0.00004 & 0.30 & 0.03 & 0.03 &  0.01 & 0.04 & 0.00003 & 0.30 & 0.02 & 0.01 & 0.006 & 0.03  \\
  $n{\scriptstyle_{eb}} / n{\scriptstyle_{eh}}$  &   0.002 & 9.86 & 0.95 & 0.69 &  0.31 & 1.13 &   0.003 & 9.20 & 0.72 & 0.45 &  0.21 & 0.79  \\
  \hline
  & \multicolumn{6}{ c }{\textit{Criteria LM: \totalnfitslMf~VDFs}} & \multicolumn{6}{ ||c| }{\textit{Criteria HM: \totalnfitshMf~VDFs}} \\
  \hline
  $n{\scriptstyle_{p}}$                          &    0.10 & 76.2 & 14.5 & 11.4 &  6.06 & 20.6 &    0.80 & 45.3 & 17.1 & 13.2 &  7.31 & 26.5  \\
  $n{\scriptstyle_{\alpha}}$                     &    0.02 & 4.75 & 0.44 & 0.25 &  0.12 & 0.65 &    0.06 & 2.16 & 0.56 & 0.42 &  0.22 & 0.84  \\
  $n{\scriptstyle_{i}}$                          &    0.18 & 98.8 & 15.0 & 11.1 &  6.93 & 19.2 &    2.78 & 50.5 & 18.5 & 15.3 &  7.84 & 29.0  \\
  $n{\scriptstyle_{ec}}$                         &    0.30 & 55.3 & 13.4 & 11.2 &  6.49 & 18.5 &    2.50 & 39.0 & 16.0 & 14.3 &  7.99 & 23.5  \\
  $n{\scriptstyle_{eh}}$                         &   0.002 & 6.87 & 0.48 & 0.33 &  0.19 & 0.59 &   0.004 & 6.55 & 0.73 & 0.55 &  0.35 & 0.91  \\
  $n{\scriptstyle_{eb}}$                         &  0.0009 & 3.50 & 0.22 & 0.16 &  0.09 & 0.27 &    0.01 & 2.62 & 0.35 & 0.26 &  0.15 & 0.46  \\
  $n{\scriptstyle_{eff}}$                        &   0.004 & 56.9 & 14.0 & 11.7 &  6.81 & 19.3 &    0.12 & 40.1 & 16.8 & 14.8 &  8.50 & 24.3  \\
  $n{\scriptstyle_{eh}} / n{\scriptstyle_{ec}}$  &  0.0002 & 0.30 & 0.05 & 0.03 &  0.02 & 0.06 &   0.001 & 0.23 & 0.05 & 0.04 &  0.02 & 0.07  \\
  $n{\scriptstyle_{eb}} / n{\scriptstyle_{ec}}$  & 0.00003 & 0.30 & 0.03 & 0.02 & 0.007 & 0.03 &  0.0005 & 0.21 & 0.03 & 0.02 &  0.01 & 0.05  \\
  $n{\scriptstyle_{eb}} / n{\scriptstyle_{eh}}$  &   0.002 & 9.86 & 0.82 & 0.50 &  0.24 & 0.98 &    0.01 & 9.18 & 0.78 & 0.50 &  0.28 & 0.87  \\
  \hline
  & \multicolumn{6}{ c }{\textit{Criteria PE: \totalnfitsQpe~VDFs}} & \multicolumn{6}{ ||c| }{\textit{Criteria PA: \totalnfitsQpa~VDFs}} \\
  \hline
  $n{\scriptstyle_{p}}$                          &    0.10 & 76.2 & 17.0 & 13.8 &  8.18 & 23.0 &    0.95 & 56.2 & 9.45 & 6.09 &  3.29 & 11.0  \\
  $n{\scriptstyle_{\alpha}}$                     &    0.02 & 4.75 & 0.52 & 0.39 &  0.16 & 0.75 &    0.03 & 2.00 & 0.24 & 0.13 &  0.10 & 0.29  \\
  $n{\scriptstyle_{i}}$                          &    0.18 & 98.8 & 17.9 & 12.9 &  8.32 & 23.7 &    1.44 & 52.7 & 9.48 & 7.07 &  3.91 & 13.4  \\
  $n{\scriptstyle_{ec}}$                         &    0.30 & 55.3 & 15.5 & 13.8 &  8.09 & 22.1 &    0.97 & 40.0 & 9.22 & 6.79 &  3.70 & 12.2  \\
  $n{\scriptstyle_{eh}}$                         &   0.002 & 6.55 & 0.59 & 0.43 &  0.24 & 0.72 &   0.002 & 6.87 & 0.31 & 0.23 &  0.14 & 0.38  \\
  $n{\scriptstyle_{eb}}$                         &   0.001 & 3.50 & 0.26 & 0.18 &  0.10 & 0.31 &  0.0009 & 3.33 & 0.19 & 0.14 &  0.09 & 0.26  \\
  $n{\scriptstyle_{eff}}$                        &   0.004 & 56.9 & 16.2 & 14.4 &  8.47 & 22.8 &    0.05 & 40.2 & 9.60 & 7.22 &  3.96 & 12.5  \\
  $n{\scriptstyle_{eh}} / n{\scriptstyle_{ec}}$  &  0.0005 & 0.30 & 0.05 & 0.03 &  0.02 & 0.06 &  0.0002 & 0.30 & 0.05 & 0.03 &  0.02 & 0.06  \\
  $n{\scriptstyle_{eb}} / n{\scriptstyle_{ec}}$  & 0.00003 & 0.28 & 0.02 & 0.02 & 0.007 & 0.03 & 0.00004 & 0.30 & 0.03 & 0.03 &  0.01 & 0.04  \\
  $n{\scriptstyle_{eb}} / n{\scriptstyle_{eh}}$  &   0.002 & 9.86 & 0.77 & 0.49 &  0.21 & 0.89 &   0.005 & 9.65 & 0.94 & 0.65 &  0.35 & 1.13  \\
  \hline
  \enddata
  \tablenotetext{a}{Header symbols match that of Table \ref{tab:Temperatures}}
  \tablecomments{For symbol definitions, see Appendix \ref{app:Definitions}.}
\end{deluxetable*}

\begin{figure*}
  \centering
    {\includegraphics[trim = 0mm 0mm 0mm 0mm, clip, height=180mm]{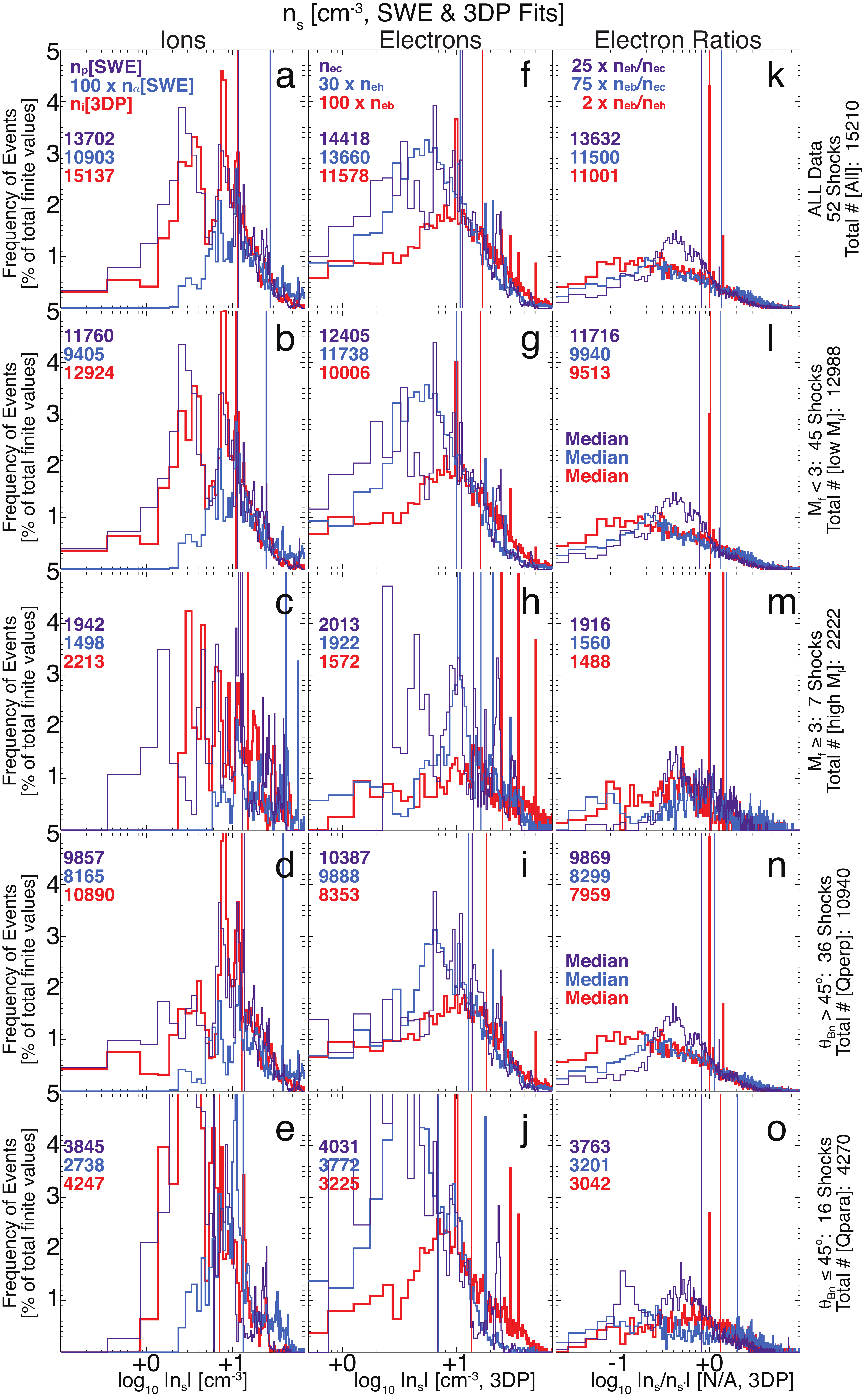}}
    \caption{Densities [$cm^{-3}$] and density ratios with a similar format to that of Figure \ref{fig:Density} but for other selection criteria.  The top row is the same as that in Figure \ref{fig:Density} for reference but the next four rows are the same that in Figure \ref{fig:ExtraTemperatures}.}
    \label{fig:ExtraDensity}
\end{figure*}

\clearpage
\startlongtable  
\begin{deluxetable*}{| l | c | c | c | c | c | c || c | c | c | c | c | c |}
  \tabletypesize{\footnotesize}    
  \tablecaption{Electron Beta Parameters \label{tab:ExtraBetas}}
  \tablehead{\colhead{$\beta{\scriptstyle_{s,j}}$ [N/A]} & \colhead{$X{\scriptstyle_{min}}$}\tablenotemark{a} & \colhead{$X{\scriptstyle_{max}}$} & \colhead{$\bar{X}$} & \colhead{$\tilde{X}$} & \colhead{$X{\scriptstyle_{25\%}}$} & \colhead{$X{\scriptstyle_{75\%}}$} & \colhead{$X{\scriptstyle_{min}}$} & \colhead{$X{\scriptstyle_{max}}$} & \colhead{$\bar{X}$} & \colhead{$\tilde{X}$} & \colhead{$X{\scriptstyle_{25\%}}$} & \colhead{$X{\scriptstyle_{75\%}}$}}
  \startdata
  & \multicolumn{6}{ c }{\textit{Criteria UP: \totalnfitsups~VDFs}} & \multicolumn{6}{ ||c| }{\textit{Criteria DN: \totalnfitsdns~VDFs}} \\
  \hline
  $\beta{\scriptstyle_{ec, \parallel}}$     &    0.05 &  809 & 4.60 & 1.21 & 0.64 & 2.36 &    0.06 & 3313 & 2.89 & 0.91 & 0.51 & 1.70  \\
  $\beta{\scriptstyle_{ec, \perp}}$         &    0.05 &  811 & 4.45 & 1.20 & 0.58 & 2.34 &    0.04 & 3268 & 2.81 & 0.85 & 0.44 & 1.63  \\
  $\beta{\scriptstyle_{ec, tot}}$           &    0.05 &  811 & 4.50 & 1.21 & 0.59 & 2.34 &    0.05 & 3283 & 2.84 & 0.86 & 0.47 & 1.65  \\
  $\beta{\scriptstyle_{eh, \parallel}}$     &  0.0007 &  160 & 0.77 & 0.17 & 0.08 & 0.31 &  0.0001 &  375 & 0.27 & 0.08 & 0.03 & 0.14  \\
  $\beta{\scriptstyle_{eh, \perp}}$         &   0.002 &  162 & 0.78 & 0.17 & 0.08 & 0.32 &  0.0008 &  378 & 0.28 & 0.08 & 0.03 & 0.15  \\
  $\beta{\scriptstyle_{eh, tot}}$           &   0.001 &  161 & 0.78 & 0.17 & 0.08 & 0.32 &  0.0009 &  377 & 0.27 & 0.08 & 0.03 & 0.15  \\
  $\beta{\scriptstyle_{eb, \parallel}}$     &  0.0001 & 33.7 & 0.20 & 0.10 & 0.06 & 0.18 & 0.00002 & 33.2 & 0.07 & 0.03 & 0.01 & 0.06  \\
  $\beta{\scriptstyle_{eb, \perp}}$         &  0.0005 & 31.8 & 0.19 & 0.09 & 0.05 & 0.15 & 0.00003 & 46.4 & 0.07 & 0.03 & 0.01 & 0.06  \\
  $\beta{\scriptstyle_{eb, tot}}$           &  0.0005 & 32.4 & 0.20 & 0.09 & 0.05 & 0.16 & 0.00003 & 42.0 & 0.07 & 0.03 & 0.02 & 0.06  \\
  $\beta{\scriptstyle_{eff, \parallel}}$    &   0.004 &  977 & 5.44 & 1.43 & 0.81 & 2.66 &  0.0009 & 3721 & 3.20 & 1.01 & 0.57 & 1.84  \\
  $\beta{\scriptstyle_{eff, \perp}}$        &   0.002 &  983 & 5.30 & 1.41 & 0.74 & 2.63 &  0.0010 & 3693 & 3.11 & 0.94 & 0.51 & 1.77  \\
  $\beta{\scriptstyle_{eff, tot}}$          &   0.003 &  981 & 5.34 & 1.42 & 0.76 & 2.63 &  0.0009 & 3702 & 3.14 & 0.96 & 0.53 & 1.79  \\
  \hline
  & \multicolumn{6}{ c }{\textit{Criteria LM: \totalnfitslMf~VDFs}} & \multicolumn{6}{ ||c| }{\textit{Criteria HM: \totalnfitshMf~VDFs}} \\
  \hline
  $\beta{\scriptstyle_{ec, \parallel}}$     &    0.05 & 3313 & 2.30 & 0.93 & 0.56 & 1.83 &    0.19 & 1084 & 11.7 & 1.78 & 0.71 & 6.77  \\
  $\beta{\scriptstyle_{ec, \perp}}$         &    0.04 & 3268 & 2.23 & 0.87 & 0.49 & 1.81 &    0.09 & 1080 & 11.3 & 1.70 & 0.66 & 6.67  \\
  $\beta{\scriptstyle_{ec, tot}}$           &    0.05 & 3283 & 2.26 & 0.88 & 0.51 & 1.82 &    0.13 & 1081 & 11.5 & 1.72 & 0.69 & 6.67  \\
  $\beta{\scriptstyle_{eh, \parallel}}$     &  0.0001 &  375 & 0.28 & 0.09 & 0.04 & 0.18 &  0.0003 &  160 & 1.72 & 0.27 & 0.11 & 1.11  \\
  $\beta{\scriptstyle_{eh, \perp}}$         &   0.001 &  378 & 0.28 & 0.10 & 0.04 & 0.18 &  0.0008 &  162 & 1.74 & 0.28 & 0.11 & 1.08  \\
  $\beta{\scriptstyle_{eh, tot}}$           &  0.0010 &  377 & 0.28 & 0.10 & 0.04 & 0.18 &  0.0009 &  161 & 1.73 & 0.27 & 0.11 & 1.09  \\
  $\beta{\scriptstyle_{eb, \parallel}}$     & 0.00002 & 33.2 & 0.09 & 0.05 & 0.02 & 0.10 &  0.0002 & 33.7 & 0.39 & 0.13 & 0.03 & 0.35  \\
  $\beta{\scriptstyle_{eb, \perp}}$         & 0.00003 & 46.4 & 0.08 & 0.04 & 0.02 & 0.08 &  0.0007 & 31.8 & 0.40 & 0.11 & 0.03 & 0.38  \\
  $\beta{\scriptstyle_{eb, tot}}$           & 0.00003 & 42.0 & 0.08 & 0.05 & 0.02 & 0.09 &  0.0005 & 32.4 & 0.40 & 0.12 & 0.03 & 0.38  \\
  $\beta{\scriptstyle_{eff, \parallel}}$    &  0.0009 & 3721 & 2.62 & 1.06 & 0.66 & 2.10 &   0.007 & 1153 & 13.5 & 2.13 & 0.88 & 7.68  \\
  $\beta{\scriptstyle_{eff, \perp}}$        &  0.0010 & 3693 & 2.55 & 0.98 & 0.58 & 2.06 &   0.005 & 1153 & 13.2 & 2.03 & 0.79 & 7.77  \\
  $\beta{\scriptstyle_{eff, tot}}$          &  0.0009 & 3702 & 2.57 & 1.00 & 0.61 & 2.07 &   0.008 & 1153 & 13.3 & 2.06 & 0.82 & 7.74  \\
  \hline
  & \multicolumn{6}{ c }{\textit{Criteria PE: \totalnfitsQpe~VDFs}} & \multicolumn{6}{ ||c| }{\textit{Criteria PA: \totalnfitsQpa~VDFs}} \\
  \hline
  $\beta{\scriptstyle_{ec, \parallel}}$     &    0.05 & 3313 & 4.49 & 1.05 & 0.57 & 2.43 &    0.11 & 15.8 & 1.36 & 0.85 & 0.59 & 1.52  \\
  $\beta{\scriptstyle_{ec, \perp}}$         &    0.04 & 3268 & 4.37 & 0.99 & 0.50 & 2.43 &    0.06 & 15.2 & 1.27 & 0.65 & 0.48 & 1.41  \\
  $\beta{\scriptstyle_{ec, tot}}$           &    0.05 & 3283 & 4.41 & 1.01 & 0.52 & 2.43 &    0.08 & 15.4 & 1.30 & 0.73 & 0.52 & 1.43  \\
  $\beta{\scriptstyle_{eh, \parallel}}$     &  0.0001 &  375 & 0.61 & 0.12 & 0.05 & 0.25 &  0.0007 & 2.98 & 0.14 & 0.09 & 0.04 & 0.14  \\
  $\beta{\scriptstyle_{eh, \perp}}$         &  0.0008 &  378 & 0.62 & 0.12 & 0.05 & 0.25 &   0.002 & 3.42 & 0.15 & 0.09 & 0.05 & 0.15  \\
  $\beta{\scriptstyle_{eh, tot}}$           &  0.0009 &  377 & 0.62 & 0.12 & 0.05 & 0.25 &   0.002 & 3.27 & 0.14 & 0.09 & 0.04 & 0.15  \\
  $\beta{\scriptstyle_{eb, \parallel}}$     & 0.00002 & 33.7 & 0.15 & 0.05 & 0.02 & 0.12 &  0.0001 & 1.60 & 0.08 & 0.05 & 0.03 & 0.09  \\
  $\beta{\scriptstyle_{eb, \perp}}$         & 0.00003 & 46.4 & 0.14 & 0.05 & 0.02 & 0.11 &   0.002 & 1.92 & 0.07 & 0.04 & 0.02 & 0.07  \\
  $\beta{\scriptstyle_{eb, tot}}$           & 0.00003 & 42.0 & 0.14 & 0.05 & 0.02 & 0.11 &   0.001 & 1.82 & 0.08 & 0.05 & 0.02 & 0.08  \\
  $\beta{\scriptstyle_{eff, \parallel}}$    &  0.0009 & 3721 & 5.16 & 1.23 & 0.68 & 2.69 &    0.01 & 19.2 & 1.55 & 0.97 & 0.70 & 1.68  \\
  $\beta{\scriptstyle_{eff, \perp}}$        &  0.0010 & 3693 & 5.05 & 1.16 & 0.62 & 2.67 &    0.01 & 19.5 & 1.46 & 0.79 & 0.58 & 1.64  \\
  $\beta{\scriptstyle_{eff, tot}}$          &  0.0009 & 3702 & 5.08 & 1.18 & 0.64 & 2.68 &    0.01 & 19.4 & 1.49 & 0.86 & 0.62 & 1.66  \\
  \hline
  \enddata
  \tablenotetext{a}{Header symbols match that of Table \ref{tab:Temperatures}}
  \tablecomments{For symbol definitions, see Appendix \ref{app:Definitions}.}
\end{deluxetable*}

\begin{figure*}
  \centering
    {\includegraphics[trim = 0mm 0mm 0mm 0mm, clip, height=180mm]{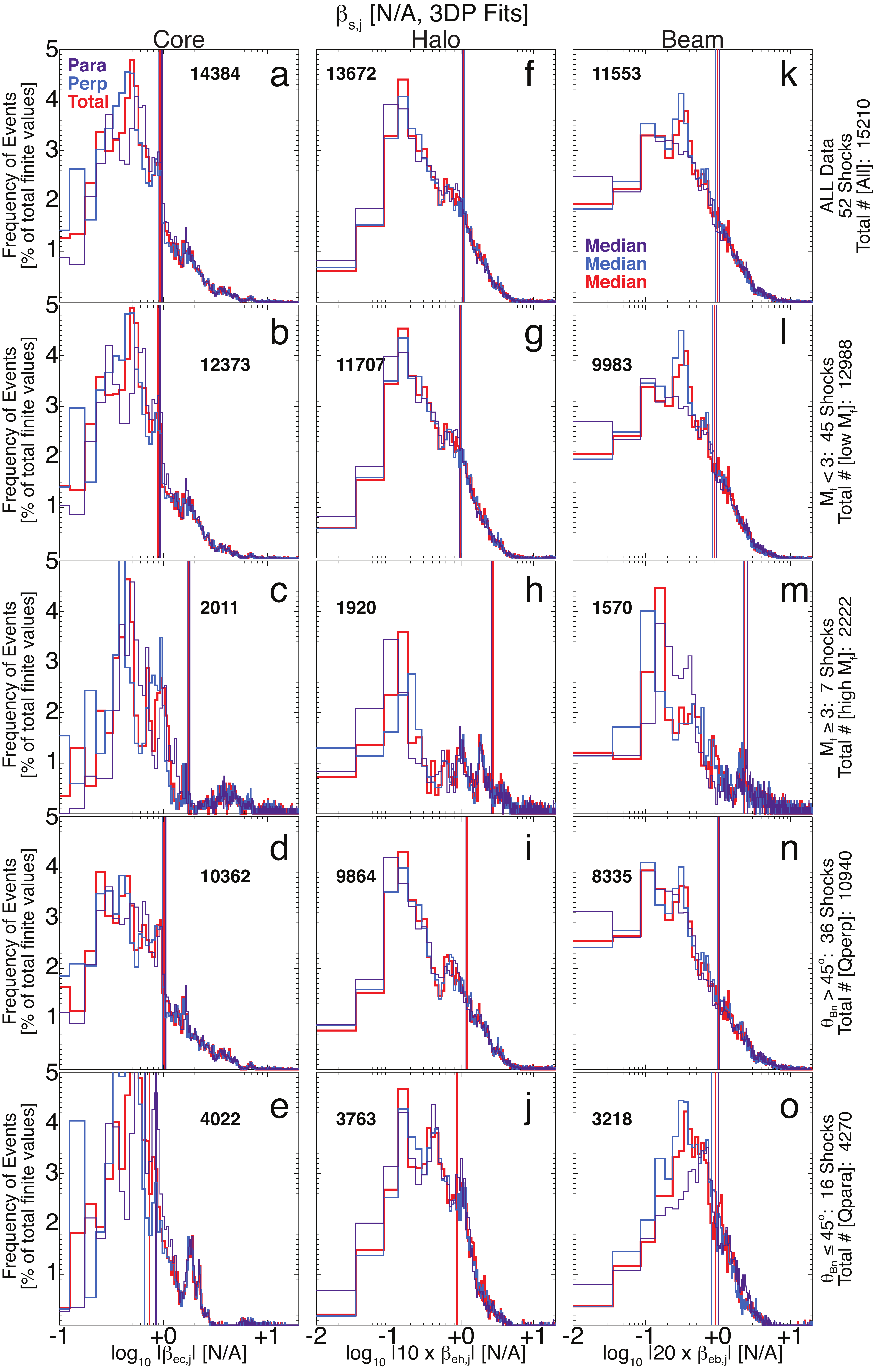}}
    \caption{The same format as Figures \ref{fig:ExtraTemperatures} and \ref{fig:ExtraDensity} except for electron betas [N/A].  Similar to Figure \ref{fig:Betas}, all $\beta{\scriptstyle_{eh, j}}$ and $\beta{\scriptstyle_{eb, j}}$ values were offset by constant factors of 10 and 20, respectively, to reduce the horizontal axis dynamic range.}
    \label{fig:ExtraBetas}
\end{figure*}

\clearpage
\startlongtable  
\begin{deluxetable*}{| l | c | c | c | c | c | c || c | c | c | c | c | c |}
  \tabletypesize{\footnotesize}    
  \tablecaption{Electron Temperature Ratio Parameters \label{tab:ExtraTempRatios}}
  \tablehead{\colhead{Ratio} & \colhead{$X{\scriptstyle_{min}}$}\tablenotemark{a} & \colhead{$X{\scriptstyle_{max}}$} & \colhead{$\bar{X}$} & \colhead{$\tilde{X}$} & \colhead{$X{\scriptstyle_{25\%}}$} & \colhead{$X{\scriptstyle_{75\%}}$} & \colhead{$X{\scriptstyle_{min}}$} & \colhead{$X{\scriptstyle_{max}}$} & \colhead{$\bar{X}$} & \colhead{$\tilde{X}$} & \colhead{$X{\scriptstyle_{25\%}}$} & \colhead{$X{\scriptstyle_{75\%}}$}}
  \startdata
  & \multicolumn{6}{ c }{\textit{Criteria UP: \totalnfitsups~VDFs}} & \multicolumn{6}{ ||c| }{\textit{Criteria DN: \totalnfitsdns~VDFs}} \\
  \hline
  $\tensor*{ \mathcal{T}  }{^{eh}_{ec}}{\scriptstyle_{\parallel}}$ & 0.49 & 17.9 & 3.56 & 3.29 & 2.52 & 4.49 & 0.17 & 11.8 & 2.79 & 2.84 & 1.93 & 3.65  \\
  $\tensor*{ \mathcal{T}  }{^{eh}_{ec}}{\scriptstyle_{    \perp}}$ & 0.69 & 17.9 & 3.92 & 3.71 & 2.74 & 4.64 & 0.41 & 17.5 & 3.03 & 3.04 & 2.27 & 3.73  \\
  $\tensor*{ \mathcal{T}  }{^{eh}_{ec}}{\scriptstyle_{      tot}}$ & 0.64 & 16.3 & 3.79 & 3.54 & 2.69 & 4.59 & 0.34 & 12.8 & 2.93 & 2.97 & 2.16 & 3.68  \\
  $\tensor*{ \mathcal{T}  }{^{eb}_{ec}}{\scriptstyle_{\parallel}}$ & 0.62 & 25.7 & 3.24 & 3.08 & 2.46 & 3.83 & 0.23 & 14.6 & 2.57 & 2.67 & 1.77 & 3.33  \\
  $\tensor*{ \mathcal{T}  }{^{eb}_{ec}}{\scriptstyle_{    \perp}}$ & 0.90 & 24.6 & 3.15 & 2.79 & 2.29 & 3.61 & 0.46 & 19.2 & 2.67 & 2.60 & 1.97 & 3.23  \\
  $\tensor*{ \mathcal{T}  }{^{eb}_{ec}}{\scriptstyle_{      tot}}$ & 0.86 & 25.0 & 3.18 & 2.87 & 2.45 & 3.59 & 0.42 & 14.3 & 2.63 & 2.61 & 1.97 & 3.22  \\
  $\tensor*{ \mathcal{T}  }{^{eb}_{eh}}{\scriptstyle_{\parallel}}$ & 0.15 & 5.97 & 1.04 & 0.91 & 0.68 & 1.25 & 0.15 & 6.12 & 1.07 & 0.94 & 0.71 & 1.33  \\
  $\tensor*{ \mathcal{T}  }{^{eb}_{eh}}{\scriptstyle_{    \perp}}$ & 0.16 & 6.15 & 0.93 & 0.77 & 0.56 & 1.07 & 0.13 & 7.11 & 0.96 & 0.83 & 0.66 & 1.14  \\
  $\tensor*{ \mathcal{T}  }{^{eb}_{eh}}{\scriptstyle_{      tot}}$ & 0.20 & 6.08 & 0.96 & 0.81 & 0.62 & 1.12 & 0.17 & 4.91 & 0.98 & 0.88 & 0.69 & 1.15  \\
  $\tensor*{ \mathcal{T} }{^{eh}_{eff}}{\scriptstyle_{\parallel}}$ & 0.50 & 17.4 & 3.07 & 2.87 & 2.20 & 3.88 & 0.17 & 10.6 & 2.60 & 2.64 & 1.80 & 3.39  \\
  $\tensor*{ \mathcal{T} }{^{eh}_{eff}}{\scriptstyle_{    \perp}}$ & 0.72 & 16.8 & 3.33 & 3.21 & 2.40 & 4.03 & 0.43 & 15.8 & 2.81 & 2.80 & 2.11 & 3.46  \\
  $\tensor*{ \mathcal{T} }{^{eh}_{eff}}{\scriptstyle_{      tot}}$ & 0.65 & 15.2 & 3.24 & 3.07 & 2.33 & 3.98 & 0.37 & 11.8 & 2.73 & 2.75 & 2.02 & 3.43  \\
  $\tensor*{ \mathcal{T} }{^{eb}_{eff}}{\scriptstyle_{\parallel}}$ & 0.36 & 19.6 & 2.78 & 2.68 & 2.16 & 3.26 & 0.24 & 14.2 & 2.40 & 2.49 & 1.65 & 3.09  \\
  $\tensor*{ \mathcal{T} }{^{eb}_{eff}}{\scriptstyle_{    \perp}}$ & 0.66 & 18.7 & 2.70 & 2.39 & 2.00 & 3.03 & 0.46 & 18.5 & 2.48 & 2.36 & 1.83 & 3.02  \\
  $\tensor*{ \mathcal{T} }{^{eb}_{eff}}{\scriptstyle_{      tot}}$ & 0.71 & 19.0 & 2.73 & 2.46 & 2.14 & 3.05 & 0.43 & 13.4 & 2.44 & 2.41 & 1.84 & 3.02  \\
  \hline
  & \multicolumn{6}{ c }{\textit{Criteria LM: \totalnfitslMf~VDFs}} & \multicolumn{6}{ ||c| }{\textit{Criteria HM: \totalnfitshMf~VDFs}} \\
  \hline
  $\tensor*{ \mathcal{T}  }{^{eh}_{ec}}{\scriptstyle_{\parallel}}$ & 0.28 & 17.9 & 3.10 & 2.97 & 2.17 & 4.00 & 0.17 & 9.76 & 3.22 & 3.20 & 2.53 & 3.72  \\
  $\tensor*{ \mathcal{T}  }{^{eh}_{ec}}{\scriptstyle_{    \perp}}$ & 0.57 & 17.9 & 3.38 & 3.25 & 2.37 & 4.11 & 0.41 & 11.2 & 3.52 & 3.34 & 2.84 & 3.94  \\
  $\tensor*{ \mathcal{T}  }{^{eh}_{ec}}{\scriptstyle_{      tot}}$ & 0.50 & 16.3 & 3.27 & 3.17 & 2.32 & 4.07 & 0.34 & 9.89 & 3.41 & 3.29 & 2.77 & 3.86  \\
  $\tensor*{ \mathcal{T}  }{^{eb}_{ec}}{\scriptstyle_{\parallel}}$ & 0.26 & 16.6 & 2.94 & 2.92 & 2.22 & 3.59 & 0.23 & 25.7 & 2.35 & 2.25 & 1.53 & 2.98  \\
  $\tensor*{ \mathcal{T}  }{^{eb}_{ec}}{\scriptstyle_{    \perp}}$ & 0.48 & 19.2 & 2.93 & 2.73 & 2.18 & 3.42 & 0.46 & 24.6 & 2.58 & 2.37 & 1.80 & 3.17  \\
  $\tensor*{ \mathcal{T}  }{^{eb}_{ec}}{\scriptstyle_{      tot}}$ & 0.47 & 15.1 & 2.92 & 2.79 & 2.33 & 3.39 & 0.42 & 25.0 & 2.49 & 2.26 & 1.79 & 3.10  \\
  $\tensor*{ \mathcal{T}  }{^{eb}_{eh}}{\scriptstyle_{\parallel}}$ & 0.15 & 6.12 & 1.09 & 0.97 & 0.72 & 1.36 & 0.20 & 5.82 & 0.83 & 0.75 & 0.55 & 0.95  \\
  $\tensor*{ \mathcal{T}  }{^{eb}_{eh}}{\scriptstyle_{    \perp}}$ & 0.13 & 7.11 & 0.97 & 0.83 & 0.63 & 1.16 & 0.16 & 5.33 & 0.78 & 0.74 & 0.57 & 0.91  \\
  $\tensor*{ \mathcal{T}  }{^{eb}_{eh}}{\scriptstyle_{      tot}}$ & 0.17 & 6.08 & 1.00 & 0.88 & 0.67 & 1.18 & 0.20 & 5.49 & 0.79 & 0.73 & 0.59 & 0.91  \\
  $\tensor*{ \mathcal{T} }{^{eh}_{eff}}{\scriptstyle_{\parallel}}$ & 0.29 & 17.4 & 2.79 & 2.70 & 2.00 & 3.61 & 0.17 & 8.78 & 2.84 & 2.83 & 2.26 & 3.36  \\
  $\tensor*{ \mathcal{T} }{^{eh}_{eff}}{\scriptstyle_{    \perp}}$ & 0.56 & 16.8 & 3.03 & 2.94 & 2.17 & 3.73 & 0.43 & 10.6 & 3.07 & 2.96 & 2.55 & 3.49  \\
  $\tensor*{ \mathcal{T} }{^{eh}_{eff}}{\scriptstyle_{      tot}}$ & 0.56 & 15.2 & 2.94 & 2.88 & 2.12 & 3.67 & 0.37 & 8.89 & 2.98 & 2.89 & 2.46 & 3.44  \\
  $\tensor*{ \mathcal{T} }{^{eb}_{eff}}{\scriptstyle_{\parallel}}$ & 0.29 & 14.8 & 2.64 & 2.65 & 2.01 & 3.22 & 0.24 & 19.6 & 2.05 & 1.97 & 1.38 & 2.61  \\
  $\tensor*{ \mathcal{T} }{^{eb}_{eff}}{\scriptstyle_{    \perp}}$ & 0.48 & 18.7 & 2.63 & 2.41 & 1.97 & 3.08 & 0.46 & 18.7 & 2.21 & 2.12 & 1.63 & 2.73  \\
  $\tensor*{ \mathcal{T} }{^{eb}_{eff}}{\scriptstyle_{      tot}}$ & 0.48 & 13.7 & 2.63 & 2.48 & 2.08 & 3.09 & 0.43 & 19.0 & 2.15 & 2.01 & 1.61 & 2.66  \\
  \hline
  & \multicolumn{6}{ c }{\textit{Criteria PE: \totalnfitsQpe~VDFs}} & \multicolumn{6}{ ||c| }{\textit{Criteria PA: \totalnfitsQpa~VDFs}} \\
  \hline
  $\tensor*{ \mathcal{T}  }{^{eh}_{ec}}{\scriptstyle_{\parallel}}$ & 0.17 & 14.4 & 3.17 & 3.03 & 2.29 & 3.96 & 0.28 & 17.9 & 2.97 & 2.97 & 1.92 & 4.00  \\
  $\tensor*{ \mathcal{T}  }{^{eh}_{ec}}{\scriptstyle_{    \perp}}$ & 0.41 & 15.8 & 3.41 & 3.23 & 2.47 & 4.05 & 0.70 & 17.9 & 3.39 & 3.39 & 2.40 & 4.22  \\
  $\tensor*{ \mathcal{T}  }{^{eh}_{ec}}{\scriptstyle_{      tot}}$ & 0.34 & 12.7 & 3.32 & 3.17 & 2.44 & 4.01 & 0.50 & 16.3 & 3.21 & 3.28 & 2.27 & 4.14  \\
  $\tensor*{ \mathcal{T}  }{^{eb}_{ec}}{\scriptstyle_{\parallel}}$ & 0.23 & 25.7 & 2.89 & 2.88 & 2.09 & 3.56 & 0.46 & 14.6 & 2.76 & 2.76 & 2.10 & 3.46  \\
  $\tensor*{ \mathcal{T}  }{^{eb}_{ec}}{\scriptstyle_{    \perp}}$ & 0.46 & 24.6 & 2.97 & 2.74 & 2.19 & 3.46 & 0.52 & 19.2 & 2.64 & 2.56 & 1.92 & 3.20  \\
  $\tensor*{ \mathcal{T}  }{^{eb}_{ec}}{\scriptstyle_{      tot}}$ & 0.42 & 25.0 & 2.94 & 2.81 & 2.25 & 3.42 & 0.53 & 13.5 & 2.67 & 2.60 & 2.16 & 3.21  \\
  $\tensor*{ \mathcal{T}  }{^{eb}_{eh}}{\scriptstyle_{\parallel}}$ & 0.15 & 5.97 & 1.04 & 0.92 & 0.67 & 1.29 & 0.17 & 6.12 & 1.09 & 0.94 & 0.75 & 1.33  \\
  $\tensor*{ \mathcal{T}  }{^{eb}_{eh}}{\scriptstyle_{    \perp}}$ & 0.13 & 7.11 & 0.98 & 0.83 & 0.62 & 1.18 & 0.16 & 4.84 & 0.85 & 0.77 & 0.60 & 0.98  \\
  $\tensor*{ \mathcal{T}  }{^{eb}_{eh}}{\scriptstyle_{      tot}}$ & 0.17 & 6.08 & 0.99 & 0.85 & 0.66 & 1.19 & 0.25 & 4.58 & 0.92 & 0.84 & 0.66 & 1.05  \\
  $\tensor*{ \mathcal{T} }{^{eh}_{eff}}{\scriptstyle_{\parallel}}$ & 0.17 & 13.0 & 2.83 & 2.74 & 2.11 & 3.55 & 0.29 & 17.4 & 2.71 & 2.69 & 1.73 & 3.62  \\
  $\tensor*{ \mathcal{T} }{^{eh}_{eff}}{\scriptstyle_{    \perp}}$ & 0.43 & 14.4 & 3.02 & 2.92 & 2.29 & 3.65 & 0.73 & 16.8 & 3.07 & 3.05 & 2.14 & 3.78  \\
  $\tensor*{ \mathcal{T} }{^{eh}_{eff}}{\scriptstyle_{      tot}}$ & 0.37 & 11.1 & 2.95 & 2.86 & 2.24 & 3.60 & 0.56 & 15.2 & 2.93 & 2.96 & 2.03 & 3.73  \\
  $\tensor*{ \mathcal{T} }{^{eb}_{eff}}{\scriptstyle_{\parallel}}$ & 0.24 & 19.6 & 2.58 & 2.61 & 1.90 & 3.20 & 0.47 & 14.2 & 2.51 & 2.49 & 1.87 & 3.10  \\
  $\tensor*{ \mathcal{T} }{^{eb}_{eff}}{\scriptstyle_{    \perp}}$ & 0.46 & 18.7 & 2.65 & 2.43 & 1.98 & 3.08 & 0.53 & 18.7 & 2.38 & 2.24 & 1.74 & 2.84  \\
  $\tensor*{ \mathcal{T} }{^{eb}_{eff}}{\scriptstyle_{      tot}}$ & 0.43 & 19.0 & 2.62 & 2.49 & 2.02 & 3.08 & 0.58 & 13.1 & 2.42 & 2.30 & 1.94 & 2.87  \\
  \hline
  \enddata
  \tablenotetext{a}{Header symbols match that of Table \ref{tab:Temperatures}}
  \tablecomments{For symbol definitions, see Appendix \ref{app:Definitions}.}
\end{deluxetable*}

\startlongtable  
\begin{deluxetable*}{| l | c | c | c | c | c | c || c | c | c | c | c | c |}
  \tabletypesize{\footnotesize}    
  \tablecaption{Electron Temperature Anisotropy Parameters \label{tab:ExtraTempAnisotropies}}
  \tablehead{\colhead{Anisotropy} & \colhead{$X{\scriptstyle_{min}}$}\tablenotemark{a} & \colhead{$X{\scriptstyle_{max}}$} & \colhead{$\bar{X}$} & \colhead{$\tilde{X}$} & \colhead{$X{\scriptstyle_{25\%}}$} & \colhead{$X{\scriptstyle_{75\%}}$} & \colhead{$X{\scriptstyle_{min}}$} & \colhead{$X{\scriptstyle_{max}}$} & \colhead{$\bar{X}$} & \colhead{$\tilde{X}$} & \colhead{$X{\scriptstyle_{25\%}}$} & \colhead{$X{\scriptstyle_{75\%}}$}}
  \startdata
  & \multicolumn{6}{ c }{\textit{Criteria UP: \totalnfitsups~VDFs}} & \multicolumn{6}{ ||c| }{\textit{Criteria DN: \totalnfitsdns~VDFs}} \\
  \hline
  $\mathcal{A}{\scriptstyle_{ec}}$  & 0.50 & 1.15 & 0.94 & 0.98 & 0.91 & 1.00 & 0.38 & 1.56 & 0.92 & 0.98 & 0.90 & 1.01  \\
  $\mathcal{A}{\scriptstyle_{eh}}$  & 0.31 & 10.9 & 1.05 & 1.03 & 0.95 & 1.12 & 0.24 & 15.0 & 1.07 & 1.04 & 0.96 & 1.12  \\
  $\mathcal{A}{\scriptstyle_{eb}}$  & 0.25 & 15.2 & 0.95 & 0.90 & 0.75 & 1.07 & 0.13 & 14.1 & 1.04 & 0.96 & 0.82 & 1.13  \\
  $\mathcal{A}{\scriptstyle_{eff}}$ & 0.45 & 2.52 & 0.95 & 0.97 & 0.92 & 1.00 & 0.35 & 2.80 & 0.93 & 0.98 & 0.91 & 1.01  \\
  \hline
  & \multicolumn{6}{ c }{\textit{Criteria LM: \totalnfitslMf~VDFs}} & \multicolumn{6}{ ||c| }{\textit{Criteria HM: \totalnfitshMf~VDFs}} \\
  \hline
  $\mathcal{A}{\scriptstyle_{ec}}$  & 0.38 & 1.56 & 0.93 & 0.98 & 0.91 & 1.01 & 0.42 & 1.21 & 0.91 & 0.96 & 0.85 & 0.99  \\
  $\mathcal{A}{\scriptstyle_{eh}}$  & 0.28 & 15.0 & 1.06 & 1.04 & 0.95 & 1.13 & 0.24 & 7.03 & 1.06 & 1.03 & 0.98 & 1.08  \\
  $\mathcal{A}{\scriptstyle_{eb}}$  & 0.13 & 15.2 & 0.99 & 0.92 & 0.77 & 1.09 & 0.33 & 3.65 & 1.08 & 1.02 & 0.85 & 1.19  \\
  $\mathcal{A}{\scriptstyle_{eff}}$ & 0.35 & 2.80 & 0.94 & 0.98 & 0.92 & 1.01 & 0.40 & 1.96 & 0.92 & 0.97 & 0.87 & 1.00  \\
  \hline
  & \multicolumn{6}{ c }{\textit{Criteria PE: \totalnfitsQpe~VDFs}} & \multicolumn{6}{ ||c| }{\textit{Criteria PA: \totalnfitsQpa~VDFs}} \\
  \hline
  $\mathcal{A}{\scriptstyle_{ec}}$  & 0.38 & 1.51 & 0.94 & 0.98 & 0.93 & 1.01 & 0.44 & 1.56 & 0.89 & 0.96 & 0.80 & 1.01  \\
  $\mathcal{A}{\scriptstyle_{eh}}$  & 0.24 & 15.0 & 1.05 & 1.03 & 0.95 & 1.11 & 0.31 & 5.44 & 1.09 & 1.04 & 0.96 & 1.15  \\
  $\mathcal{A}{\scriptstyle_{eb}}$  & 0.13 & 14.1 & 1.04 & 0.96 & 0.81 & 1.15 & 0.18 & 15.2 & 0.89 & 0.88 & 0.69 & 1.01  \\
  $\mathcal{A}{\scriptstyle_{eff}}$ & 0.35 & 2.80 & 0.95 & 0.98 & 0.94 & 1.01 & 0.45 & 1.50 & 0.90 & 0.95 & 0.83 & 1.01  \\
  \hline
  \enddata
  \tablenotetext{a}{Header symbols match that of Table \ref{tab:Temperatures}}
  \tablecomments{For symbol definitions, see Appendix \ref{app:Definitions}.}
\end{deluxetable*}

\begin{figure*}
  \centering
    {\includegraphics[trim = 0mm 0mm 0mm 0mm, clip, height=170mm, angle=0]{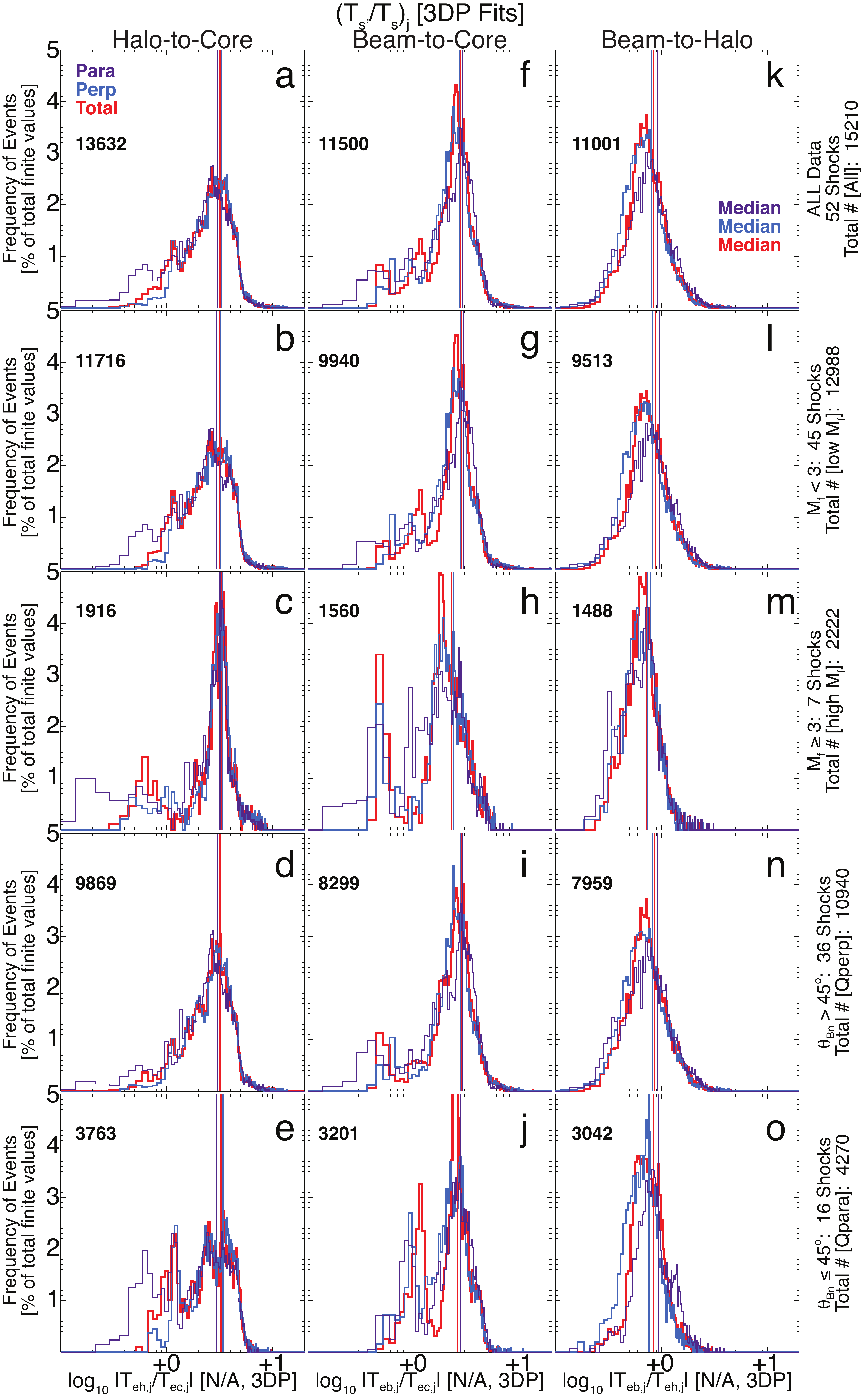}}
    \caption{The same format as Figures \ref{fig:ExtraTemperatures} and \ref{fig:ExtraDensity} except for electron temperature ratios [N/A].}
    \label{fig:ExtraTempRatios}
\end{figure*}

\begin{figure*}
  \centering
    {\includegraphics[trim = 0mm 0mm 0mm 0mm, clip, height=150mm]{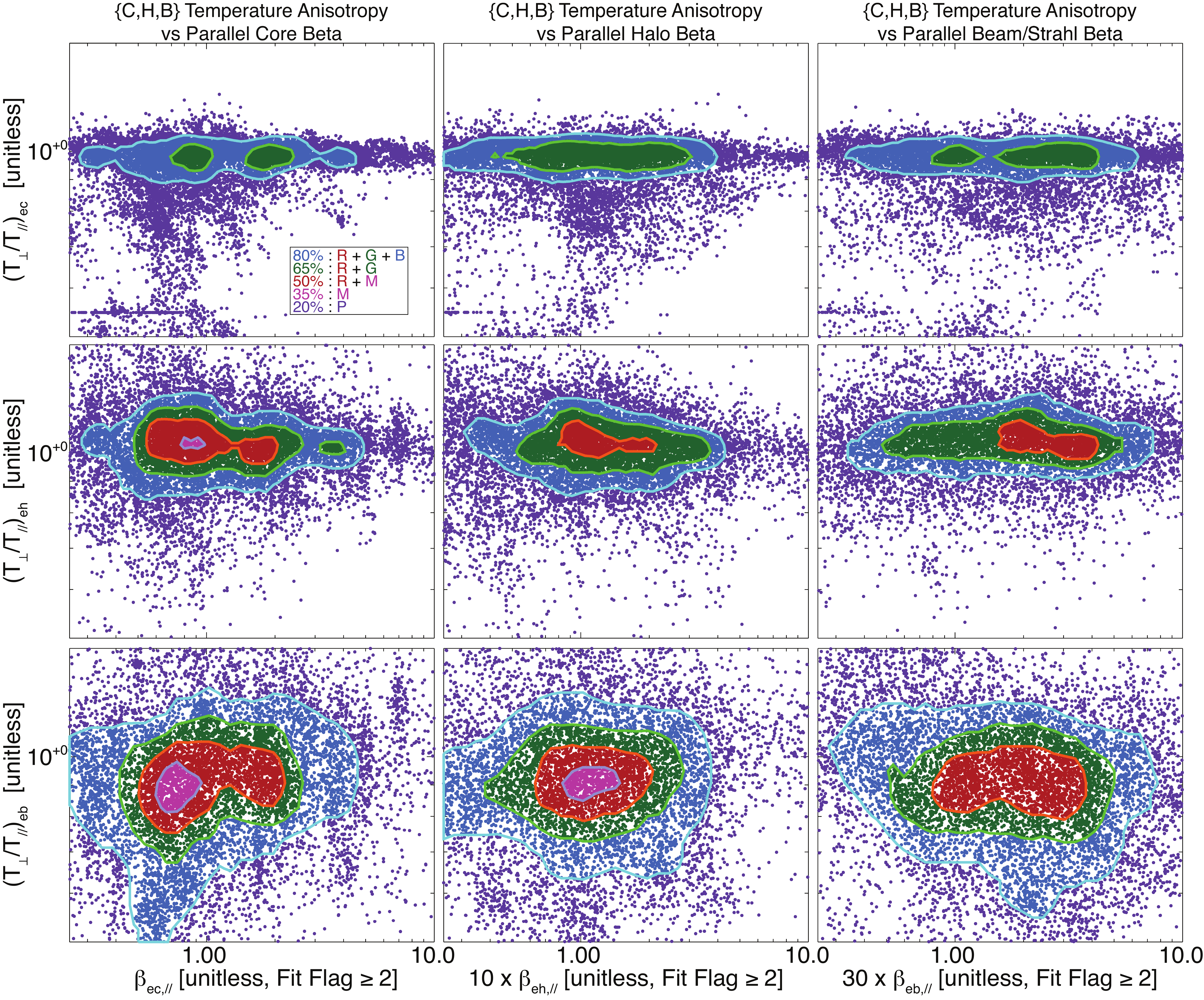}}
    \caption{Temperature anisotropy, $\mathcal{A}{\scriptstyle_{es}}$ [N/A], versus parallel electron beta, $\beta{\scriptstyle_{s, \parallel}}$ [N/A], of electron component $s$.  The color-coded contours (legend in lower right-hand corner of upper left-hand panel) are generated in the same fashion as those in Figure \ref{fig:HeatFluxvsCoreParaBeta}.  Note that the halo and beam/strahl beta values have been increased by factors of 10 and 30, respectively, to maintain a uniform horizontal axis scale all columns.}
    \label{fig:ExtraTempAnisotropies}
\end{figure*}

\clearpage
\startlongtable  
\begin{deluxetable*}{| l | c | c | c | c | c | c || c | c | c | c | c | c |}
  \tabletypesize{\footnotesize}    
  \tablecaption{Coulomb Collision Rates [\# per week] \label{tab:ExtraCoulombCollisionRates}}
  \tablehead{\colhead{$\nu{\scriptstyle_{ss'}}$} & \colhead{$X{\scriptstyle_{min}}$}\tablenotemark{a} & \colhead{$X{\scriptstyle_{max}}$} & \colhead{$\bar{X}$} & \colhead{$\tilde{X}$} & \colhead{$X{\scriptstyle_{25\%}}$} & \colhead{$X{\scriptstyle_{75\%}}$} & \colhead{$X{\scriptstyle_{min}}$} & \colhead{$X{\scriptstyle_{max}}$} & \colhead{$\bar{X}$} & \colhead{$\tilde{X}$} & \colhead{$X{\scriptstyle_{25\%}}$} & \colhead{$X{\scriptstyle_{75\%}}$}}
  \startdata
  & \multicolumn{6}{ c }{\textit{Criteria UP: \totalnfitsups~VDFs}} & \multicolumn{6}{ ||c| }{\textit{Criteria DN: \totalnfitsdns~VDFs}} \\
  \hline
  $\nu{\scriptstyle_{ecc}}$           &    0.05 & 16.4 & 4.68 & 3.90 &  2.00 & 6.68 &    0.05 & 22.3 & 6.01 & 5.93 &  2.66 & 8.99  \\
  $\nu{\scriptstyle_{ehh}}$           & 0.00007 & 1.91 & 0.05 & 0.02 &  0.01 & 0.05 & 0.00006 & 3.14 & 0.06 & 0.03 &  0.02 & 0.06  \\
  $\nu{\scriptstyle_{ebb}}$           & 0.00007 & 0.58 & 0.03 & 0.02 & 0.008 & 0.04 & 0.00007 & 1.42 & 0.03 & 0.02 & 0.008 & 0.03  \\
  $\nu{\scriptstyle_{ehc}}$           &    0.02 & 40.9 & 1.95 & 1.40 &  0.66 & 2.33 &    0.03 & 49.3 & 2.68 & 2.09 &  0.94 & 3.07  \\
  $\nu{\scriptstyle_{ebc}}$           &    0.03 & 44.3 & 3.77 & 1.66 &  0.87 & 3.50 &    0.02 & 57.8 & 5.36 & 2.58 &  1.34 & 4.27  \\
  $\nu{\scriptstyle_{ehb}}$           & 0.00009 & 4.54 & 0.07 & 0.03 &  0.02 & 0.06 & 0.00008 & 8.76 & 0.07 & 0.04 &  0.02 & 0.08  \\
  $\nu{\scriptstyle_{ecp}}$           &   0.009 & 9.60 & 2.57 & 2.17 &  1.15 & 3.53 &    0.03 & 14.7 & 3.30 & 3.18 &  1.20 & 4.86  \\
  $\nu{\scriptstyle_{ehp}}$           &   0.004 & 5.86 & 0.47 & 0.35 &  0.17 & 0.65 &   0.008 & 16.1 & 0.76 & 0.60 &  0.29 & 0.93  \\
  $\nu{\scriptstyle_{ebp}}$           &   0.004 & 4.88 & 0.55 & 0.40 &  0.20 & 0.68 &    0.01 & 8.78 & 0.82 & 0.65 &  0.37 & 1.01  \\
  $\nu{\scriptstyle_{ec \alpha}}$     &    0.02 & 2.59 & 0.24 & 0.15 &  0.09 & 0.26 &   0.009 & 3.11 & 0.43 & 0.30 &  0.17 & 0.60  \\
  $\nu{\scriptstyle_{eh \alpha}}$     &  0.0008 & 0.61 & 0.05 & 0.02 &  0.01 & 0.06 &   0.002 & 1.53 & 0.11 & 0.07 &  0.03 & 0.14  \\
  $\nu{\scriptstyle_{eb \alpha}}$     &   0.003 & 0.41 & 0.05 & 0.03 &  0.02 & 0.07 &   0.004 & 1.55 & 0.12 & 0.09 &  0.04 & 0.16  \\
  $\nu{\scriptstyle_{p p}}$           &   0.001 & 3.97 & 0.65 & 0.36 &  0.12 & 0.91 &  0.0001 & 2.29 & 0.43 & 0.28 &  0.09 & 0.58  \\
  $\nu{\scriptstyle_{\alpha \alpha}}$ &   0.002 & 1.60 & 0.16 & 0.08 &  0.03 & 0.21 &   0.004 & 0.34 & 0.05 & 0.03 &  0.02 & 0.05  \\
  $\nu{\scriptstyle_{p \alpha}}$      &  0.0003 & 0.52 & 0.03 & 0.02 & 0.007 & 0.03 &  0.0003 & 0.10 & 0.02 & 0.01 & 0.008 & 0.02  \\
 \hline
  & \multicolumn{6}{ c }{\textit{Criteria LM: \totalnfitslMf~VDFs}} & \multicolumn{6}{ ||c| }{\textit{Criteria HM: \totalnfitshMf~VDFs}} \\
 \hline
  $\nu{\scriptstyle_{ecc}}$           &    0.05 & 22.3 & 5.62 & 4.76 &  2.51 & 8.21 &    0.45 & 10.4 & 4.35 & 5.13 &  1.20 & 6.43  \\
  $\nu{\scriptstyle_{ehh}}$           & 0.00006 & 3.14 & 0.06 & 0.03 &  0.01 & 0.06 &  0.0002 & 1.44 & 0.04 & 0.03 &  0.02 & 0.06  \\
  $\nu{\scriptstyle_{ebb}}$           & 0.00007 & 1.42 & 0.03 & 0.02 & 0.008 & 0.03 &  0.0005 & 0.80 & 0.04 & 0.02 &  0.01 & 0.05  \\
  $\nu{\scriptstyle_{ehc}}$           &    0.02 & 49.3 & 2.51 & 1.82 &  0.82 & 2.95 &    0.07 & 9.34 & 1.51 & 1.54 &  0.64 & 2.12  \\
  $\nu{\scriptstyle_{ebc}}$           &    0.02 & 57.8 & 4.73 & 2.16 &  1.12 & 3.83 &    0.16 & 29.1 & 4.38 & 2.29 &  0.86 & 4.18  \\
  $\nu{\scriptstyle_{ehb}}$           & 0.00008 & 8.76 & 0.07 & 0.03 &  0.02 & 0.07 & 0.00010 & 2.09 & 0.07 & 0.04 &  0.02 & 0.09  \\
  $\nu{\scriptstyle_{ecp}}$           &   0.009 & 14.7 & 3.07 & 2.60 &  1.25 & 4.50 &    0.12 & 7.48 & 2.48 & 2.99 &  0.52 & 3.82  \\
  $\nu{\scriptstyle_{ehp}}$           &   0.004 & 16.1 & 0.66 & 0.49 &  0.24 & 0.83 &   0.009 & 3.32 & 0.47 & 0.48 &  0.19 & 0.66  \\
  $\nu{\scriptstyle_{ebp}}$           &   0.004 & 8.78 & 0.69 & 0.51 &  0.30 & 0.86 &    0.04 & 7.13 & 0.78 & 0.59 &  0.24 & 1.12  \\
  $\nu{\scriptstyle_{ec \alpha}}$     &   0.009 & 3.11 & 0.35 & 0.23 &  0.12 & 0.50 &    0.05 & 0.99 & 0.30 & 0.25 &  0.15 & 0.43  \\
  $\nu{\scriptstyle_{eh \alpha}}$     &  0.0008 & 1.53 & 0.08 & 0.04 &  0.02 & 0.10 &   0.006 & 0.80 & 0.06 & 0.04 &  0.02 & 0.08  \\
  $\nu{\scriptstyle_{eb \alpha}}$     &   0.003 & 0.92 & 0.08 & 0.05 &  0.02 & 0.12 &   0.003 & 1.55 & 0.12 & 0.07 &  0.04 & 0.14  \\
  $\nu{\scriptstyle_{p p}}$           &  0.0001 & 3.97 & 0.51 & 0.28 &  0.11 & 0.68 &   0.005 & 3.09 & 0.69 & 0.61 &  0.06 & 0.94  \\
  $\nu{\scriptstyle_{\alpha \alpha}}$ &   0.002 & 1.60 & 0.12 & 0.06 &  0.02 & 0.11 &   0.005 & 0.44 & 0.14 & 0.11 &  0.05 & 0.24  \\
  $\nu{\scriptstyle_{p \alpha}}$      &  0.0003 & 0.52 & 0.02 & 0.01 & 0.008 & 0.02 &   0.001 & 0.15 & 0.03 & 0.02 & 0.006 & 0.05  \\
 \hline
  & \multicolumn{6}{ c }{\textit{Criteria PE: \totalnfitsQpe~VDFs}} & \multicolumn{6}{ ||c| }{\textit{Criteria PA: \totalnfitsQpa~VDFs}} \\
 \hline
  $\nu{\scriptstyle_{ecc}}$           &    0.05 & 21.6 & 5.96 & 5.85 &  3.04 & 8.51 &    0.26 & 22.3 & 4.12 & 3.46 &  1.00 & 5.59  \\
  $\nu{\scriptstyle_{ehh}}$           & 0.00006 & 1.91 & 0.06 & 0.03 &  0.02 & 0.07 & 0.00008 & 3.14 & 0.04 & 0.02 &  0.01 & 0.04  \\
  $\nu{\scriptstyle_{ebb}}$           & 0.00007 & 1.03 & 0.03 & 0.02 & 0.008 & 0.03 &  0.0001 & 1.42 & 0.02 & 0.02 & 0.008 & 0.03  \\
  $\nu{\scriptstyle_{ehc}}$           &    0.02 & 49.3 & 2.62 & 1.99 &  0.95 & 2.98 &    0.05 & 40.9 & 1.74 & 1.13 &  0.59 & 2.24  \\
  $\nu{\scriptstyle_{ebc}}$           &    0.02 & 57.8 & 5.01 & 2.49 &  1.32 & 4.31 &    0.09 & 46.4 & 3.84 & 1.52 &  0.75 & 3.27  \\
  $\nu{\scriptstyle_{ehb}}$           & 0.00008 & 4.54 & 0.08 & 0.04 &  0.02 & 0.08 & 0.00009 & 8.76 & 0.06 & 0.02 &  0.01 & 0.05  \\
  $\nu{\scriptstyle_{ecp}}$           &   0.009 & 12.1 & 3.31 & 3.21 &  1.72 & 4.60 &    0.11 & 14.7 & 2.15 & 1.49 &  0.56 & 2.78  \\
  $\nu{\scriptstyle_{ehp}}$           &   0.004 & 16.1 & 0.71 & 0.57 &  0.28 & 0.89 &    0.01 & 6.65 & 0.42 & 0.30 &  0.16 & 0.57  \\
  $\nu{\scriptstyle_{ebp}}$           &   0.004 & 8.78 & 0.79 & 0.61 &  0.36 & 0.98 &    0.02 & 6.56 & 0.47 & 0.35 &  0.20 & 0.61  \\
  $\nu{\scriptstyle_{ec \alpha}}$     &   0.009 & 3.11 & 0.40 & 0.26 &  0.14 & 0.56 &    0.02 & 1.26 & 0.20 & 0.14 &  0.10 & 0.24  \\
  $\nu{\scriptstyle_{eh \alpha}}$     &  0.0008 & 1.53 & 0.09 & 0.06 &  0.03 & 0.12 &   0.001 & 1.37 & 0.05 & 0.03 &  0.02 & 0.05  \\
  $\nu{\scriptstyle_{eb \alpha}}$     &   0.003 & 1.55 & 0.10 & 0.07 &  0.03 & 0.14 &   0.004 & 0.58 & 0.05 & 0.03 &  0.02 & 0.06  \\
  $\nu{\scriptstyle_{p p}}$           &  0.0001 & 3.97 & 0.61 & 0.37 &  0.16 & 0.80 &   0.001 & 1.98 & 0.33 & 0.15 &  0.02 & 0.45  \\
  $\nu{\scriptstyle_{\alpha \alpha}}$ &   0.002 & 1.60 & 0.13 & 0.06 &  0.02 & 0.15 &   0.005 & 0.23 & 0.06 & 0.06 &  0.02 & 0.08  \\
  $\nu{\scriptstyle_{p \alpha}}$      &  0.0003 & 0.52 & 0.02 & 0.02 & 0.009 & 0.03 &  0.0003 & 0.08 & 0.01 & 0.01 & 0.003 & 0.02  \\
 \hline
  \enddata
  \tablenotetext{a}{Header symbols match that of Table \ref{tab:Temperatures}}
  \tablecomments{For symbol definitions, see Appendix \ref{app:Definitions}.}
\end{deluxetable*}

\phantomsection   
\section{Previous Electron Studies}  \label{app:PreviousElectronStudies}

\indent  In this appendix, we summarize, by way of tables, the observations of previous electron velocity moments near 1 AU similar to the appendices in \citet[][]{wilsoniii18b}.  The symbols/parameters are same as elsewhere herein.

\startlongtable  
\begin{deluxetable*}{| l | c | c | c | c | c |}
  \tabletypesize{\footnotesize}    
  \tablecaption{Measurements of Electron Temperatures [eV] at 1 AU \label{tab:PrevTemperatures}}
  \tablehead{\colhead{Reference} & \colhead{Parameter} & \colhead{Spacecraft} & \colhead{Notes} & \colhead{$X{\scriptstyle_{min}}$--$X{\scriptstyle_{max}}$} & \colhead{$\bar{\mathbf{X}}$}\tablenotemark{a}}
  \startdata
  \multirow{2}{*}{\citet[][]{skoug00a}}                   & $T{\scriptstyle_{ec, tot}}$                                           &         ACE          &    SW\tablenotemark{b}   & $\sim$3--60          &                   \\
                                                          & $T{\scriptstyle_{eh, tot}}$                                           &                      &                          & $\sim$26--560        &                   \\
  \hline
  \citet[][]{masters11a}                                  & $\Delta T{\scriptstyle_{e, tot}}$                                     &       Cassini        &   KBS\tablenotemark{c}   & $\sim$10--120        &                   \\
  \hline
  \multirow{2}{*}{\citet[][]{lefebvre07a}}                & $\langle T{\scriptstyle_{e, tot}} \rangle{\scriptstyle_{up}}$         &       Cluster        &   BS\tablenotemark{d}    & $\sim$14.3--22.8     &                   \\
                                                          & $\langle T{\scriptstyle_{e, tot}} \rangle{\scriptstyle_{dn}}$         &                      &                          & $\sim$25.8--90.8     &                   \\
  \multirow{4}{*}{\citet[][]{schwartz11a}}                & $\langle T{\scriptstyle_{e, \parallel}} \rangle{\scriptstyle_{up}}$   &                      &                          & $\sim$17.2--51.7     &                   \\
                                                          & $\langle T{\scriptstyle_{e, \perp}} \rangle{\scriptstyle_{up}}$       &                      &                          & $\sim$12.9--34.5     &                   \\
                                                          & $\langle T{\scriptstyle_{e, \parallel}} \rangle{\scriptstyle_{dn}}$   &                      &                          & $\sim$73.2--103.4    &                   \\
                                                          & $\langle T{\scriptstyle_{e, \perp}} \rangle{\scriptstyle_{dn}}$       &                      &                          & $\sim$73.2--103.4    &                   \\
  \citet[][]{vinas10a}                                    & $T{\scriptstyle_{eb, tot}}$                                           &                      &            SW            & $\sim$20--50         &                   \\
  \hline
  \multirow{6}{*}{\citet[][]{hull98a}}                    & $\langle T{\scriptstyle_{e, tot}} \rangle{\scriptstyle_{up}}$         &       Galileo        &            BS            &                      & $\sim$12.9        \\
                                                          & $\langle T{\scriptstyle_{ec, \parallel}} \rangle{\scriptstyle_{up}}$  &                      &                          &                      & $\sim$16.4        \\
                                                          & $\langle T{\scriptstyle_{ec, \perp}} \rangle{\scriptstyle_{up}}$      &                      &                          &                      & $\sim$11.2        \\
                                                          & $\langle T{\scriptstyle_{e, tot}} \rangle{\scriptstyle_{dn}}$         &                      &                          &                      & $\sim$14.6        \\
                                                          & $\langle T{\scriptstyle_{ec, \parallel}} \rangle{\scriptstyle_{dn}}$  &                      &                          &                      & $\sim$16.4        \\
                                                          & $\langle T{\scriptstyle_{ec, \perp}} \rangle{\scriptstyle_{dn}}$      &                      &                          &                      & $\sim$13.8        \\
  \hline
  \multirow{2}{*}{\citet[][]{pilipp90a}}                  & $T{\scriptstyle_{ec, tot}}$                                           &     Helios 1 \& 2    & Slow SW\tablenotemark{e} & $\sim$7--13          &                   \\
                                                          & $T{\scriptstyle_{ec, tot}}$                                           &                      &          Fast SW         & $\sim$6--9           &                   \\
  \hline
  \multirow{2}{*}{\citet[][]{feldman73b}}                 & $\langle T{\scriptstyle_{e, tot}} \rangle{\scriptstyle_{up}}$         &        Imp 6         &            BS            & $\sim$4.3--24.0      &                   \\
                                                          & $\langle T{\scriptstyle_{ec, tot}} \rangle{\scriptstyle_{up}}$        &                      &                          & $\sim$2.6--22.4      &                   \\
  \multirow{2}{*}{\citet[][]{feldman75a}}                 & $T{\scriptstyle_{ec, tot}}$                                           &      Imp 7 \& 8      &            SW            &                      & $\sim$10.7--10.9  \\
                                                          & $T{\scriptstyle_{eh, tot}}$                                           &                      &                          &                      & $\sim$58--60      \\
  \multirow{2}{*}{\citet[][]{feldman78b}}                 & $T{\scriptstyle_{ec, tot}}$                                           &    Imp 6, 7, \& 8    &          Fast SW         &                      & $\sim$56          \\
                                                          & $T{\scriptstyle_{eh, tot}}$                                           &                      &                          &                      & $\sim$7.3         \\
  \multirow{2}{*}{\citet[][]{feldman79b}}                 & $T{\scriptstyle_{ec, tot}}$                                           &                      &            SW            & $\sim$7.8--16        &                   \\
                                                          & $T{\scriptstyle_{eh, \parallel}}$                                     &                      &                          & $\sim$49--83         &                   \\
  \hline
  \citet[][]{feldman83a}                                  & $\langle T{\scriptstyle_{ec, tot}} \rangle{\scriptstyle_{dn}}$        &       ISEE 2         &            BS            & $\sim$23--139        &                   \\
  \citet[][]{feldman83b}                                  & $\langle T{\scriptstyle_{e, tot}} \rangle{\scriptstyle_{up}}$         &       ISEE 3         &   IPS\tablenotemark{f}   & $\sim$8.6--19.8      &                   \\
  \multirow{2}{*}{\citet[][]{hull00a}}                    & $\langle T{\scriptstyle_{e, tot}} \rangle{\scriptstyle_{up}}$         &       ISEE 1         &            BS            & $\sim$6.0--31.9      & $\sim$14.6        \\
                                                          & $\Delta T{\scriptstyle_{e, tot}}$                                     &                      &                          & $<$10 to $>$200      & $\sim$30          \\
  \multirow{3}{*}{\citet[][]{hull00b}}                    & $\Delta T{\scriptstyle_{e, tot}}$                                     &                      &                          & $\sim$7--205         &                   \\
                                                          & $\Delta T{\scriptstyle_{e, \parallel}}$                               &                      &                          & $\sim$6--205         &                   \\
                                                          & $\Delta T{\scriptstyle_{e, \perp}}$                                   &                      &                          & $\sim$8--200         &                   \\
  \citet[][]{schwartz88a}                                 & $\Delta T{\scriptstyle_{e, tot}}$                                     &       ISEE 3         &         BS \& IPS        & $\sim$8.6--198       &                   \\
  \multirow{2}{*}{\citet[][]{thomsen85a}}                 & $\Delta T{\scriptstyle_{e, \parallel}}$                               &      ISEE 1 \& 2     &            BS            & $\sim$14--41         &                   \\
                                                          & $\Delta T{\scriptstyle_{e, \perp}}$                                   &                      &                          & $\sim$13--52         &                   \\
  \multirow{3}{*}{\citet[][]{thomsen87b}}                 & $\Delta T{\scriptstyle_{e, tot}}$                                     &                      &                          & $\sim$9.5--198       &                   \\
                                                          & $\langle T{\scriptstyle_{e, tot}} \rangle{\scriptstyle_{up}}$         &                      &                          & $\sim$5.2--31.9      &                   \\
                                                          & $\langle T{\scriptstyle_{e, tot}} \rangle{\scriptstyle_{dn}}$         &                      &                          & $\sim$28--224        &                   \\
  \citet[][]{thomsen93a}                                  & $\Delta T{\scriptstyle_{e, tot}}$                                     &        ISEE 2        &                          & $\sim$7.8--172       &                   \\
  \hline
  \multirow{2}{*}{\citet[][]{chen18a}}                    & $T{\scriptstyle_{e, \parallel}}$                                      &         MMS          &            BS            & $\sim$25--210        &                   \\
                                                          & $T{\scriptstyle_{e, \perp}}$                                          &                      &                          & $\sim$25--150        &                   \\
  \hline
  \multirow{2}{*}{\citet[][]{wilsoniii14a, wilsoniii14b}} & $\langle T{\scriptstyle_{e, tot}} \rangle{\scriptstyle_{up}}$         &        THEMIS        &            BS            & $\sim$7.9--31.2      &                   \\
                                                          & $\langle T{\scriptstyle_{e, tot}} \rangle{\scriptstyle_{dn}}$         &                      &                          & $\sim$30.5--81.5     &                   \\
  \hline
  \multirow{2}{*}{\citet[][]{maksimovic97a}}              & $T{\scriptstyle_{ec, tot}}$                                           &       Ulysses        &            SW            & $\sim$4.6--15.5      &                   \\
                                                          & $T{\scriptstyle_{eh, tot}}$                                           &       Ulysses        &                          & $\sim$49--86         &                   \\
  \hline
  \multirow{2}{*}{\citet[][]{hull01a}}                    & $\langle T{\scriptstyle_{e, tot}} \rangle{\scriptstyle_{up}}$         &      \emph{Wind}     &            BS            &                      & $\sim$12.1        \\
                                                          & $\langle T{\scriptstyle_{e, tot}} \rangle{\scriptstyle_{dn}}$         &                      &                          &                      & $\sim$29.3        \\
  \multirow{2}{*}{\citet[][]{maksimovic05a}}              & $T{\scriptstyle_{ec, tot}}$                                           &                      &            SW            & $\sim$6.5--10        &                   \\
                                                          & $T{\scriptstyle_{eh, tot}}$                                           &                      &                          & $\sim$14--43         &                   \\
  \multirow{2}{*}{\citet[][]{tao16b}}                     & $T{\scriptstyle_{eh, tot}}$                                           &                      &                          & $\sim$21--62         &                   \\
                                                          & $T{\scriptstyle_{eb, tot}}$                                           &                      &                          & $\sim$23--68         &                   \\
  \citet[][]{ogilvie00a}                                  & $T{\scriptstyle_{eb, tot}}$                                           &                      &                          & $\sim$100--150       &                   \\
  \citet[][]{fitzenreiter03a}                             & $\Delta T{\scriptstyle_{e, tot}}$                                     &                      &           IPS            & $\sim$4.3--41.6      &                   \\
  \multirow{2}{*}{\citet[][]{pulupa10a}}                  & $\langle T{\scriptstyle_{e, tot}} \rangle{\scriptstyle_{up}}$         &                      &                          & $\sim$4.1--36.8      &                   \\
                                                          & $\langle T{\scriptstyle_{e, tot}} \rangle{\scriptstyle_{dn}}$         &                      &                          & $\sim$7.3--60.2      &                   \\
  \multirow{9}{*}{\citet[][]{wilsoniii09a}}               & $T{\scriptstyle_{e, tot}}$                                            &                      &                          & $\sim$11--76         &                   \\
                                                          & $T{\scriptstyle_{e, \parallel}}$                                      &                      &                          & $\sim$10--80         &                   \\
                                                          & $T{\scriptstyle_{e, \perp}}$                                          &                      &                          & $\sim$11--75         &                   \\
                                                          & $T{\scriptstyle_{ec, tot}}$                                           &                      &                          & $\sim$9--38          &                   \\
                                                          & $T{\scriptstyle_{ec, \parallel}}$                                     &                      &                          & $\sim$9--37          &                   \\
                                                          & $T{\scriptstyle_{ec, \perp}}$                                         &                      &                          & $\sim$10--38         &                   \\
                                                          & $T{\scriptstyle_{eh, tot}}$                                           &                      &                          & $\sim$43--175        &                   \\
                                                          & $T{\scriptstyle_{eh, \parallel}}$                                     &                      &                          & $\sim$39--190        &                   \\
                                                          & $T{\scriptstyle_{eh, \perp}}$                                         &                      &                          & $\sim$44--189        &                   \\
  \multirow{9}{*}{\citet[][]{wilsoniii10a}}               & $T{\scriptstyle_{e, tot}}$                                            &                      &                          & $\sim$10--64         &                   \\
                                                          & $T{\scriptstyle_{e, \parallel}}$                                      &                      &                          & $\sim$9.8--90        &                   \\
                                                          & $T{\scriptstyle_{e, \perp}}$                                          &                      &                          & $\sim$9.7--90        &                   \\
                                                          & $T{\scriptstyle_{ec, tot}}$                                           &                      &                          & $\sim$6--37          &                   \\
                                                          & $T{\scriptstyle_{ec, \parallel}}$                                     &                      &                          & $\sim$6--55          &                   \\
                                                          & $T{\scriptstyle_{ec, \perp}}$                                         &                      &                          & $\sim$6--38          &                   \\
                                                          & $T{\scriptstyle_{eh, tot}}$                                           &                      &                          & $\sim$35--220        &                   \\
                                                          & $T{\scriptstyle_{eh, \parallel}}$                                     &                      &                          & $\sim$35--250        &                   \\
                                                          & $T{\scriptstyle_{eh, \perp}}$                                         &                      &                          & $\sim$35--240        &                   \\
  \multirow{5}{*}{\citet[][]{wilsoniii12c}}               & $T{\scriptstyle_{e, tot}}$                                            &                      &                          & $\sim$18.9--60.6     & $\sim$41.6        \\
                                                          & $T{\scriptstyle_{ec, tot}}$                                           &                      &                          & $\sim$16.6--38.7     & $\sim$31.1        \\
                                                          & $T{\scriptstyle_{ec, \parallel}}$                                     &                      &                          & $\sim$16.5--42.2     & $\sim$30.2        \\
                                                          & $T{\scriptstyle_{eh, \parallel}}$                                     &                      &                          & $\sim$117--290       & $\sim$208         \\
                                                          & $T{\scriptstyle_{eh, \perp}}$                                         &                      &                          & $\sim$113--294       & $\sim$201         \\
  \multirow{9}{*}{\citet[][]{wilsoniii13a}}               & $T{\scriptstyle_{e, tot}}$                                            &                      &                          & $\sim$26--64         &                   \\
                                                          & $T{\scriptstyle_{e, \parallel}}$                                      &                      &                          & $\sim$24--64         &                   \\
                                                          & $T{\scriptstyle_{e, \perp}}$                                          &                      &                          & $\sim$25--65         &                   \\
                                                          & $T{\scriptstyle_{ec, tot}}$                                           &                      &                          & $\sim$24--54         &                   \\
                                                          & $T{\scriptstyle_{ec, \parallel}}$                                     &                      &                          & $\sim$23--57         &                   \\
                                                          & $T{\scriptstyle_{ec, \perp}}$                                         &                      &                          & $\sim$23--55         &                   \\
                                                          & $T{\scriptstyle_{eh, tot}}$                                           &                      &                          & $\sim$160--300       &                   \\
                                                          & $T{\scriptstyle_{eh, \parallel}}$                                     &                      &                          & $\sim$115--280       &                   \\
                                                          & $T{\scriptstyle_{eh, \perp}}$                                         &                      &                          & $\sim$160--315       &                   \\
  \hline
  \enddata
  \tablenotetext{a}{mean or average}
  \tablenotetext{b}{SW $\equiv$ Solar Wind, a generic term for ambient/all solar wind conditions}
  \tablenotetext{c}{Kronian bow shock}
  \tablenotetext{d}{terrestrial bow shock}
  \tablenotetext{e}{Fast and Slow SW are typically defined as bulk flow speed above or below, respectively, some threshold (typically $\sim$350--500 $km \ s^{-1}$)}
  \tablenotetext{f}{interplanetary shock}
  \tablenotetext{g}{IFS $\equiv$ terrestrial ion foreshock}
  \tablecomments{OMNI is a dataset comprised of multiple spacecraft from SPDF/CDAWeb, where All refers to 1963--Present and Late to 1978--Present.  For symbol definitions, see Appendix \ref{app:Definitions}.}
\end{deluxetable*}

\startlongtable  
\begin{deluxetable*}{| l | c | c | c | c | c |}
  \tabletypesize{\footnotesize}    
  \tablecaption{Measurements of Electron Densities [$cm^{-3}$] at 1 AU \label{tab:PrevDensities}}
  \tablehead{\colhead{Reference} & \colhead{Parameter} & \colhead{Spacecraft} & \colhead{Notes} & \colhead{$X{\scriptstyle_{min}}$--$X{\scriptstyle_{max}}$} & \colhead{$\bar{\mathbf{X}}$}}
  \startdata
  \multirow{3}{*}{\citet[][]{skoug00a}}                   & $n{\scriptstyle_{ec}}$                                                &         ACE          &        SW and ICME       &                        & $\sim$7.6--10.2     \\
                                                          & $n{\scriptstyle_{eh}}$                                                &                      &                          &                        & $\sim$0.11--0.19    \\
                                                          & $n{\scriptstyle_{eh}} / n{\scriptstyle_{ec}}$                         &                      &                          &                        & $\sim$0.027--0.028  \\
  \hline
  \multirow{2}{*}{\citet[][]{lefebvre07a}}                & $\langle n{\scriptstyle_{e}} \rangle{\scriptstyle_{up}}$              &       Cluster        &            BS            &                        & $\sim$6.6--11.0     \\
                                                          & $\langle n{\scriptstyle_{e}} \rangle{\scriptstyle_{dn}}$              &                      &                          &                        & $\sim$19.3--37.8    \\
  \multirow{3}{*}{\citet[][]{vinas10a}}                   & $n{\scriptstyle_{e}}$                                                 &                      &            SW            & $\sim$14--19           &                     \\
                                                          & $n{\scriptstyle_{eb}}$                                                &                      &                          & $\sim$0.05--0.20       &                     \\
                                                          & $n{\scriptstyle_{eb}} / n{\scriptstyle_{e}}$                          &                      &                          & $\sim$0.0025--0.02     &                     \\
  \hline
  \multirow{2}{*}{\citet[][]{hull98a}}                    & $\langle n{\scriptstyle_{e}} \rangle{\scriptstyle_{up}}$              &       Galileo        &            BS            &                        & $\sim$6.7           \\
                                                          & $\langle n{\scriptstyle_{e}} \rangle{\scriptstyle_{dn}}$              &                      &                          &                        & $\sim$8.5           \\
  \hline
  \multirow{10}{*}{\citet[][]{stverak09a}}                & $n{\scriptstyle_{ec}}$                                                &     Helios 1 \&      &          Slow SW         & $\sim$7--10            &                     \\
                                                          & $n{\scriptstyle_{eh}}$                                                &       Cluster        &                          & $\sim$0.20--0.33       &                     \\
                                                          & $n{\scriptstyle_{eb}}$                                                &                      &                          & $\sim$0.18--0.28       &                     \\
                                                          & $n{\scriptstyle_{eh}} / n{\scriptstyle_{e}}$                          &                      &                          & $\sim$0.038--0.045     &                     \\
                                                          & $n{\scriptstyle_{eb}} / n{\scriptstyle_{e}}$                          &                      &                          & $\sim$0.029--0.039     &                     \\
                                                          & $n{\scriptstyle_{ec}}$                                                &                      &          Fast SW         & $\sim$4--6             &                     \\
                                                          & $n{\scriptstyle_{eh}}$                                                &                      &                          & $\sim$0.20--0.36       &                     \\
                                                          & $n{\scriptstyle_{eb}}$                                                &                      &                          & $\sim$0.17--0.29       &                     \\
                                                          & $n{\scriptstyle_{eh}} / n{\scriptstyle_{e}}$                          &                      &                          & $\sim$0.041--0.071     &                     \\
                                                          & $n{\scriptstyle_{eb}} / n{\scriptstyle_{e}}$                          &                      &                          & $\sim$0.025--0.051     &                     \\
  \hline
  \multirow{3}{*}{\citet[][]{feldman75a}}                 & $n{\scriptstyle_{e}}$                                                 &         Imp 7        &            SW            &                        & $\sim$9.0--11.3     \\
                                                          & $n{\scriptstyle_{eh}}$                                                &       and Imp 8      &                          &                        & $\sim$0.31--0.56    \\
                                                          & $n{\scriptstyle_{eh}} / n{\scriptstyle_{e}}$                          &                      &                          &                        & $\sim$0.033--0.071  \\
  \citet[][]{feldman79b}                                  & $n{\scriptstyle_{eh}} / n{\scriptstyle_{e}}$                          &    Imp 6, 7, \& 8    &                          & $\sim$0.015--0.075     &                     \\
  \hline
  \multirow{3}{*}{\citet[][]{feldman83b}}                 & $\langle n{\scriptstyle_{e}} \rangle{\scriptstyle_{up}}$              &        ISEE 3        &           IPS            & $\sim$2.0--19.1        &                     \\
                                                          & $\langle n{\scriptstyle_{e}} \rangle{\scriptstyle_{dn}}$              &                      &                          & $\sim$10.0--23.0       &                     \\
                                                          & $\mathcal{R}{\scriptstyle_{ne}}$                                      &                      &                          & $\sim$1.2--4.2         &                     \\
  \citet[][]{hull00a}                                     & $\langle n{\scriptstyle_{e}} \rangle{\scriptstyle_{up}}$              &        ISEE 1        &            BS            & $<$1 to $>$55          & $\sim$10            \\
  \multirow{2}{*}{\citet[][]{phillips89a}}                & $n{\scriptstyle_{e}}$                                                 &        ISEE 3        &            SW            & $<$1 to $>$30          &                     \\
                                                          & $n{\scriptstyle_{ec}}$                                                &                      &                          & $\sim$1--30            &                     \\
  \hline
  \multirow{2}{*}{\citet[][]{maksimovic97a}}              & $n{\scriptstyle_{ec}}$                                                &       Ulysses        &                          & $\sim$0.49--4.81       &                     \\
                                                          & $n{\scriptstyle_{eh}}$                                                &                      &                          & $\sim$0.06--0.18       &                     \\
  \hline
  \multirow{2}{*}{\citet[][]{hull01a}}                    & $\langle n{\scriptstyle_{e}} \rangle{\scriptstyle_{up}}$              &      \emph{Wind}     &            BS            &                        & $\sim$17            \\
                                                          & $\langle n{\scriptstyle_{e}} \rangle{\scriptstyle_{dn}}$              &                      &                          &                        & $\sim$48            \\
  \multirow{5}{*}{\citet[][]{maksimovic05a}}              & $n{\scriptstyle_{e}}$                                                 &                      &            SW            & $\sim$2.7--4.0         &                     \\
                                                          & $n{\scriptstyle_{eh}}$                                                &                      &                          & $\sim$0.23--0.38       &                     \\
                                                          & $n{\scriptstyle_{ec}} / n{\scriptstyle_{e}}$                          &                      &                          & $\sim$0.80--0.99       &                     \\
                                                          & $n{\scriptstyle_{eh}} / n{\scriptstyle_{e}}$                          &                      &                          & $\sim$0.075--0.11      &                     \\
                                                          & $n{\scriptstyle_{eb}} / n{\scriptstyle_{e}}$                          &                      &                          & $\sim$0.0015--0.02     &                     \\
  \citet[][]{nieveschinchilla08a}                         & $n{\scriptstyle_{e}}$                                                 &                      &           ICME           & $\sim$0.5--40          &                     \\
  \multirow{2}{*}{\citet[][]{pulupa10a}}                  & $\langle n{\scriptstyle_{e}} \rangle{\scriptstyle_{up}}$              &                      &           IPS            & $\sim$1--24            &                     \\
                                                          & $\langle n{\scriptstyle_{e}} \rangle{\scriptstyle_{dn}}$              &                      &                          & $\sim$3--50            &                     \\
  \citet[][]{salem01a}                                    & $n{\scriptstyle_{e}}$                                                 &                      &            SW            & $\sim$2--90            &                     \\
  \multirow{3}{*}{\citet[][]{tao16a}}                     & $n{\scriptstyle_{eh}}$                                                &                      &                          & $\sim$0.018--0.29      &                     \\
                                                          & $n{\scriptstyle_{eb}}$                                                &                      &                          & $\sim$0.0017--0.08     &                     \\
                                                          & $n{\scriptstyle_{eb}} / n{\scriptstyle_{eh}}$                         &                      &                          & $\sim$0.025--0.88      &                     \\
  \multirow{2}{*}{\citet[][]{wilsoniii09a}}               & $n{\scriptstyle_{ec}}$                                                &                      &           IPS            & $\sim$4.7--10.4        &                     \\
                                                          & $n{\scriptstyle_{eh}}$                                                &                      &                          & $\sim$0.023--0.051     &                     \\
  \multirow{2}{*}{\citet[][]{wilsoniii10a}}               & $n{\scriptstyle_{ec}}$                                                &                      &                          & $\sim$3--25            &                     \\
                                                          & $n{\scriptstyle_{eh}}$                                                &                      &                          & $\sim$0.03--1.10       &                     \\
  \hline
  \enddata
  \tablecomments{Definitions/Symbols are the same as in Table \ref{tab:PrevTemperatures}.  For symbol definitions, see Appendix \ref{app:Definitions}.}
\end{deluxetable*}

\startlongtable  
\begin{deluxetable*}{| l | c | c | c | c | c |}
  \tabletypesize{\small}    
  \tablecaption{Measurements of Electron Temperature Ratios at 1 AU \label{tab:PrevTempRatios}}
  \tablehead{\colhead{Reference} & \colhead{Parameter} & \colhead{Spacecraft} & \colhead{Notes} & \colhead{$X{\scriptstyle_{min}}$--$X{\scriptstyle_{max}}$} & \colhead{$\bar{\mathbf{X}}$}}
  \startdata
  \citet[][]{skoug00a}                                    & $\tensor*{ \mathcal{T}  }{^{eh}_{ec}}{\scriptstyle_{      tot}}$      &         ACE          &            SW            & $\sim$2--40            & $\sim$7.25          \\
  \hline
  \citet[][]{feldman75a}                                  & $\tensor*{ \mathcal{T}  }{^{eh}_{ec}}{\scriptstyle_{      tot}}$      &      Imp 7 \& 8      &            SW            &                        & $\sim$5.5--7.2      \\
  \hline
  \citet[][]{bame79a}                                     & $\mathcal{R}{\scriptstyle_{Te, tot}}$                                 &      ISEE 1 \& 2     &            BS            & $\sim$1.3--9.5         & $\sim$2.7           \\
  \citet[][]{thomsen85a}                                  & $\mathcal{R}{\scriptstyle_{Te, \perp}}$                               &                      &                          & $\sim$1.7--3.5         &                     \\
  \citet[][]{thomsen87b}                                  & $\mathcal{R}{\scriptstyle_{Te, tot}}$                                 &                      &                          & $\sim$1.0--19.6        &                     \\
  \citet[][]{feldman83b}                                  & $\mathcal{R}{\scriptstyle_{Te, tot}}$                                 &        ISEE 3        &           IPS            & $\sim$1.0--3.0         &                     \\
  \hline
  \citet[][]{pulupa10a}                                   & $\mathcal{R}{\scriptstyle_{Te, tot}}$                                 &      \emph{Wind}     &           IPS            & $\sim$1.0--4.0         &                     \\
  \multirow{2}{*}{\citet[][]{wilsoniii09a}}               & $\tensor*{ \mathcal{T}  }{^{eh}_{ec}}{\scriptstyle_{\parallel}}$      &                      &                          & $\sim$3.5--12.8        &                     \\
                                                          & $\tensor*{ \mathcal{T}  }{^{eh}_{ec}}{\scriptstyle_{    \perp}}$      &                      &                          & $\sim$4.4--10.9        &                     \\
  \multirow{2}{*}{\citet[][]{wilsoniii12c}}               & $\tensor*{ \mathcal{T}  }{^{eh}_{ec}}{\scriptstyle_{\parallel}}$      &                      &                          & $\sim$3.82--8.38       & $\sim$6.78          \\
                                                          & $\tensor*{ \mathcal{T}  }{^{eh}_{ec}}{\scriptstyle_{    \perp}}$      &                      &                          & $\sim$5.53--7.10       & $\sim$6.61          \\
  \hline
  \enddata
  \tablecomments{Definitions/Symbols are the same as in Tables \ref{tab:PrevTemperatures}.  For symbol definitions, see Appendix \ref{app:Definitions}.}
\end{deluxetable*}

\begin{deluxetable*}{| l | c | c | c | c | c |}
\startlongtable  
  \tabletypesize{\small}    
  \tablecaption{Measurements of Electron betas at 1 AU \label{tab:PrevBetas}}
  \tablehead{\colhead{Reference} & \colhead{Parameter} & \colhead{Spacecraft} & \colhead{Notes} & \colhead{$X{\scriptstyle_{min}}$--$X{\scriptstyle_{max}}$} & \colhead{$\bar{\mathbf{X}}$}}
  \startdata
  \citet[][]{lacombe14a}                                  & $\beta{\scriptstyle_{e, tot}}$                                        &       Cluster        &            SW            & $\sim$0.09--25         &                     \\
  \citet[][]{lacombe17a}                                  & $\beta{\scriptstyle_{e, \parallel}}$                                  &                      &                          & $\sim$0.08--3.9        &                     \\
  \multirow{2}{*}{\citet[][]{lefebvre07a}}                & $\langle \beta{\scriptstyle_{e, tot}} \rangle{\scriptstyle_{up}}$     &                      &                          &                        & $\sim$0.45--5.99    \\
                                                          & $\langle \beta{\scriptstyle_{e, tot}} \rangle{\scriptstyle_{dn}}$     &                      &                          &                        & $\sim$0.63--3.40    \\
  \multirow{2}{*}{\citet[][]{vinas10a}}                   & $\beta{\scriptstyle_{e, \parallel}}$                                  &                      &                          & $\sim$0.4--1.0         &                     \\
                                                          & $\beta{\scriptstyle_{eb, \parallel}}$                                 &                      &                          & $\sim$1.0--4.0         &                     \\
  \hline
  \multirow{2}{*}{\citet[][]{hull98a}}                    & $\langle \beta{\scriptstyle_{e, tot}} \rangle{\scriptstyle_{up}}$     &       Galileo        &            BS            &                        & $\sim$0.46          \\
                                                          & $\langle \beta{\scriptstyle_{e, tot}} \rangle{\scriptstyle_{dn}}$     &                      &                          &                        & $\sim$0.41          \\
  \hline
  \multirow{4}{*}{\citet[][]{stverak08a}}                 & $\beta{\scriptstyle_{ec, \parallel}}$                                 &      Helios I \&     &          Slow SW         & $\sim$0.04--40         &                     \\
                                                          & $\beta{\scriptstyle_{eh, \parallel}}$                                 &       Cluster        &                          & $\sim$0.002--15        &                     \\
                                                          & $\beta{\scriptstyle_{ec, \parallel}}$                                 &                      &          Fast SW         & $\sim$0.025--1.2       &                     \\
                                                          & $\beta{\scriptstyle_{eh, \parallel}}$                                 &                      &                          & $\sim$0.002--4.0       &                     \\
  \hline
  \citet[][]{lazar17b}                                    & $\beta{\scriptstyle_{eh, \parallel}}$                                 & HCU\tablenotemark{a} &            SW            & $\sim$0.001--80        &                     \\
  \hline
  \citet[][]{hull00a}                                     & $\langle \beta{\scriptstyle_{e, tot}} \rangle{\scriptstyle_{up}}$     &        ISEE 1        &            BS            & $\sim$0.1--15.8        &                     \\
  \hline
  \multirow{2}{*}{\citet[][]{wilsoniii14a, wilsoniii14b}} & $\langle \beta{\scriptstyle_{e, tot}} \rangle{\scriptstyle_{up}}$     &        THEMIS        &            BS            &                        & $\sim$0.39--17.2    \\
                                                          & $\langle \beta{\scriptstyle_{e, tot}} \rangle{\scriptstyle_{dn}}$     &                      &                          &                        & $\sim$0.64--5.97    \\
  \hline
  \multirow{2}{*}{\citet[][]{adrian16a}}                  & $\beta{\scriptstyle_{e, \parallel}}$                                  &      \emph{Wind}     &          Slow SW         & $\sim$0.02 to $>$10    &                     \\
                                                          & $\beta{\scriptstyle_{e, \parallel}}$                                  &                      &          Fast SW         & $\sim$0.05 to $>$10    &                     \\
  \citet[][]{bale13a}                                     & $\beta{\scriptstyle_{e, tot}}$                                        &                      &            SW            & $\sim$0.01 to $>$100   &                     \\
  \citet[][]{chen16b}                                     & $\beta{\scriptstyle_{e, \parallel}}$                                  &                      &                          & $\sim$0.03 to $>$100   &                     \\
  \multirow{2}{*}{\citet[][]{hull01a}}                    & $\langle \beta{\scriptstyle_{e, tot}} \rangle{\scriptstyle_{up}}$     &                      &            BS            &                        & $\sim$1.5           \\
                                                          & $\langle \beta{\scriptstyle_{e, tot}} \rangle{\scriptstyle_{dn}}$     &                      &                          &                        & $\sim$1.4           \\
  \citet[][]{wilsoniii09a}                                & $\beta{\scriptstyle_{ec, \parallel}}$                                 &                      &           IPS            & $\sim$0.70--1.16       &                     \\
  \multirow{2}{*}{\citet[][]{wilsoniii10a}}               & $\beta{\scriptstyle_{e, tot}}$                                        &                      &                          & $\sim$0.55--11.5       &                     \\
                                                          & $\beta{\scriptstyle_{ec, \parallel}}$                                 &                      &                          & $\sim$0.1--8.0         &                     \\
  \citet[][]{wilsoniii12c}                                & $\beta{\scriptstyle_{ec, \parallel}}$                                 &                      &                          & $\sim$0.52--1.80       & $\sim$1.35          \\
  \citet[][]{wilsoniii13a}                                & $\beta{\scriptstyle_{ec, \parallel}}$                                 &                      &                          & $\sim$0.20--1.05       &                     \\
  \multirow{2}{*}{\citet[][]{wilsoniii13b}}               & $\beta{\scriptstyle_{ec, tot}}$                                       &                      &           IFS            & $\sim$0.1 to $>$100    &                     \\
                                                          & $\beta{\scriptstyle_{eh, tot}}$                                       &                      &                          & $\sim$0.1 to $>$400    &                     \\
  \multirow{9}{*}{\citet[][]{wilsoniii18b}}               & $\beta{\scriptstyle_{e, tot}}$                                        &                      &            SW            & $\sim$0.006--8870      & $\sim$2.31          \\
                                                          & $\beta{\scriptstyle_{e, \parallel}}$                                  &                      &                          & $\sim$0.005--8848      &                     \\
                                                          & $\beta{\scriptstyle_{e, \perp}}$                                      &                      &                          & $\sim$0.007--8914      &                     \\
                                                          & $\beta{\scriptstyle_{e, tot}}$                                        &                      &          Slow SW         & $\sim$0.01--4329       & $\sim$3.35          \\
                                                          & $\beta{\scriptstyle_{e, \parallel}}$                                  &                      &                          & $\sim$0.01--4328       & $\sim$3.33          \\
                                                          & $\beta{\scriptstyle_{e, \perp}}$                                      &                      &                          & $\sim$0.01--4332       & $\sim$3.41          \\
                                                          & $\beta{\scriptstyle_{e, tot}}$                                        &                      &          Fast SW         & $\sim$0.02--680        & $\sim$1.05          \\
                                                          & $\beta{\scriptstyle_{e, \parallel}}$                                  &                      &                          & $\sim$0.02--665        & $\sim$1.00          \\
                                                          & $\beta{\scriptstyle_{e, \perp}}$                                      &                      &                          & $\sim$0.02--710        & $\sim$1.16          \\
  \hline
  \enddata
  \tablenotetext{a}{HCU $\equiv$ Helios 1, Cluster, and Ulysses spacecraft}
  \tablecomments{Definitions/Symbols are the same as in Table \ref{tab:PrevTemperatures}.  For symbol definitions, see Appendix \ref{app:Definitions}.}
\end{deluxetable*}

\startlongtable  
\begin{deluxetable*}{| l | c | c | c | c | c |}
  \tabletypesize{\small}    
  \tablecaption{Measurements of Electron Temperature Anisotropies at 1 AU \label{tab:PrevTempAnisotropies}}
  \tablehead{\colhead{Reference} & \colhead{Parameter} & \colhead{Spacecraft} & \colhead{Notes} & \colhead{$X{\scriptstyle_{min}}$--$X{\scriptstyle_{max}}$} & \colhead{$\bar{\mathbf{X}}$}}
  \startdata
  \citet[][]{lacombe14a}                                  & $\mathcal{A}{\scriptstyle_{e}}$                                       &       Cluster        &            SW            & $\sim$0.5--1.1         &                     \\
  \citet[][]{lacombe17a}                                  & $\mathcal{A}{\scriptstyle_{e}}$                                       &                      &                          & $\sim$0.57--1.0        &                     \\
  \multirow{2}{*}{\citet[][]{vinas10a}}                   & $\mathcal{A}{\scriptstyle_{e}}$                                       &                      &                          & $\sim$1.0              &                     \\
                                                          & $\mathcal{A}{\scriptstyle_{eb}}$                                      &                      &                          & $\sim$0.5--4.0         &                     \\
  \multirow{2}{*}{\citet[][]{lefebvre07a}}                & $\langle \mathcal{A}{\scriptstyle_{e}} \rangle{\scriptstyle_{up}}$    &                      &            BS            &                        & $\sim$0.61--0.93    \\
                                                          & $\langle \mathcal{A}{\scriptstyle_{e}} \rangle{\scriptstyle_{dn}}$    &                      &                          &                        & $\sim$0.91--1.07    \\
  \hline
  \citet[][]{lazar17b}                                    & $\mathcal{A}{\scriptstyle_{eh}}$                                      & HCU\tablenotemark{a} &            SW            & $\sim$0.25--1.75       &                     \\
  \multirow{4}{*}{\citet[][]{stverak08a}}                 & $\mathcal{A}{\scriptstyle_{ec}}$                                      & HC\tablenotemark{b}  &          Slow SW         & $\sim$0.4--1.5         &                     \\
                                                          & $\mathcal{A}{\scriptstyle_{eh}}$                                      &                      &                          & $\sim$0.5--1.5         &                     \\
                                                          & $\mathcal{A}{\scriptstyle_{ec}}$                                      &                      &          Fast SW         & $\sim$0.45--1.1        &                     \\
                                                          & $\mathcal{A}{\scriptstyle_{eh}}$                                      &                      &                          & $\sim$0.55--1.2        &                     \\
  \multirow{3}{*}{\citet[][]{schwenn90a}}                 & $\mathcal{A}{\scriptstyle_{e}}$                                       & HCI\tablenotemark{c} &            SW            &                        & $\sim$0.83          \\
                                                          & $\mathcal{A}{\scriptstyle_{e}}$                                       &                      &          Slow SW         &                        & $\sim$0.62          \\
                                                          & $\mathcal{A}{\scriptstyle_{e}}$                                       &                      &          Fast SW         &                        & $\sim$0.85          \\
  \hline
  \citet[][]{feldman73b}                                  & $\langle \mathcal{A}{\scriptstyle_{e}} \rangle{\scriptstyle_{up}}$    &         Imp 6        &            BS            & $\sim$0.67--1.00       &                     \\
  \multirow{2}{*}{\citet[][]{feldman75a}}                 & $\mathcal{A}{\scriptstyle_{ec}}$                                      &      Imp 7 \& 8      &            SW            &                        & $\sim$1.06--1.10    \\
                                                          & $\mathcal{A}{\scriptstyle_{eh}}$                                      &                      &                          &                        & $\sim$1.22--1.31    \\
  \multirow{3}{*}{\citet[][]{feldman78b}}                 & $\mathcal{A}{\scriptstyle_{e}}$                                       &    Imp 6, 7, \& 8    &          Fast SW         &                        & $\sim$0.67          \\
                                                          & $\mathcal{A}{\scriptstyle_{ec}}$                                      &                      &                          &                        & $\sim$0.80          \\
                                                          & $\mathcal{A}{\scriptstyle_{eh}}$                                      &                      &                          &                        & $\sim$0.50          \\
  \multirow{3}{*}{\citet[][]{feldman79b}}                 & $\mathcal{A}{\scriptstyle_{e}}$                                       &                      &            SW            & $<$0.69 to $>$0.95     &                     \\
                                                          & $\mathcal{A}{\scriptstyle_{ec}}$                                      &                      &                          & $\sim$0.79 to $>$0.97  &                     \\
                                                          & $\mathcal{A}{\scriptstyle_{eh}}$                                      &                      &                          & $<$0.50 to $>$0.87     &                     \\
  \hline
  \multirow{2}{*}{\citet[][]{phillips89a}}                & $\mathcal{A}{\scriptstyle_{ec}}$                                      &        ISEE 3        &            SW            & $\sim$0.30 to $>$1.0   &                     \\
                                                          & $\mathcal{A}{\scriptstyle_{eh}}$                                      &                      &                          & $\sim$0.25 to $>$1.0   &                     \\
  \citet[][]{phillips89b}                                 & $\mathcal{A}{\scriptstyle_{ec}}$                                      &                      &                          & $\sim$0.7--1.16        &                     \\
  \hline
  \multirow{2}{*}{\citet[][]{adrian16a}}                  & $\mathcal{A}{\scriptstyle_{e}}$                                       &      \emph{Wind}     &          Slow SW         & $\sim$0.4--2.0         &                     \\
                                                          & $\mathcal{A}{\scriptstyle_{e}}$                                       &                      &          Fast SW         & $\sim$0.3--3.0         &                     \\
  \citet[][]{crooker03a}                                  & $\mathcal{A}{\scriptstyle_{eh}}$                                      &                      &            SW            & $\sim$0.1--100         &                     \\
  \citet[][]{salem03a}                                    & $\mathcal{A}{\scriptstyle_{e}}$                                       &                      &                          & $\sim$0.6--1.1         &                     \\
  \multirow{3}{*}{\citet[][]{wilsoniii13b}}               & $\mathcal{A}{\scriptstyle_{e}}$                                       &                      &           IFS            & $\sim$0.50--2.00       &                     \\
                                                          & $\mathcal{A}{\scriptstyle_{ec}}$                                      &                      &                          & $\sim$0.40--1.50       &                     \\
                                                          & $\mathcal{A}{\scriptstyle_{eh}}$                                      &                      &                          & $\sim$0.30--1.40       &                     \\
  \citet[][]{pulupa10a}                                   & $\langle \mathcal{A}{\scriptstyle_{e}} \rangle{\scriptstyle_{up}}$    &                      &           IPS            & $\sim$0.50--1.10       &                     \\
  \multirow{2}{*}{\citet[][]{wilsoniii09a}}               & $\mathcal{A}{\scriptstyle_{ec}}$                                      &                      &                          & $\sim$0.70--1.11       &                     \\
                                                          & $\mathcal{A}{\scriptstyle_{eh}}$                                      &                      &                          & $\sim$0.55--1.14       &                     \\
  \multirow{3}{*}{\citet[][]{wilsoniii10a}}               & $\mathcal{A}{\scriptstyle_{e}}$                                       &                      &                          & $\sim$0.55--1.30       &                     \\
                                                          & $\mathcal{A}{\scriptstyle_{ec}}$                                      &                      &                          & $\sim$0.55--1.35       &                     \\
                                                          & $\mathcal{A}{\scriptstyle_{eh}}$                                      &                      &                          & $\sim$0.55--1.80       &                     \\
  \multirow{2}{*}{\citet[][]{wilsoniii12c}}               & $\mathcal{A}{\scriptstyle_{ec}}$                                      &                      &                          & $\sim$0.81--1.09       & $\sim$0.97          \\
                                                          & $\mathcal{A}{\scriptstyle_{eh}}$                                      &                      &                          & $\sim$0.81--1.29       & $\sim$0.995         \\
  \multirow{3}{*}{\citet[][]{wilsoniii13a}}               & $\mathcal{A}{\scriptstyle_{e}}$                                       &                      &                          & $\sim$0.75--1.31       &                     \\
                                                          & $\mathcal{A}{\scriptstyle_{ec}}$                                      &                      &                          & $\sim$0.73--1.30       &                     \\
                                                          & $\mathcal{A}{\scriptstyle_{eh}}$                                      &                      &                          & $\sim$0.75--1.76       &                     \\
  \hline
  \enddata
  \tablenotetext{a}{HCU $\equiv$ Helios 1, Cluster, and Ulysses spacecraft}
  \tablenotetext{b}{HC $\equiv$ Helios 1 and Cluster}
  \tablenotetext{c}{HCI $\equiv$ Helios 1 \& 2 and Imp 7 \& 8}
  \tablecomments{Definitions/Symbols are the same as in Table \ref{tab:PrevTemperatures}.  For symbol definitions, see Appendix \ref{app:Definitions}.}
\end{deluxetable*}


\begin{thebibliography}{}
\expandafter\ifx\csname natexlab\endcsname\relax\def\natexlab#1{#1}\fi
\providecommand{\url}[1]{\href{#1}{#1}}
\providecommand{\dodoi}[1]{doi:~\href{http://doi.org/#1}{\nolinkurl{#1}}}
\providecommand{\doeprint}[1]{\href{http://ascl.net/#1}{\nolinkurl{http://ascl.net/#1}}}
\providecommand{\doarXiv}[1]{\href{https://arxiv.org/abs/#1}{\nolinkurl{https://arxiv.org/abs/#1}}}

\bibitem[{{Adrian} {et~al.}(2016){Adrian}, {Vi{\~n}as}, {Moya}, \&
  {Wendel}}]{adrian16a}
{Adrian}, M.~L., {Vi{\~n}as}, A.~F., {Moya}, P.~S., \& {Wendel}, D.~E. 2016,
  Astrophys. J., 833, 49, \dodoi{10.3847/1538-4357/833/1/49}

\bibitem[{{Anderson} {et~al.}(2012){Anderson}, {Skoug}, {Steinberg}, \&
  {McComas}}]{anderson12a}
{Anderson}, B.~R., {Skoug}, R.~M., {Steinberg}, J.~T., \& {McComas}, D.~J.
  2012, J. Geophys. Res., 117, A04107, \dodoi{10.1029/2011JA017269}

\bibitem[{{Anderson}(1981)}]{anderson81a}
{Anderson}, K.~A. 1981, J. Geophys. Res., 86, 4445,
  \dodoi{10.1029/JA086iA06p04445}

\bibitem[{{Anderson} {et~al.}(1979){Anderson}, {Lin}, {Martel}, {Lin}, {Parks},
  \& {Reme}}]{anderson79a}
{Anderson}, K.~A., {Lin}, R.~P., {Martel}, F., {et~al.} 1979, Geophys. Res.
  Lett., 6, 401, \dodoi{10.1029/GL006i005p00401}

\bibitem[{{Bale} {et~al.}(2013){Bale}, {Pulupa}, {Salem}, {Chen}, \&
  {Quataert}}]{bale13a}
{Bale}, S.~D., {Pulupa}, M., {Salem}, C., {Chen}, C.~H.~K., \& {Quataert}, E.
  2013, Astrophys. J. Lett., 769, L22, \dodoi{10.1088/2041-8205/769/2/L22}

\bibitem[{{Ball} \& {Melrose}(2001)}]{ball01a}
{Ball}, L., \& {Melrose}, D.~B. 2001, Publ. Astron. Soc. Aust., 18, 361,
  \dodoi{10.1071/AS01047}

\bibitem[{{Bame} {et~al.}(1979){Bame}, {Asbridge}, {Gosling}, {Halbig},
  {Paschmann}, {Sckopke}, \& {Rosenbauer}}]{bame79a}
{Bame}, S.~J., {Asbridge}, J.~R., {Gosling}, J.~T., {et~al.} 1979, Space Sci.
  Rev., 23, 75, \dodoi{10.1007/BF00174112}

\bibitem[{{Caprioli} \& {Spitkovsky}(2014)}]{caprioli14a}
{Caprioli}, D., \& {Spitkovsky}, A. 2014, Astrophys. J., 783, 91,
  \dodoi{10.1088/0004-637X/783/2/91}

\bibitem[{{Chen} {et~al.}(2016){Chen}, {Matteini}, {Schekochihin}, {Stevens},
  {Salem}, {Maruca}, {Kunz}, \& {Bale}}]{chen16b}
{Chen}, C.~H.~K., {Matteini}, L., {Schekochihin}, A.~A., {et~al.} 2016,
  Astrophys. J. Lett., 825, L26, \dodoi{10.3847/2041-8205/825/2/L26}

\bibitem[{{Chen} {et~al.}(2018){Chen}, {Wang}, {Wilson III}, {Schwartz},
  {Gershman}, {Bessho}, {Malaspina}, {Wilder}, {Moore}, {Giles}, {Ergun},
  {Hesse}, {Lai}, {Russell}, {Strangeway}, {Torbert}, {Vi{\~n}as}, {Burch},
  {Lee}, {Pollock}, {Dorelli}, {Paterson}, {Goodrich}, {Lavraud},
  {Khotyaintsev}, {Lindqvist}, {Le}, \& {Avanov}}]{chen18a}
{Chen}, L.-J., {Wang}, S., {Wilson III}, L.~B., {et~al.} 2018, Phys. Rev.
  Lett., 120, 225101, \dodoi{10.1103/PhysRevLett.120.225101}

\bibitem[{{Coroniti}(1970)}]{coroniti70b}
{Coroniti}, F.~V. 1970, J. Plasma Phys., 4, 265,
  \dodoi{10.1017/S0022377800004992}

\bibitem[{{Crooker} {et~al.}(2003){Crooker}, {Larson}, {Kahler}, {Lamassa}, \&
  {Spence}}]{crooker03a}
{Crooker}, N.~U., {Larson}, D.~E., {Kahler}, S.~W., {Lamassa}, S.~M., \&
  {Spence}, H.~E. 2003, Geophys. Res. Lett., 30, 120000,
  \dodoi{10.1029/2003GL017036}

\bibitem[{{Crooker} \& {Pagel}(2008)}]{crooker08a}
{Crooker}, N.~U., \& {Pagel}, C. 2008, J. Geophys. Res., 113, 2106,
  \dodoi{10.1029/2007JA012421}

\bibitem[{{Feldman} {et~al.}(1983{\natexlab{a}}){Feldman}, {Anderson}, {Bame},
  {Gosling}, {Zwickl}, \& {Smith}}]{feldman83b}
{Feldman}, W.~C., {Anderson}, R.~C., {Bame}, S.~J., {et~al.}
  1983{\natexlab{a}}, J. Geophys. Res., 88, 9949,
  \dodoi{10.1029/JA088iA12p09949}

\bibitem[{{Feldman} {et~al.}(1979){Feldman}, {Asbridge}, {Bame}, \&
  {Gosling}}]{feldman79b}
{Feldman}, W.~C., {Asbridge}, J.~R., {Bame}, S.~J., \& {Gosling}, J.~T. 1979,
  J. Geophys. Res., 84, 7371, \dodoi{10.1029/JA084iA12p07371}

\bibitem[{{Feldman} {et~al.}(1978){Feldman}, {Asbridge}, {Bame}, {Gosling}, \&
  {Lemons}}]{feldman78b}
{Feldman}, W.~C., {Asbridge}, J.~R., {Bame}, S.~J., {Gosling}, J.~T., \&
  {Lemons}, D.~S. 1978, J. Geophys. Res., 83, 5285,
  \dodoi{10.1029/JA083iA11p05285}

\bibitem[{{Feldman} {et~al.}(1973){Feldman}, {Asbridge}, {Bame}, \&
  {Montgomery}}]{feldman73b}
{Feldman}, W.~C., {Asbridge}, J.~R., {Bame}, S.~J., \& {Montgomery}, M.~D.
  1973, J. Geophys. Res., 78, 3697, \dodoi{10.1029/JA078i019p03697}

\bibitem[{{Feldman} {et~al.}(1975){Feldman}, {Asbridge}, {Bame}, {Montgomery},
  \& {Gary}}]{feldman75a}
{Feldman}, W.~C., {Asbridge}, J.~R., {Bame}, S.~J., {Montgomery}, M.~D., \&
  {Gary}, S.~P. 1975, J. Geophys. Res., 80, 4181,
  \dodoi{10.1029/JA080i031p04181}

\bibitem[{{Feldman} {et~al.}(1983{\natexlab{b}}){Feldman}, {Anderson}, {Bame},
  {Gary}, {Gosling}, {McComas}, {Thomsen}, {Paschmann}, \&
  {Hoppe}}]{feldman83a}
{Feldman}, W.~C., {Anderson}, R.~C., {Bame}, S.~J., {et~al.}
  1983{\natexlab{b}}, J. Geophys. Res., 88, 96, \dodoi{10.1029/JA088iA01p00096}

\bibitem[{{Fitzenreiter} {et~al.}(2003){Fitzenreiter}, {Ogilvie}, {Bale}, \&
  {Vi{\~n}as}}]{fitzenreiter03a}
{Fitzenreiter}, R.~J., {Ogilvie}, K.~W., {Bale}, S.~D., \& {Vi{\~n}as}, A.~F.
  2003, J. Geophys. Res., 108, 1415, \dodoi{10.1029/2003JA009865}

\bibitem[{{Gary} {et~al.}(1994){Gary}, {Scime}, {Phillips}, \&
  {Feldman}}]{gary94a}
{Gary}, S.~P., {Scime}, E.~E., {Phillips}, J.~L., \& {Feldman}, W.~C. 1994, J.
  Geophys. Res., 99, 23391, \dodoi{10.1029/94JA02067}

\bibitem[{{Gary} {et~al.}(1999){Gary}, {Skoug}, \& {Daughton}}]{gary99a}
{Gary}, S.~P., {Skoug}, R.~M., \& {Daughton}, W. 1999, Phys. Plasmas, 6, 2607,
  \dodoi{10.1063/1.873532}

\bibitem[{{Gershman} {et~al.}(2015){Gershman}, {Dorelli}, {F.-Vi{\~n}as}, \&
  {Pollock}}]{gershman15a}
{Gershman}, D.~J., {Dorelli}, J.~C., {F.-Vi{\~n}as}, A., \& {Pollock}, C.~J.
  2015, J. Geophys. Res., 120, 6633, \dodoi{10.1002/2014JA020775}

\bibitem[{{Goodrich} {et~al.}(2018){Goodrich}, {Ergun}, {Schwartz}, {Wilson
  III}, {Newman}, {Wilder}, {Holmes}, {Johlander}, {Burch}, {Torbert},
  {Khotyaintsev}, {Lindqvist}, {Russell}, {Gershman}, {Giles}, \&
  {Andersson}}]{goodrich18c}
{Goodrich}, K.~A., {Ergun}, R.~E., {Schwartz}, S.~J., {et~al.} 2018, J.
  Geophys. Res., 123, 9430, \dodoi{10.1029/2018JA025830}

\bibitem[{{Goodrich} {et~al.}(2019){Goodrich}, {Ergun}, {Schwartz}, {Wilson
  III}, {Johlander}, {Newman}, {Wilder}, {Holmes}, {Burch}, {Torbert},
  {Khotyaintsev}, {Lindqvist}, {Strangeway}, {Russell}, {Gershman}, \&
  {Giles}}]{goodrich19a}
---. 2019, J. Geophys. Res., 124, 1855, \dodoi{10.1029/2018JA026436}

\bibitem[{{Graham} {et~al.}(2018){Graham}, {Rae}, {Owen}, \&
  {Walsh}}]{graham18a}
{Graham}, G.~A., {Rae}, I.~J., {Owen}, C.~J., \& {Walsh}, A.~P. 2018,
  Astrophys. J., 855, 40, \dodoi{10.3847/1538-4357/aaaf1b}

\bibitem[{{Graham} {et~al.}(2017){Graham}, {Rae}, {Owen}, {Walsh}, {Arridge},
  {Gilbert}, {Lewis}, {Jones}, {Forsyth}, {Coates}, \& {Waite}}]{graham17a}
{Graham}, G.~A., {Rae}, I.~J., {Owen}, C.~J., {et~al.} 2017, J. Geophys. Res.,
  122, 3858, \dodoi{10.1002/2016JA023656}

\bibitem[{{Gurgiolo} \& {Goldstein}(2016)}]{gurgiolo16a}
{Gurgiolo}, C., \& {Goldstein}, M.~L. 2016, Ann. Geophys., 34, 1175,
  \dodoi{10.5194/angeo-34-1175-2016}

\bibitem[{{Gurgiolo} {et~al.}(2012){Gurgiolo}, {Goldstein}, {Vi{\~n}as}, \&
  {Fazakerley}}]{gurgiolo12a}
{Gurgiolo}, C., {Goldstein}, M.~L., {Vi{\~n}as}, A.~F., \& {Fazakerley}, A.~N.
  2012, Ann. Geophys., 30, 163, \dodoi{10.5194/angeo-30-163-2012}

\bibitem[{{Harten} \& {Clark}(1995)}]{harten95a}
{Harten}, R., \& {Clark}, K. 1995, Space Sci. Rev., 71, 23,
  \dodoi{10.1007/BF00751324}

\bibitem[{{Hernandez} \& {Marsch}(1985)}]{hernandez85a}
{Hernandez}, R., \& {Marsch}, E. 1985, J. Geophys. Res., 90, 11062,
  \dodoi{10.1029/JA090iA11p11062}

\bibitem[{{Hinton}(1984)}]{hinton84a}
{Hinton}, F.~L. 1984, in Basic Plasma Physics: Selected Chapters, Handbook of
  Plasma Physics, Volume 1, ed. A.~A. {Galeev} \& R.~N. {Sudan}, 147

\bibitem[{{Hobara} {et~al.}(2010){Hobara}, {Balikhin}, {Krasnoselskikh},
  {Gedalin}, \& {Yamagishi}}]{hobara10a}
{Hobara}, Y., {Balikhin}, M., {Krasnoselskikh}, V., {Gedalin}, M., \&
  {Yamagishi}, H. 2010, J. Geophys. Res., 115, 11106,
  \dodoi{10.1029/2010JA015659}

\bibitem[{{Horaites} {et~al.}(2018){Horaites}, {Boldyrev}, {Wilson III},
  {Vi{\~n}as}, \& {Merka}}]{horaites18a}
{Horaites}, K., {Boldyrev}, S., {Wilson III}, L.~B., {Vi{\~n}as}, A.~F., \&
  {Merka}, J. 2018, Mon. Not. Roy. Astron. Soc., 474, 115,
  \dodoi{10.1093/mnras/stx2555}

\bibitem[{{Hull} \& {Scudder}(2000)}]{hull00b}
{Hull}, A.~J., \& {Scudder}, J.~D. 2000, J. Geophys. Res., 105, 27323,
  \dodoi{10.1029/2000JA900105}

\bibitem[{{Hull} {et~al.}(2000){Hull}, {Scudder}, {Fitzenreiter}, {Ogilvie},
  {Newbury}, \& {Russell}}]{hull00a}
{Hull}, A.~J., {Scudder}, J.~D., {Fitzenreiter}, R.~J., {et~al.} 2000, J.
  Geophys. Res., 105, 20957, \dodoi{10.1029/2000JA900049}

\bibitem[{{Hull} {et~al.}(1998){Hull}, {Scudder}, {Frank}, {Paterson}, \&
  {Kivelson}}]{hull98a}
{Hull}, A.~J., {Scudder}, J.~D., {Frank}, L.~A., {Paterson}, W.~R., \&
  {Kivelson}, M.~G. 1998, J. Geophys. Res., 103, 2041,
  \dodoi{10.1029/97JA03058}

\bibitem[{{Hull} {et~al.}(2001){Hull}, {Scudder}, {Larson}, \& {Lin}}]{hull01a}
{Hull}, A.~J., {Scudder}, J.~D., {Larson}, D.~E., \& {Lin}, R. 2001, J.
  Geophys. Res., 106, 15711, \dodoi{10.1029/2001JA900001}

\bibitem[{{Kajdi{\v c}} {et~al.}(2016){Kajdi{\v c}}, {Alexandrova},
  {Maksimovic}, {Lacombe}, \& {Fazakerley}}]{kajdic16a}
{Kajdi{\v c}}, P., {Alexandrova}, O., {Maksimovic}, M., {Lacombe}, C., \&
  {Fazakerley}, A.~N. 2016, Astrophys. J., 833, 172,
  \dodoi{10.3847/1538-4357/833/2/172}

\bibitem[{{Kasper} \& {Klein}(2019)}]{kasper19a}
{Kasper}, J.~C., \& {Klein}, K.~G. 2019, Astrophys. J. Lett., 877, L35,
  \dodoi{10.3847/2041-8213/ab1de5}

\bibitem[{{Kasper} {et~al.}(2006){Kasper}, {Lazarus}, {Steinberg}, {Ogilvie},
  \& {Szabo}}]{kasper06a}
{Kasper}, J.~C., {Lazarus}, A.~J., {Steinberg}, J.~T., {Ogilvie}, K.~W., \&
  {Szabo}, A. 2006, J. Geophys. Res., 111, 3105, \dodoi{10.1029/2005JA011442}

\bibitem[{{Kasper} {et~al.}(2013){Kasper}, {Maruca}, {Stevens}, \&
  {Zaslavsky}}]{kasper13a}
{Kasper}, J.~C., {Maruca}, B.~A., {Stevens}, M.~L., \& {Zaslavsky}, A. 2013,
  Phys. Rev. Lett., 110, 091102, \dodoi{10.1103/PhysRevLett.110.091102}

\bibitem[{{Kasper} {et~al.}(2012){Kasper}, {Stevens}, {Korreck}, {Maruca},
  {Kiefer}, {Schwadron}, \& {Lepri}}]{kasper12a}
{Kasper}, J.~C., {Stevens}, M.~L., {Korreck}, K.~E., {et~al.} 2012, Astrophys.
  J., 745, 162, \dodoi{10.1088/0004-637X/745/2/162}

\bibitem[{{Kasper} {et~al.}(2017){Kasper}, {Klein}, {Weber}, {Maksimovic},
  {Zaslavsky}, {Bale}, {Maruca}, {Stevens}, \& {Case}}]{kasper17a}
{Kasper}, J.~C., {Klein}, K.~G., {Weber}, T., {et~al.} 2017, ArXiv e-prints.
\newblock \doarXiv{1708.05683}

\bibitem[{{Kellogg}(1962)}]{kellogg62a}
{Kellogg}, P.~J. 1962, J. Geophys. Res., 67, 3805,
  \dodoi{10.1029/JZ067i010p03805}

\bibitem[{{Krall} \& {Trivelpiece}(1973)}]{krall73a}
{Krall}, N.~A., \& {Trivelpiece}, A.~W. 1973, {Principles of plasma physics}

\bibitem[{{Krauss-Varban} \& {Wu}(1989)}]{kraussvarban89b}
{Krauss-Varban}, D., \& {Wu}, C.~S. 1989, J. Geophys. Res., 94, 15367,
  \dodoi{10.1029/JA094iA11p15367}

\bibitem[{{Lacombe} {et~al.}(2017){Lacombe}, {Alexandrova}, \&
  {Matteini}}]{lacombe17a}
{Lacombe}, C., {Alexandrova}, O., \& {Matteini}, L. 2017, Astrophys. J., 848,
  45, \dodoi{10.3847/1538-4357/aa8c06}

\bibitem[{{Lacombe} {et~al.}(2014){Lacombe}, {Alexandrova}, {Matteini},
  {Santol{\'{\i}}k}, {Cornilleau-Wehrlin}, {Mangeney}, {de Conchy}, \&
  {Maksimovic}}]{lacombe14a}
{Lacombe}, C., {Alexandrova}, O., {Matteini}, L., {et~al.} 2014, Astrophys. J.,
  796, 5, \dodoi{10.1088/0004-637X/796/1/5}

\bibitem[{{Lazar} {et~al.}(2017){Lazar}, {Shaaban}, {Poedts}, \& {{\v
  S}tver{\'a}k}}]{lazar17b}
{Lazar}, M., {Shaaban}, S.~M., {Poedts}, S., \& {{\v S}tver{\'a}k}, {\v S}.
  2017, Mon. Not. Roy. Astron. Soc., 464, 564, \dodoi{10.1093/mnras/stw2336}

\bibitem[{{Lefebvre} {et~al.}(2007){Lefebvre}, {Schwartz}, {Fazakerley}, \&
  {D\'{e}cr\'{e}au}}]{lefebvre07a}
{Lefebvre}, B., {Schwartz}, S.~J., {Fazakerley}, A.~F., \& {D\'{e}cr\'{e}au},
  P. 2007, J. Geophys. Res., 112, 9212, \dodoi{10.1029/2007JA012277}

\bibitem[{{Lepping} {et~al.}(1995){Lepping}, {Ac{\~u}na}, {Burlaga}, {Farrell},
  {Slavin}, {Schatten}, {Mariani}, {Ness}, {Neubauer}, {Whang}, {Byrnes},
  {Kennon}, {Panetta}, {Scheifele}, \& {Worley}}]{lepping95}
{Lepping}, R.~P., {Ac{\~u}na}, M.~H., {Burlaga}, L.~F., {et~al.} 1995, Space
  Sci. Rev., 71, 207, \dodoi{10.1007/BF00751330}

\bibitem[{{Leroy} \& {Mangeney}(1984)}]{leroy84a}
{Leroy}, M.~M., \& {Mangeney}, A. 1984, Ann. Geophys., 2, 449

\bibitem[{{Lever} {et~al.}(2001){Lever}, {Quest}, \& {Shapiro}}]{lever01}
{Lever}, E.~L., {Quest}, K.~B., \& {Shapiro}, V.~D. 2001, Geophys. Res. Lett.,
  28, 1367, \dodoi{10.1029/2000GL012516}

\bibitem[{{Lin}(1998)}]{lin98a}
{Lin}, R.~P. 1998, Space Sci. Rev., 86, 61, \dodoi{10.1023/A:1005048428480}

\bibitem[{{Lin} {et~al.}(1995){Lin}, {Anderson}, {Ashford}, {Carlson},
  {Curtis}, {Ergun}, {Larson}, {McFadden}, {McCarthy}, {Parks}, {R\`{e}me},
  {Bosqued}, {Coutelier}, {Cotin}, {D'Uston}, {Wenzel}, {Sanderson}, {Henrion},
  {Ronnet}, \& {Paschmann}}]{lin95a}
{Lin}, R.~P., {Anderson}, K.~A., {Ashford}, S., {et~al.} 1995, Space Sci. Rev.,
  71, 125, \dodoi{10.1007/BF00751328}

\bibitem[{{Livadiotis}(2015)}]{livadiotis15a}
{Livadiotis}, G. 2015, J. Geophys. Res., 120, 1607,
  \dodoi{10.1002/2014JA020825}

\bibitem[{{Livadiotis}(2017)}]{livadiotis17b}
---. 2017, J. Phys. Conf. Ser., 900, 012014,
  \dodoi{10.1088/1742-6596/900/1/012014}

\bibitem[{{Maksimovic} {et~al.}(1997){Maksimovic}, {Pierrard}, \&
  {Riley}}]{maksimovic97a}
{Maksimovic}, M., {Pierrard}, V., \& {Riley}, P. 1997, Geophys. Res. Lett., 24,
  1151, \dodoi{10.1029/97GL00992}

\bibitem[{{Maksimovic} {et~al.}(2005){Maksimovic}, {Zouganelis}, {Chaufray},
  {Issautier}, {Scime}, {Littleton}, {Marsch}, {McComas}, {Salem}, {Lin}, \&
  {Elliott}}]{maksimovic05a}
{Maksimovic}, M., {Zouganelis}, I., {Chaufray}, J.-Y., {et~al.} 2005, J.
  Geophys. Res., 110, 9104, \dodoi{10.1029/2005JA011119}

\bibitem[{{Marsch}(2006)}]{marsch06a}
{Marsch}, E. 2006, Living Reviews in Solar Physics, 3, 1

\bibitem[{{Marsch} \& {Livi}(1985)}]{marsch85a}
{Marsch}, E., \& {Livi}, S. 1985, Phys. Fluids, 28, 1379,
  \dodoi{10.1063/1.864971}

\bibitem[{{Maruca} \& {Kasper}(2013)}]{maruca13a}
{Maruca}, B.~A., \& {Kasper}, J.~C. 2013, Adv. Space Res., 52, 723,
  \dodoi{10.1016/j.asr.2013.04.006}

\bibitem[{{Maruca} {et~al.}(2011){Maruca}, {Kasper}, \& {Bale}}]{maruca11a}
{Maruca}, B.~A., {Kasper}, J.~C., \& {Bale}, S.~D. 2011, Phys. Rev. Lett., 107,
  201101, \dodoi{10.1103/PhysRevLett.107.201101}

\bibitem[{{Masters} {et~al.}(2011){Masters}, {Schwartz}, {Henley}, {Thomsen},
  {Zieger}, {Coates}, {Achilleos}, {Mitchell}, {Hansen}, \&
  {Dougherty}}]{masters11a}
{Masters}, A., {Schwartz}, S.~J., {Henley}, E.~M., {et~al.} 2011, J. Geophys.
  Res., 116, A10107, \dodoi{10.1029/2011JA016941}

\bibitem[{{Mazelle} {et~al.}(2010){Mazelle}, {Lemb\`{e}ge}, {Morgenthaler},
  {Meziane}, {Horbury}, {G\'{e}not}, {Lucek}, \& {Dandouras}}]{mazelle10a}
{Mazelle}, C., {Lemb\`{e}ge}, B., {Morgenthaler}, A., {et~al.} 2010, Proc. 12th
  Intl. Solar Wind Conf., 1216, 471, \dodoi{10.1063/1.3395905}

\bibitem[{{Nicolaou} {et~al.}(2018){Nicolaou}, {Livadiotis}, {Owen},
  {Verscharen}, \& {Wicks}}]{nicolaou18a}
{Nicolaou}, G., {Livadiotis}, G., {Owen}, C.~J., {Verscharen}, D., \& {Wicks},
  R.~T. 2018, Astrophys. J., 864, 3, \dodoi{10.3847/1538-4357/aad45d}

\bibitem[{{Nieves-Chinchilla} \& {Vi{\~n}as}(2008)}]{nieveschinchilla08a}
{Nieves-Chinchilla}, T., \& {Vi{\~n}as}, A.~F. 2008, J. Geophys. Res., 113,
  A02105, \dodoi{10.1029/2007JA012703}

\bibitem[{{Ogilvie} {et~al.}(2000){Ogilvie}, {Fitzenreiter}, \&
  {Desch}}]{ogilvie00a}
{Ogilvie}, K.~W., {Fitzenreiter}, R., \& {Desch}, M. 2000, J. Geophys. Res.,
  105, 27277, \dodoi{10.1029/2000JA000131}

\bibitem[{{Ogilvie} {et~al.}(1995){Ogilvie}, {Chornay}, {Fritzenreiter},
  {Hunsaker}, {Keller}, {Lobell}, {Miller}, {Scudder}, {Sittler}, {Torbert},
  {Bodet}, {Needell}, {Lazarus}, {Steinberg}, {Tappan}, {Mavretic}, \&
  {Gergin}}]{ogilvie95}
{Ogilvie}, K.~W., {Chornay}, D.~J., {Fritzenreiter}, R.~J., {et~al.} 1995,
  Space Sci. Rev., 71, 55, \dodoi{10.1007/BF00751326}

\bibitem[{{Pagel} {et~al.}(2005{\natexlab{a}}){Pagel}, {Crooker}, \&
  {Larson}}]{pagel05b}
{Pagel}, C., {Crooker}, N.~U., \& {Larson}, D.~E. 2005{\natexlab{a}}, Geophys.
  Res. Lett., 32, 14105, \dodoi{10.1029/2005GL023043}

\bibitem[{{Pagel} {et~al.}(2005{\natexlab{b}}){Pagel}, {Crooker}, {Larson},
  {Kahler}, \& {Owens}}]{pagel05a}
{Pagel}, C., {Crooker}, N.~U., {Larson}, D.~E., {Kahler}, S.~W., \& {Owens},
  M.~J. 2005{\natexlab{b}}, J. Geophys. Res., 110, 1103,
  \dodoi{10.1029/2004JA010767}

\bibitem[{{Pagel} {et~al.}(2007){Pagel}, {Gary}, {de Koning}, {Skoug}, \&
  {Steinberg}}]{pagel07}
{Pagel}, C., {Gary}, S.~P., {de Koning}, C.~A., {Skoug}, R.~M., \& {Steinberg},
  J.~T. 2007, J. Geophys. Res., 112, 4103, \dodoi{10.1029/2006JA011967}

\bibitem[{{Park} {et~al.}(2015){Park}, {Caprioli}, \& {Spitkovsky}}]{park15a}
{Park}, J., {Caprioli}, D., \& {Spitkovsky}, A. 2015, Phys. Rev. Lett., 114,
  085003, \dodoi{10.1103/PhysRevLett.114.085003}

\bibitem[{{Paschmann} \& {Daly}(1998)}]{paschmann98a}
{Paschmann}, G., \& {Daly}, P.~W. 1998, ISSI Sci. Rep. Ser., 1

\bibitem[{{Petschek}(1958)}]{petschek58}
{Petschek}, H.~E. 1958, Rev. Mod. Phys., 30, 966,
  \dodoi{10.1103/RevModPhys.30.966}

\bibitem[{{Phillips} {et~al.}(1989{\natexlab{a}}){Phillips}, {Gosling},
  {McComas}, {Bame}, {Gary}, \& {Smith}}]{phillips89a}
{Phillips}, J.~L., {Gosling}, J.~T., {McComas}, D.~J., {et~al.}
  1989{\natexlab{a}}, J. Geophys. Res., 94, 6563,
  \dodoi{10.1029/JA094iA06p06563}

\bibitem[{{Phillips} {et~al.}(1989{\natexlab{b}}){Phillips}, {Gosling},
  {McComas}, {Bame}, \& {Smith}}]{phillips89b}
{Phillips}, J.~L., {Gosling}, J.~T., {McComas}, D.~J., {Bame}, S.~J., \&
  {Smith}, E.~J. 1989{\natexlab{b}}, J. Geophys. Res., 94, 13377,
  \dodoi{10.1029/JA094iA10p13377}

\bibitem[{{Pilipp} {et~al.}(1990){Pilipp}, {Muehlhaeuser}, {Miggenrieder},
  {Rosenbauer}, \& {Schwenn}}]{pilipp90a}
{Pilipp}, W.~G., {Muehlhaeuser}, K., {Miggenrieder}, H., {Rosenbauer}, H., \&
  {Schwenn}, R. 1990, J. Geophys. Res., 95, 6305,
  \dodoi{10.1029/JA095iA05p06305}

\bibitem[{{Pulupa} {et~al.}(2010){Pulupa}, {Bale}, \& {Kasper}}]{pulupa10a}
{Pulupa}, M.~P., {Bale}, S.~D., \& {Kasper}, J.~C. 2010, J. Geophys. Res., 115,
  4106, \dodoi{10.1029/2009JA014680}

\bibitem[{{Pulupa} {et~al.}(2014){Pulupa}, {Bale}, {Salem}, \&
  {Horaites}}]{pulupa14a}
{Pulupa}, M.~P., {Bale}, S.~D., {Salem}, C., \& {Horaites}, K. 2014, J.
  Geophys. Res., 119, 647, \dodoi{10.1002/2013JA019359}

\bibitem[{{Roberg-Clark} {et~al.}(2016){Roberg-Clark}, {Drake}, {Reynolds}, \&
  {Swisdak}}]{robergclark16a}
{Roberg-Clark}, G.~T., {Drake}, J.~F., {Reynolds}, C.~S., \& {Swisdak}, M.
  2016, Astrophys. J. Lett., 830, L9, \dodoi{10.3847/2041-8205/830/1/L9}

\bibitem[{{Roberg-Clark} {et~al.}(2018){Roberg-Clark}, {Drake}, {Swisdak}, \&
  {Reynolds}}]{robergclark18b}
{Roberg-Clark}, G.~T., {Drake}, J.~F., {Swisdak}, M., \& {Reynolds}, C.~S.
  2018, Astrophys. J., 867, 154, \dodoi{10.3847/1538-4357/aae393}

\bibitem[{{Sagdeev}(1966)}]{sagdeev66}
{Sagdeev}, R.~Z. 1966, Rev. Plasma Phys., 4, 23

\bibitem[{{Salem} {et~al.}(2001){Salem}, {Bosqued}, {Larson}, {Mangeney},
  {Maksimovic}, {Perche}, {Lin}, \& {Bougeret}}]{salem01a}
{Salem}, C., {Bosqued}, J.-M., {Larson}, D.~E., {et~al.} 2001, J. Geophys.
  Res., 106, 21701, \dodoi{10.1029/2001JA900031}

\bibitem[{{Salem} {et~al.}(2003){Salem}, {Hubert}, {Lacombe}, {Bale},
  {Mangeney}, {Larson}, \& {Lin}}]{salem03a}
{Salem}, C., {Hubert}, D., {Lacombe}, C., {et~al.} 2003, Astrophys. J., 585,
  1147, \dodoi{10.1086/346185}

\bibitem[{{Schunk}(1975)}]{schunk75a}
{Schunk}, R.~W. 1975, Planet. Space Sci., 23, 437,
  \dodoi{10.1016/0032-0633(75)90118-X}

\bibitem[{{Schunk}(1977)}]{schunk77a}
---. 1977, Rev. Geophys. Space Phys., 15, 429, \dodoi{10.1029/RG015i004p00429}

\bibitem[{{Schwartz} {et~al.}(2011){Schwartz}, {Henley}, {Mitchell}, \&
  {Krasnoselskikh}}]{schwartz11a}
{Schwartz}, S.~J., {Henley}, E., {Mitchell}, J., \& {Krasnoselskikh}, V. 2011,
  Phys. Rev. Lett., 107, 215002, \dodoi{10.1103/PhysRevLett.107.215002}

\bibitem[{{Schwartz} \& {Marsch}(1983)}]{schwartz83b}
{Schwartz}, S.~J., \& {Marsch}, E. 1983, J. Geophys. Res., 88, 9919,
  \dodoi{10.1029/JA088iA12p09919}

\bibitem[{{Schwartz} {et~al.}(1988){Schwartz}, {Thomsen}, {Bame}, \&
  {Stansberry}}]{schwartz88a}
{Schwartz}, S.~J., {Thomsen}, M.~F., {Bame}, S.~J., \& {Stansberry}, J. 1988,
  J. Geophys. Res., 93, 12923, \dodoi{10.1029/JA093iA11p12923}

\bibitem[{{Schwenn}(1990)}]{schwenn90a}
{Schwenn}, R. 1990, {Large-Scale Structure of the Interplanetary Medium}, ed.
  R.~{Schwenn} \& E.~{Marsch}, 99

\bibitem[{{Scime} {et~al.}(1994){Scime}, {Phillips}, \& {Bame}}]{scime94a}
{Scime}, E.~E., {Phillips}, J.~L., \& {Bame}, S.~J. 1994, J. Geophys. Res., 99,
  14769, \dodoi{10.1029/94JA00489}

\bibitem[{{Scudder} \& {Karimabadi}(2013)}]{scudder13a}
{Scudder}, J.~D., \& {Karimabadi}, H. 2013, Astrophys. J., 770, 26,
  \dodoi{10.1088/0004-637X/770/1/26}

\bibitem[{{Scudder} \& {Olbert}(1979)}]{scudder79a}
{Scudder}, J.~D., \& {Olbert}, S. 1979, J. Geophys. Res., 84, 6603,
  \dodoi{10.1029/JA084iA11p06603}

\bibitem[{{Shizgal}(2018)}]{shizgal18a}
{Shizgal}, B.~D. 2018, Phys. Rev. E, 97, 052144,
  \dodoi{10.1103/PhysRevE.97.052144}

\bibitem[{{Skoug} {et~al.}(2000){Skoug}, {Feldman}, {Gosling}, {McComas}, \&
  {Smith}}]{skoug00a}
{Skoug}, R.~M., {Feldman}, W.~C., {Gosling}, J.~T., {McComas}, D.~J., \&
  {Smith}, C.~W. 2000, J. Geophys. Res., 105, 23069,
  \dodoi{10.1029/2000JA000017}

\bibitem[{{Song} {et~al.}(1997){Song}, {Zhang}, \& {Paschmann}}]{song97a}
{Song}, P., {Zhang}, X.~X., \& {Paschmann}, G. 1997, Planet. Space Sci., 45,
  255, \dodoi{10.1016/S0032-0633(96)00087-6}

\bibitem[{{Spitzer} \& {H{\"a}rm}(1953)}]{spitzer53a}
{Spitzer}, L., \& {H{\"a}rm}, R. 1953, Phys. Rev., 89, 977,
  \dodoi{10.1103/PhysRev.89.977}

\bibitem[{{\v{S}tver\'{a}k} {et~al.}(2009){\v{S}tver\'{a}k}, {Maksimovic},
  {Tr\'{a}vn{\'{\i}}\v{c}ek}, {Marsch}, {Fazakerley}, \& {Scime}}]{stverak09a}
{\v{S}tver\'{a}k}, v., {Maksimovic}, M., {Tr\'{a}vn{\'{\i}}\v{c}ek}, P.~M.,
  {et~al.} 2009, J. Geophys. Res., 114, 5104, \dodoi{10.1029/2008JA013883}

\bibitem[{{\v{S}tver\'{a}k} {et~al.}(2008){\v{S}tver\'{a}k},
  {Tr\'{a}vn{\'{\i}}\v{c}ek}, {Maksimovic}, {Marsch}, {Fazakerley}, \&
  {Scime}}]{stverak08a}
{\v{S}tver\'{a}k}, v., {Tr\'{a}vn{\'{\i}}\v{c}ek}, P., {Maksimovic}, M.,
  {et~al.} 2008, J. Geophys. Res., 113, 3103, \dodoi{10.1029/2007JA012733}

\bibitem[{{Tao} {et~al.}(2016{\natexlab{a}}){Tao}, {Wang}, {Zong}, {Li},
  {Salem}, {Wimmer-Schweingruber}, {He}, {Tu}, \& {Bale}}]{tao16a}
{Tao}, J., {Wang}, L., {Zong}, Q., {et~al.} 2016{\natexlab{a}}, Astrophys. J.,
  820, 22, \dodoi{10.3847/0004-637X/820/1/22}

\bibitem[{{Tao} {et~al.}(2016{\natexlab{b}}){Tao}, {Wang}, {Zong}, {Li},
  {Salem}, {Wimmer-Schweingruber}, {He}, {Tu}, \& {Bale}}]{tao16b}
{Tao}, J., {Wang}, L., {Zong}, Q., {et~al.} 2016{\natexlab{b}}, in American
  Institute of Physics Conference Series, Vol. 1720, American Institute of
  Physics Conference Series, 070006

\bibitem[{{Thomsen} {et~al.}(1985){Thomsen}, {Gosling}, {Bame}, \&
  {Mellott}}]{thomsen85a}
{Thomsen}, M.~F., {Gosling}, J.~T., {Bame}, S.~J., \& {Mellott}, M.~M. 1985, J.
  Geophys. Res., 90, 137, \dodoi{10.1029/JA090iA01p00137}

\bibitem[{{Thomsen} {et~al.}(1993){Thomsen}, {Gosling}, {Onsager}, \&
  {Russell}}]{thomsen93a}
{Thomsen}, M.~F., {Gosling}, J.~T., {Onsager}, T.~G., \& {Russell}, C.~T. 1993,
  J. Geophys. Res., 98, 3875, \dodoi{10.1029/92JA02560}

\bibitem[{{Thomsen} {et~al.}(1987){Thomsen}, {Stansberry}, {Bame}, {Gosling},
  \& {Mellott}}]{thomsen87b}
{Thomsen}, M.~F., {Stansberry}, J.~A., {Bame}, S.~J., {Gosling}, J.~T., \&
  {Mellott}, M.~M. 1987, J. Geophys. Res., 92, 10119,
  \dodoi{10.1029/JA092iA09p10119}

\bibitem[{{Tong} {et~al.}(2018){Tong}, {Bale}, {Salem}, \& {Pulupa}}]{tong18b}
{Tong}, Y., {Bale}, S.~D., {Salem}, C., \& {Pulupa}, M. 2018, arXiv e-prints,
  arXiv:1801.07694.
\newblock \doarXiv{1801.07694}

\bibitem[{{Tong} {et~al.}(2019{\natexlab{a}}){Tong}, {Vasko}, {Artemyev},
  {Bale}, \& {Mozer}}]{tong19b}
{Tong}, Y., {Vasko}, I.~Y., {Artemyev}, A.~V., {Bale}, S.~D., \& {Mozer}, F.~S.
  2019{\natexlab{a}}, Astrophys. J., 878, 41, \dodoi{10.3847/1538-4357/ab1f05}

\bibitem[{{Tong} {et~al.}(2019{\natexlab{b}}){Tong}, {Vasko}, {Pulupa},
  {Mozer}, {Bale}, {Artemyev}, \& {Krasnoselskikh}}]{tong19a}
{Tong}, Y., {Vasko}, I.~Y., {Pulupa}, M., {et~al.} 2019{\natexlab{b}},
  Astrophys. J. Lett., 870, L6, \dodoi{10.3847/2041-8213/aaf734}

\bibitem[{{Vandas}(2001)}]{vandas01a}
{Vandas}, M. 2001, J. Geophys. Res., 106, 1859, \dodoi{10.1029/2000JA900128}

\bibitem[{{Vasko} {et~al.}(2019){Vasko}, {Krasnoselskikh}, {Tong}, {Bale},
  {Bonnell}, \& {Mozer}}]{vasko19a}
{Vasko}, I.~Y., {Krasnoselskikh}, V., {Tong}, Y., {et~al.} 2019, Astrophys. J.
  Lett., 871, L29, \dodoi{10.3847/2041-8213/ab01bd}

\bibitem[{{Vi{\~n}as} {et~al.}(2010){Vi{\~n}as}, {Gurgiolo},
  {Nieves-Chinchilla}, {Gary}, \& {Goldstein}}]{vinas10a}
{Vi{\~n}as}, A.~F., {Gurgiolo}, C., {Nieves-Chinchilla}, T., {Gary}, S.~P., \&
  {Goldstein}, M.~L. 2010, Proc. 12th Intl. Solar Wind Conf., 1216, 265,
  \dodoi{10.1063/1.3395852}

\bibitem[{{Walsh} {et~al.}(2013){Walsh}, {Arridge}, {Masters}, {Lewis},
  {Fazakerley}, {Jones}, {Owen}, \& {Coates}}]{walsh13a}
{Walsh}, A.~P., {Arridge}, C.~S., {Masters}, A., {et~al.} 2013, Geophys. Res.
  Lett., 40, 2495, \dodoi{10.1002/grl.50529}

\bibitem[{{Wang} {et~al.}(2012){Wang}, {Lin}, {Salem}, {Pulupa}, {Larson},
  {Yoon}, \& {Luhmann}}]{wang12a}
{Wang}, L., {Lin}, R.~P., {Salem}, C., {et~al.} 2012, Astrophys. J. Lett., 753,
  L23, \dodoi{10.1088/2041-8205/753/1/L23}

\bibitem[{{Wicks} {et~al.}(2016){Wicks}, {Alexander}, {Stevens}, {Wilson III},
  {Moya}, {Vi{\~n}as}, {Jian}, {Roberts}, {O'Modhrain}, {Gilbert}, \&
  {Zurbuchen}}]{wicks16a}
{Wicks}, R.~T., {Alexander}, R.~L., {Stevens}, M.~L., {et~al.} 2016, Astrophys.
  J., 819, 6, \dodoi{10.3847/0004-637X/819/1/6}

\bibitem[{{Wilson III} {et~al.}(2007){Wilson III}, {Cattell}, {Kellogg},
  {Goetz}, {Kersten}, {Hanson}, {MacGregor}, \& {Kasper}}]{wilsoniii07a}
{Wilson III}, L.~B., {Cattell}, C., {Kellogg}, P.~J., {et~al.} 2007, Phys. Rev.
  Lett., 99, 041101, \dodoi{10.1103/PhysRevLett.99.041101}

\bibitem[{{Wilson III} {et~al.}(2009){Wilson III}, {Cattell}, {Kellogg},
  {Goetz}, {Kersten}, {Kasper}, {Szabo}, \& {Meziane}}]{wilsoniii09a}
{Wilson III}, L.~B., {Cattell}, C.~A., {Kellogg}, P.~J., {et~al.} 2009, J.
  Geophys. Res., 114, 10106, \dodoi{10.1029/2009JA014376}

\bibitem[{{Wilson III} {et~al.}(2010){Wilson III}, {Cattell}, {Kellogg},
  {Goetz}, {Kersten}, {Kasper}, {Szabo}, \& {Wilber}}]{wilsoniii10a}
---. 2010, J. Geophys. Res., 115, 12104, \dodoi{10.1029/2010JA015332}

\bibitem[{{Wilson III} {et~al.}(2017){Wilson III}, {Koval}, {Szabo}, {Stevens},
  {Kasper}, {Cattell}, \& {Krasnoselskikh}}]{wilsoniii17c}
{Wilson III}, L.~B., {Koval}, A., {Szabo}, A., {et~al.} 2017, J. Geophys. Res.,
  122, 9115, \dodoi{10.1002/2017JA024352}

\bibitem[{{Wilson III} {et~al.}(2014{\natexlab{a}}){Wilson III}, {Sibeck},
  {Breneman}, {Le Contel}, {Cully}, {Turner}, {Angelopoulos}, \&
  {Malaspina}}]{wilsoniii14a}
{Wilson III}, L.~B., {Sibeck}, D.~G., {Breneman}, A.~W., {et~al.}
  2014{\natexlab{a}}, J. Geophys. Res., 119, 6455, \dodoi{10.1002/2014JA019929}

\bibitem[{{Wilson III} {et~al.}(2014{\natexlab{b}}){Wilson III}, {Sibeck},
  {Breneman}, {Le Contel}, {Cully}, {Turner}, {Angelopoulos}, \&
  {Malaspina}}]{wilsoniii14b}
---. 2014{\natexlab{b}}, J. Geophys. Res., 119, 6475,
  \dodoi{10.1002/2014JA019930}

\bibitem[{{Wilson III} {et~al.}(2012){Wilson III}, {Koval}, {Szabo},
  {Breneman}, {Cattell}, {Goetz}, {Kellogg}, {Kersten}, {Kasper}, {Maruca}, \&
  {Pulupa}}]{wilsoniii12c}
{Wilson III}, L.~B., {Koval}, A., {Szabo}, A., {et~al.} 2012, Geophys. Res.
  Lett., 39, 8109, \dodoi{10.1029/2012GL051581}

\bibitem[{{Wilson III} {et~al.}(2013{\natexlab{a}}){Wilson III}, {Koval},
  {Szabo}, {Breneman}, {Cattell}, {Goetz}, {Kellogg}, {Kersten}, {Kasper},
  {Maruca}, \& {Pulupa}}]{wilsoniii13a}
---. 2013{\natexlab{a}}, J. Geophys. Res., 118, 5, \dodoi{10.1029/2012JA018167}

\bibitem[{{Wilson III} {et~al.}(2013{\natexlab{b}}){Wilson III}, {Koval},
  {Sibeck}, {Szabo}, {Cattell}, {Kasper}, {Maruca}, {Pulupa}, {Salem}, \&
  {Wilber}}]{wilsoniii13b}
{Wilson III}, L.~B., {Koval}, A., {Sibeck}, D.~G., {et~al.} 2013{\natexlab{b}},
  J. Geophys. Res., 118, 957, \dodoi{10.1029/2012JA018186}

\bibitem[{{Wilson III} {et~al.}(2018){Wilson III}, {Stevens}, {Kasper},
  {Klein}, {Maruca}, {Bale}, {Bowen}, {Pulupa}, \& {Salem}}]{wilsoniii18b}
{Wilson III}, L.~B., {Stevens}, M.~L., {Kasper}, J.~C., {et~al.} 2018,
  Astrophys. J. Suppl., 236, 41, \dodoi{10.3847/1538-4365/aab71c}

\bibitem[{{Wilson III} {et~al.}(2019{\natexlab{a}}){Wilson III}, {Chen},
  {Wang}, {Schwartz}, {Turner}, {Stevens}, {Kasper}, {Osmane}, {Caprioli},
  {Bale}, {Pulupa}, {Salem}, \& {Goodrich}}]{wilsoniii19a}
{Wilson III}, L.~B., {Chen}, L.-J., {Wang}, S., {et~al.} 2019{\natexlab{a}},
  Astrophys. J. Suppl., 243, \dodoi{10.3847/1538-4365/ab22bd}

\bibitem[{{Wilson III} {et~al.}(2019{\natexlab{b}}){Wilson III}, {Chen},
  {Wang}, {Schwartz}, {Turner}, {Stevens}, {Kasper}, {Osmane}, {Caprioli},
  {Bale}, {Pulupa}, {Salem}, \& {Goodrich}}]{wilsoniii19c}
---. 2019{\natexlab{b}}, Astrophys. J.

\bibitem[{{Wilson III} {et~al.}(2019{\natexlab{c}}){Wilson III}, {Chen},
  {Wang}, {Schwartz}, {Turner}, {Stevens}, {Kasper}, {Osmane}, {Caprioli},
  {Bale}, {Pulupa}, {Salem}, \& {Goodrich}}]{wilsoniii19k}
---. 2019{\natexlab{c}}, {Supplement to: Electron energy partition across
  interplanetary shocks}, 1.0,  Zenodo, \dodoi{10.5281/zenodo.2875806}.
\newblock \url{https://doi.org/10.5281/zenodo.2875806}

\bibitem[{{Wu}(1984)}]{wu84b}
{Wu}, C.~S. 1984, J. Geophys. Res., 89, 8857, \dodoi{10.1029/JA089iA10p08857}

\end{thebibliography}

\end{document}